\documentclass[preprint2]{emulateapj}
\usepackage{longtable}
\usepackage{cancel}	
\usepackage{color}
\usepackage{graphicx, subfigure}
\usepackage{placeins}
\usepackage{amsmath}
\usepackage{tabulary}
\usepackage{ulem}
\usepackage{tablefootnote}

\def\lesssim{\mathrel{\hbox{\rlap{\hbox{\lower4pt\hbox{$\sim$}}}\hbox{$<$}}}}
\def\gtrsim{\mathrel{\hbox{\rlap{\hbox{\lower4pt\hbox{$\sim$}}}\hbox{$>$}}}}
\newcommand{\gps}{\ensuremath{g_{\rm P1}}}
\newcommand{\rps}{\ensuremath{r_{\rm P1}}}
\newcommand{\ips}{\ensuremath{i_{\rm P1}}}
\newcommand{\zps}{\ensuremath{z_{\rm P1}}}
\newcommand{\yps}{\ensuremath{y_{\rm P1}}}
\newcommand{\wps}{\ensuremath{w_{\rm P1}}}
\newcommand{\grizy}{\ensuremath{grizy_{\rm P1}}}

\newcommand{\PS}{\protect \hbox {Pan-STARRS1}}

\newcommand{\degree}{\mbox{$^\circ$}}

\shorttitle{Pan-STARRS1 Surveys}
\shortauthors{K. C. Chambers et al.}

\begin{document}


\title{The Pan-STARRS1 Surveys}



\def\IfA{1}
\def\DUR{2}
\def\Harvard{3}
\def\ITC{4}
\def\MPIA{5}
\def\Princeton{6}
\def\MPE{7}
\def\Hubble{8}
\def\LBNL{9}
\def\QUB{10}
\def\GMTO{11}
\def\PSI{12}
\def\JHU{13}
\def\BYO{14}
\def\GLS{15}
\def\GAL{16}
\def\USNO{17}
\def\CIW{18}
\def\CPFellow{19}
\def\UMich{20}
\def\UCA{21}
\def\INAF{22}
\def\STScI{23}
\def\NCU{24}
\def\Google{25}
\def\CAR{26} 
\def\UMary{27}
\def\ARI{28}
\def\UE{29}
\def\ICIC{30}
\def\Spire{31}
\def\CfA{32}
\def\CNRS{33}
\def\NCSA{34}
\def\Camb{35}
\def\ICC{36}
\def\Ocean{37}
\def\NZ{38}
\def\Munich{39}
\def\SVVS{40}
\def\USRA{41} 
\def\PITT{42}

\author{ 
K. C. Chambers\altaffilmark{\IfA},
E. A. Magnier\altaffilmark{\IfA},
N. Metcalfe\altaffilmark{\DUR},
%
%
H. A. Flewelling\altaffilmark{\IfA},
M. E. Huber\altaffilmark{\IfA},
C. Z. Waters\altaffilmark{\IfA},
%
%
L. Denneau\altaffilmark{\IfA},
P. W. Draper\altaffilmark{\DUR},
D. Farrow\altaffilmark{\DUR,\MPE},
D. P. Finkbeiner\altaffilmark{\Harvard,\ITC}
% ...
}
\altaffiltext{\IfA}{Institute of Astronomy, University of Hawaii, 2680 Woodlawn Drive, Honolulu, Hawaii 96822, USA}
\altaffiltext{\DUR}{Department of Physics, Durham University, South Road, Durham DH1 3LE, UK}
\altaffiltext{\Harvard}{Department of Physics, Harvard University, Cambridge, MA 02138, USA}
\altaffiltext{\ITC}{Institute for Theory and Computation,  Harvard-Smithsonian Center for Astrophysics,   60 Garden Street, MS-51, Cambridge, MA 02138 USA} 
\altaffiltext{\MPIA}{Max-Planck-Institut f\"ur Astronomie, K\"onigstuhl 17, D-69117 Heidelberg, Germany}
\altaffiltext{\Princeton}{Department of Astrophysical Sciences, Princeton University, Princeton, NJ 08544, USA}
\altaffiltext{\MPE}{ Max-Planck Institut f\"ur extraterrestrische Physik, Giessenbachstrasse 1, D-85748 Garching, Germany}
\altaffiltext{\STScI}{Space Telescope Science Institute, 3700 San Martin Drive, Baltimore, MD 21218, USA}
\altaffiltext{\Hubble}{Hubble Fellow}
\altaffiltext{\LBNL}{Lawrence Berkeley National Laboratory, One Cyclotron
Road, Berkeley, CA 94720, USA}
\altaffiltext{\QUB}{Astrophysics Research Centre, School of Mathematics and Physics, Queens University Belfast, Belfast BT7 1NN, UK}
\altaffiltext{\GMTO}{GMTO Corp., 465 N. Halstead St. Suite 250, Pasadena, CA  91107, USA}
\altaffiltext{\JHU}{Department of Physics and Astronomy, Johns Hopkins University, 3400 North Charles Street, Baltimore, MD 21218, USA}
\altaffiltext{\BYO}{Back Yard Observatory, P.O. BOX 68856, Tucson, AZ 85737, USA}
\altaffiltext{\GLS}{G.L. Scientific, 3367 Waialae Avenue, Honolulu, HI 96816, USA}
\altaffiltext{\GAL}{deceased}
\altaffiltext{\USNO}{US Naval Observatory, Flagstaff Station, Flagstaff, AZ 86001, USA}
\altaffiltext{\CIW}{The Observatories of the Carnegie Institute of Washington, 813 Santa Barbara Street, Pasadena, CA 91101, USA}
\altaffiltext{\CPFellow}{Carnegie-Princeton Fellow}
\altaffiltext{\UMich}{Department of Astronomy, University of Michigan, USA}
\altaffiltext{\UCA}{Universit\'e C\^ote d'Azur, OCA, CNRS, Lagrange, France}
\altaffiltext{\NCU}{Graduate Institute of Astronomy, National Central University,   300 Zhonda Road, Zhongli, Taoyuan 32001, Taiwan }
\altaffiltext{\CAR}{Centre for Astrophysics Research, University of Hertfordshire, College Lane Campus, Hatfield, AL10 9AB, UK}
\altaffiltext{\UMary}{Department of Astronomy, University of Maryland, College Park, MD  20742 USA}
\altaffiltext{\UE}{Institute for Astronomy, School of Physics and Astronomy, University of Edinburgh,
Royal Observatory, Blackford Hill, Edinburgh, EH9~3HJ, UK}
\altaffiltext{\ICIC}{ICIC, Imperial College, Blackett Laboratory, London SW7 2AZ}
\altaffiltext{\ARI}{Astronomisches Rechen-Institut, Zentrum f\"ur Astronomie der Universit\"at Heidelberg, M\"onchhofstr.\ 12--14, 69120 Heidelberg, Germany}
\altaffiltext{\CfA}{Harvard-Smithsonian Center for Astrophysics, 60 Garden Street, Cambridge, MA 02138, USA}
\altaffiltext{\CNRS}{Universite de Strasbourg, CNRS, Observatoire astronomique de Strasbourg, UMR 7550, F-67000 Strasbourg, France}
\altaffiltext{\NCSA}{National Center for Supercomputing Applications, University of Illinois at Urbana-Champaign, 1205 W. Clark Street, Urbana, IL 61801, USA}
\altaffiltext{\Camb}{Institute of Astronomy, University of Cambridge, Madingley Road, Cambridge CB3 0HA, UK}
\altaffiltext{\ICC}{Institute for Computational Cosmology, Department of Physics, Durham University, South Road, Durham DH1 3LE, UK}
\altaffiltext{\PITT}{Pittsburgh Particle Physics, Astrophysics, and Cosmology Center (PITT PACC). Physics and Astronomy Department, University of Pittsburgh, Pittsburgh, PA 15260, USA}
\altaffiltext{\USRA}{Universities Space Research Association, 7178 Columbia Gateway Drive, Columbia, MD 21046}
\altaffiltext{\Ocean}{Oceanit, 828 Fort Street Mall, Suite 600, Honolulu, HI, 96813, USA}
\altaffiltext{\NZ}{School of Chemical \& Physical Sciences, Victoria University of Wellington, PO Box 600, Wellington 6140, New Zealand}
\altaffiltext{\Google}{Google Inc., 1600 Amphitheatre Pkwy, Mountain View, CA 94043, USA}
\altaffiltext{\Spire}{Spire Global, Sky Park 5,45 Finnieston Street, Glasgow, G3 8JU, UK }
\altaffiltext{\SVVS}{St. Vrain Valley School, 3180 County Road 5, Erie, CO 80516, USA}
\altaffiltext{\PSI}{Planetary Science Institute, 1700 East Fort Lowell, Suite 106, Tucson, AZ 85719, USA}
\altaffiltext{\Munich}{University Observatory, Scheinerstrasse 1, 81679 Munich, Germany}
\altaffiltext{\INAF}{INAF-Osservatorio di Capodimonte Salita Moiariello, 16 80131  Naples, Italy}

\begin{abstract}

Pan-STARRS1 has carried out a set of distinct synoptic imaging sky surveys including the $3\pi$ Steradian Survey and the Medium Deep Survey in 5 bands (\grizy).
The mean 5$\sigma$ point source limiting sensitivities in the stacked 3$\pi$ Steradian Survey in $grizy_{P1}$ 
are (23.3, 23.2, 23.1, 22.3, 21.4) respectively. 
The upper bound on the systematic uncertainty in 
the photometric calibration across the sky is 7-12 millimag depending on the bandpass. 
The systematic uncertainty of the astrometric calibration using the Gaia frame comes from a comparison of the results with Gaia: 
the standard deviation of the mean and median residuals ($ \Delta ra, \Delta dec $ ) are
(2.3, 1.7) milliarcsec, and (3.1, 4.8) milliarcsec respectively.
The Pan-STARRS system and the design of the PS1 surveys are described and an overview of the resulting image and catalog data products and their basic characteristics are described
together with a summary of important results. 
The images, reduced data products, and derived data products from the Pan-STARRS1 surveys
are available to the community from the Mikulski Archive for Space Telescopes (MAST) at STScI.

\end{abstract}

\keywords{astronomical databases, catalogs, standards, surveys }

\section{Introduction}
\label{sec:intro} 


The Panoramic Survey Telescope and Rapid Response System (Pan-STARRS) is an innovative wide-field astronomical imaging and data processing facility developed at the University of Hawaii's Institute for Astronomy \citep{2002SPIE.4836..154K,kaiser2010}. 
The first telescope of the Pan-STARRS Observatory is the Pan-STARRS Telescope \#1, (Pan-STARRS1 or informally PS1).
The PS1 Science Consortium (PS1SC) was formed to use
and extend the Pan-STARRS System for a series of surveys to address a set of science goals and in the process the PS1SC continued the development of the Pan-STARRS System. An original goal the PS1SC set for itself was to insure the data would eventually become public. 

This is the first in a series of seven papers that describe the Pan-STARRS1 Surveys, the data reduction techniques, the photometric and astrometric calibration
of the data set, and the resulting data products. 
These papers are
intended to support the 
public release of the Pan-STARRS1 data products
from the
{\it Barbara A. Mikulski Archive for Space Telescopes}  (MAST) at the Space Telescope Science Institute.\footnote{http://panstarrs.stsci.edu/}

There are two Data Releases supported: Data Release 1, 
(DR1) containing the stacked images and the supporting
database of the $3\pi$ Steradian Survey, and Data Release 2 (DR2) containing all of the individual epoch data of the $3\pi$ Survey including forced photometry on individual images based on information from the stacked data. 
Further Data Releases will depend on the availability of resources to support them. 

This Paper (Paper I) provides an overview of the fully implemented Pan-STARRS System, the design and execution of the Pan-STARRS1 Surveys, the image and catalog data products, a discussion of the overall data quality and basic characteristics, and a summary of scientific results from the Surveys.  

\citet[][Paper II]{magnier2017c}
describes how the various data processing stages are organized and implemented
in the Imaging Processing Pipeline (IPP), including details of the 
the processing database which is a critical element in the IPP infrastructure . 

\citet[][Paper III]{waters2017}
describes the details of the pixel processing algorithms, including detrending, warping, and adding (to create stacked images) and subtracting (to create difference images) and resulting image products and their properties.

\citet[][Paper IV]{magnier2017a}
describes the details of the source detection and photometry, including point-spread-function and extended source fitting models, and the techniques for ``forced" photometry measurements. 

\citet[][Paper V]{magnier2017b}
describes the final calibration process, and the resulting photometric and astrometric quality.

\citet[][Paper VI]{flewelling2017}
describes  the details of the resulting catalog data and its organization in the Pan-STARRS1 database. 
%
%
Huber et al. 2019 (in preparation - Paper VII)
describes the Medium Deep Survey in detail, including the unique issues and data products specific to that survey. 
The Medium Deep Survey is not part of DR1 or DR2. 

%




The paper is laid out as follows. 
In Section\,\ref{sec:system}
of this paper we begin with an overview of the completed Pan-STARRS1 System,
and a brief description of its associated subsystems:
the Pan-STARRS Telescope \#1, (PS1),  the Gigapixel Camera \#1 (GPC1), the Image Processing Pipeline (IPP), hierarchical
database or Pan-STARRS Products System (PSPS), and the Science Servers:
the Moving Object Pipeline (MOPS), Transient Science Server (TSS),
Photo-Classification Server (PCS). 
Section\ \ref{sec:surveys}
describes the various Pan-STARRS1 Surveys and their characteristics;  
the details of the observing strategy and the resulting impact on the time sampling and survey depth as a function of position on the sky. 
Section\ \ref{sec:products}
provides a summary of the Pan-STARRS1 data products.
Section\ \ref{sec:calib} summarizes the overall astrometric and photometric calibration of the surveys.
Section\ \ref{sec:3pi}
provides an overview of the features and characteristics of the 3$\pi$ Survey.
Finally, a summary of the legacy science of the PS1 Science Consortium
and a brief discussion 
of the future of Pan-STARRS
is provided in Section\ \ref{sec:results}.

\section{The Pan-STARRS System}
\label{sec:system}
\subsection{Background}
\label{subsec:background}
\subsubsection{The Pan-STARRS Project}

The Panoramic Survey Telescope and Rapid Response System (Pan-STARRS) is an innovative wide-field astronomical imaging and data processing facility developed at the University of Hawaii's Institute for Astronomy \citep{2002SPIE.4836..154K,kaiser2010}. 
Approximately 80 percent of the construction and development funds came from the US Air Force Research Labs (AFRL) 
in response to a Broad Agency Announcement 
``to develop the technology to survey the sky." 
The remainder of the development funds came from NASA, the PS1 Science Consortium (PS1SC), the State of Hawaii,
and some private funds. The project's goal was originally to construct 4 separate 1.8-meter telescope units
each equipped with a 1.4 gigapixel camera, and operate them in union.  The ambitious nature and full scale cost of the 
project led to a decision to build a prototype system of a single 1.8-meter 
telescope unit. This provided an opportunity not only to test the hardware, software and design  but also to carry out 
a unique science mission.  This system, located on the island of Maui, was named Pan-STARRS1 (PS1). 



\subsubsection{ The PS1 Science Consortium}
In order to execute and deliver a competitive and scientifically interesting set of sky surveys, 
the Institute for Astronomy (IfA) of the University of Hawaii (UH) assembled the PS1 Science Consortium (PS1SC). 
This group of interested academic institutions 
established a set of science goals \citep{chamberskc2007199832},
and a Mission Concept Statement \citep{chamberskc2006199843} and 
funded the operations of PS1 for the 
purpose of executing the PS1 Science Mission 
\citep{2006amos.confE..39C,chamberskc2008199860}.
The founding institutions of the PS1SC defined 12 Key Projects to ensure that the definition of the surveys
and their implementation were shaped by science drivers covering a range of 
topics from solar system objects to the highest redshift QSOs.  
The Memorandum of Agreement of the PS1SC established that the funding for operations was provided in return for the proprietary use of the Pan-STARRS1 data for scientific purposes.  As the PS1 Mission went on, additional members were added to bring in additional resources. The member institutions of the 
PS1 Science Consortium are provided in Table\ \ref{tab:ps1sc}.


\begin{table}
\caption{PS1 Science Consortium}
\begin{center}
\begin{tabular}{l}
\hline
{Member Institution }\\
\hline
\hline
University of Hawaii Institute for Astronomy  \\
Max Planck Institute for Astronomy\\
Max Planck Institute for Extraterrestrial Physics\\
The Johns Hopkins University \\
Durham University \\
University of Edinburgh\\
Queen's University Belfast \\
Harvard-Smithsonian Center for Astrophysics \\
Las Cumbres Observatory Global Telescope Network \\
National Central University of Taiwan \\
Space Telescope Science Institute \\
National Aeronautics and Space Administration \\
National Science Foundation \\
University of Maryland \\
E\"{o}tv\"{o}s Lor\'{a}nd University \\
Los Alamos National Laboratory\\
\hline
\end{tabular}
\end{center}
\label{tab:ps1sc}
\end{table}

\subsubsection{ The PS1 Science Mission }

The PS1 Telescope began formal operations on 2010 May 13, with the start of the PS1 Science Mission, funded by the PS1SC and with K. Chambers as PI and Director of PS1.
%
%
At the beginning of the PS1 Mission, the Image Processing Pipeline (IPP) - the
software and hardware for managing and processing the data - 
was not at an advanced stage of development, nor were the characteristics of the unusual OTA devices well understood.  
Furthermore, because of the AFRL funding, the imaging data was initially required to be censored. The AFRL ``Magic" software  was devised so that the pixels surrounding 
any feature in an individual image that could be interpreted as a potential satellite streak were masked. This meant removal of pixels
in a broad streak, or elongated box,  which was large enough to prevent the determination of any orbital element of the artificial satellite before the images left the IfA servers to the consortium scientists. This  requirement hindered analysis of the very features that were triggering the censor, nearly all of which were not satellite streaks, but were inherent detector characteristics. This effectively delayed the full and rapid analysis of the pixel
data by consortium scientists until the ARFL finally dropped the requirement on 2011 Dec 12.
From that date,  all Pan-STARRS1 images, including prior data taken during commissioning and 
from the start of survey operations, were no longer subject to any such masking software. Earlier data were re-processed from the untouched original raw data without the streak removal. There is no real-time nor archival censorship of any Pan-STARRS data.
None of the data now being released in DR1 and DR2
suffer from any application of the ``Magic" streak removal software either in the individual  or in the stack images. 

At the start of the PS1 Mission the development of the IPP (software and hardware), and eventually the development of the PSPS, shifted from the Pan-STARRS Project Office (2003-2014) to the PS1 Science Consortium funded PS1 Operations team. 
The Project Office went on to develop the second Pan-STARRS facility, Pan-STARRS2.
In August of 2014 the Pan-STARRS Office closed, and the Operations team also took over
responsibility for the completion and commissioning of PS2 with the support of the NASA NEO Program, the State of Hawaii, and private funding.  No further involvement with the AFRL is expected. 

\subsubsection{ The STScI MAST Archive and Data Releases}

To fully exploit the scientific potential of the PS1 survey data the PS1SC committed to make all PS1SC data public and accessible as soon as possible, but not before one year after the end of PS1SC survey operations. The science consortium made this commitment in principle in order that the data reach as wide a usage as possible. However the original founding members of the PS1SC did not have the resources to provide and maintain a public interface server. To enable such a public release, 
the PS1SC joined forces with the Space Telescope Science Institute (STScI)
and the {\it Barbara A. Mikulski Archive for Space Telescopes}  (MAST).
The STScI joined the PS1 Science Consortium 
through a Memorandum of Agreement to 
contribute resources to the creation of an archive of the Pan-STARRS1 Data Products. Subsequently STScI obtained funding from the Gordon and Betty Moore Foundation for support of operations of this archive that now serves the entire astronomical community. 


DR1 contains the static sky and mean data (See Section \ref{sec:products}).
The individual detections and forced detections (See Section \ref{sec:products})
are in DR2.  Our intention is to support
further releases as resources allow.

\subsection{Flow of information in the Pan-STARRS System}
\label{subsec:flow}

\begin{figure}
\includegraphics[width=\columnwidth,angle=0]
{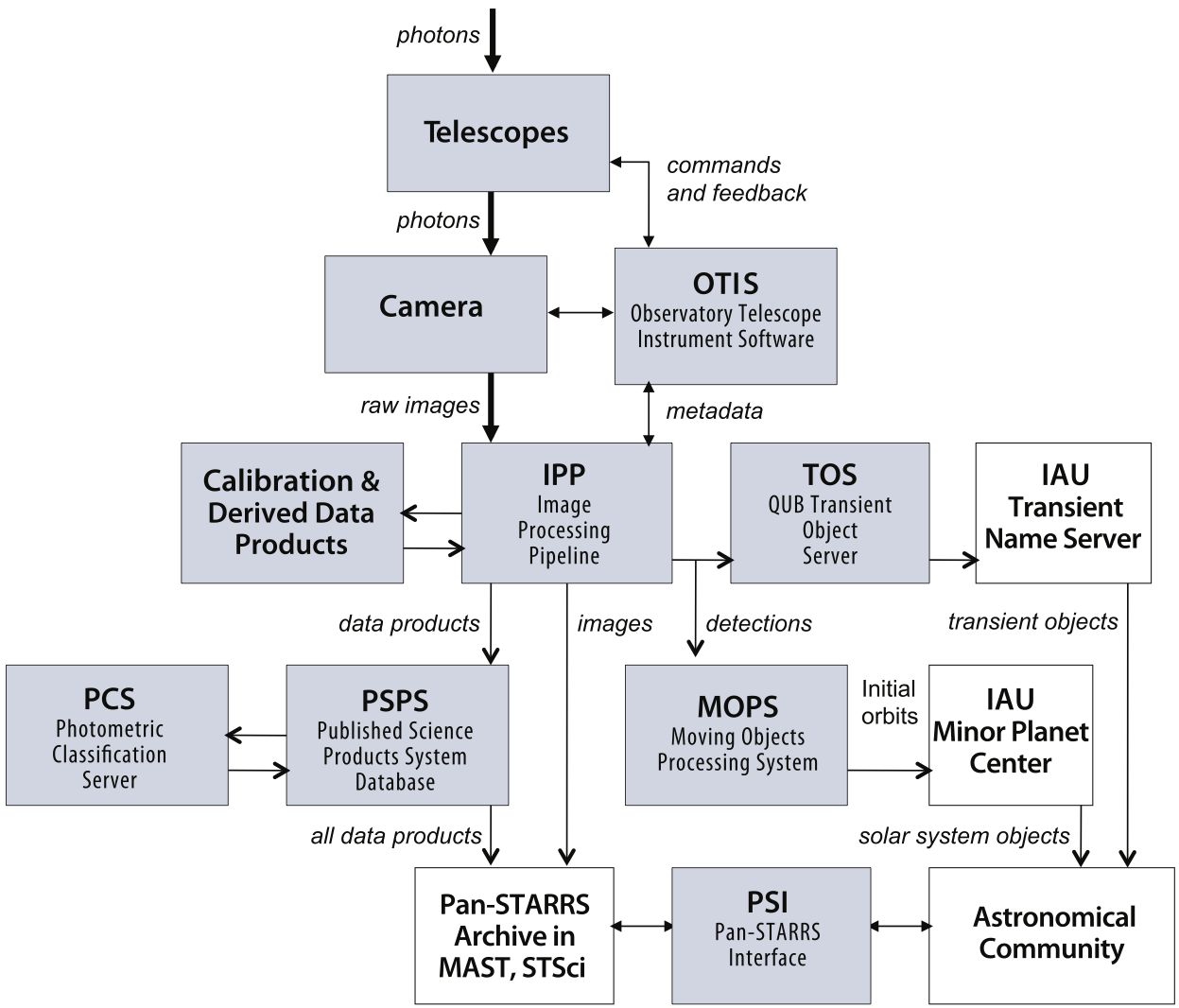}
\caption{Flow of information through the Pan-STARRS System. The various subsystems are discussed in Section\ \ref{sec:system}.\\}
\label{fig:flowchart}
\end{figure}

An overview of the flow of information through the Pan-STARRS System is shown in Figure\ \ref{fig:flowchart}.
In brief: 
photons from astronomical objects are brought to a focus by the Telescope onto the focal plane of the Gigapixel Camera \#1 (GPC1). 
As discussed below, a feedback signal is generated from selected areas of GPC1 and fed back to the telescope through the Observatory, Telescope, and Instrument Software system or OTIS, see Section\ \ref{subsec:tel}. 
During the night, as new images are downloaded, they are
processed by the IPP, see Section \ref{subsec:ipp}.
The results are passed to the Moving Object System (Section\  \ref{subsec:mops})
and the Transient Science Server (Section\ \ref{subsec:tss})
Near Earth Object (NEOs) candidates from MOPS are sent to the Minor Planet Center, and stationary transient objects are now posted on the IAU Transient Name Server\footnote{https://wis-tns.weizmann.ac.il/} for use by the community. 
Offline from nightly processing, 
the IPP uses a variety of tools for calibration (Section\ \ref{sub:dvo}).
The catalog data products produced by IPP are passed on the PSPS database
(Section \ref{subsec:psps}).
Both the PSPS database and all the image products from the IPP
are then available to the community from the 
{\it Barbara Mikulski Archive for Space Telescopes} (MAST) at STScI.

\subsection{Site}
\label{subsec:site}

The Pan-STARRS telescopes (both PS1 and PS2) are located at Haleakala Observatories (HO)
on the island of Maui on the site of the Lunar Ranging Experiment (LURE) 
\citep{1973egfs.conf..433C}.
Measurements by the HO
Differential Image Motion Monitor (DIMM)
show the site has a median image quality of 0.83 arc-seconds (the mode is 0.66 arc-seconds). On average 35\% of the nights on Haleakala are photometric, with an additional 30\% usable with very low extinction or more than 60\% of the sky clear of clouds. 
The wind pattern is predominately trade winds from the east-northeast,
with occasional ``Kona" winds from west-southwest. PS2 is 
due north of PS1, the center of the two telescope piers is separated by 20.05 meters. The domes are situated in 
the wake of the flow from trade winds into the crater wall. 
Detailed metrics of the site characteristics will be published elsewhere (Chambers, 2019 in prep). 
More recently the Daniel K. Inouye Solar Telescope (DKIST)\footnote{http://dkist.nso.edu/} 
has been erected to the south-south west of the Pan-STARRS facility. 
The ultimate impact of DKIST operations on the Pan-STARRS environment is not yet fully known, their operational plan is to manufacture ice at night for
use in the daytime cooling of DKIST, and subsequently dissipation of heat into the atmosphere at the summit. 

The International Astronomical Union has determined that the acceptable level of Radio Frequency Interference outside an observatory doing optical and infrared observations should be less than $2\mu W/m^2$ integrated over the radio spectrum. This is exceeded
at Haleakala and 
at the start of the PS1 Mission,
radio frequency interference (RFI) from various Federal and commercial transmission sites near the summit was an issue. 
However after the relocation of TV broadcasters to the Ulukalapua site in 2009, the level of RFI was reduced to the point where we see no
evidence of RFI in GPC1. However cellphone transmission, wifi transmission, and microwave ovens have a noticeable effect and are not allowed at the Observatory.

\subsection{Telescope, optics, and control system }
\label{subsec:tel}

PS1 is an alt-az telescope with an instrument rotator built by Electro Optic Systems Technologies Inc., Tucson, (EOST) with an enclosure by Electro Optic Systems Ltd. (EOS), Australia. The PS1 Dome motion closely follows the telescope through a featherweight direct coupling. The dome has four independently controllable vents for air flow through the dome. The dome slit is covered by two independently controllable shutters that can be deployed over the top on to the back side of the dome. When the moon is up the dome slit shutters are used to mitigate scattered light from the moon. 

The Observatory, Instrument, Telescope, Software (OTIS) system controls all these aspects of the Observatory and collects and stores a wide variety of auxiliary and metadata on the conditions and all the functions of the Observatory.

\begin{figure}
\includegraphics[width=\columnwidth,angle=0]
{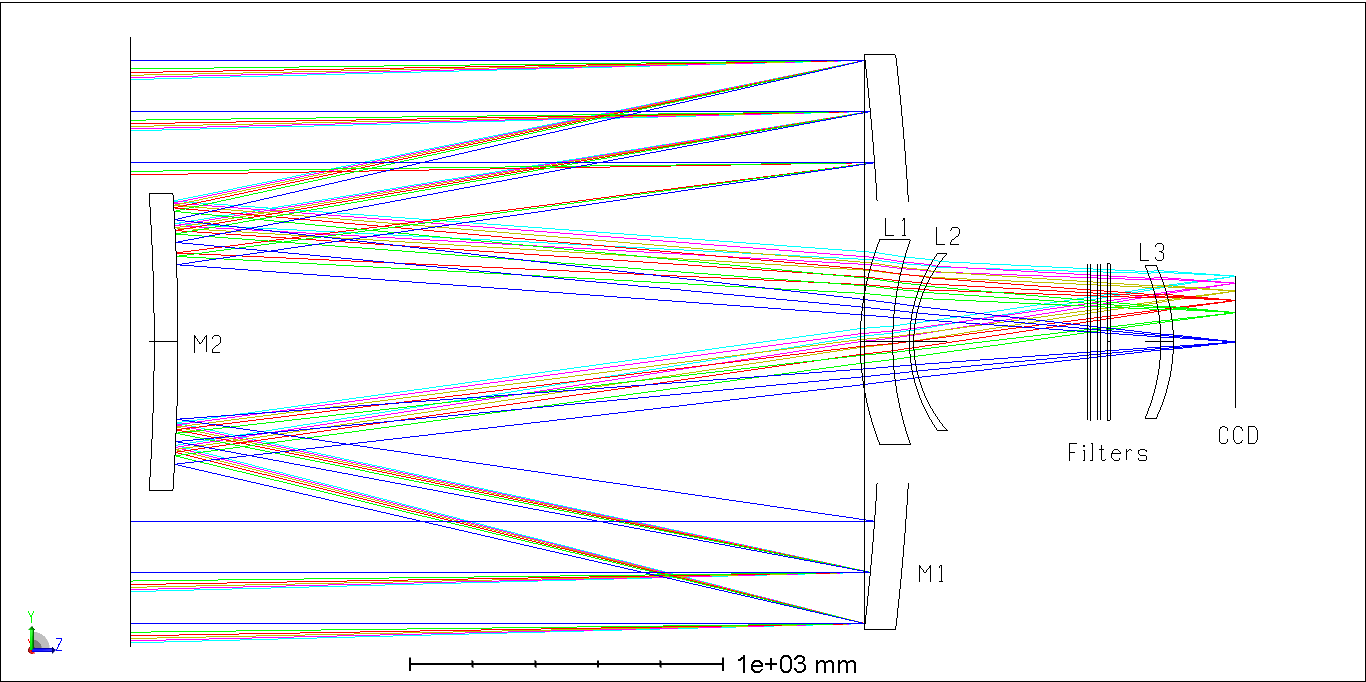}
\caption{The Pan-STARRS1 optical design \citep{2008SPIE.7012E..1KM}. The as-built design version is specified as NOADC-M-3.0 shown here. In the figure rays enter from left at the top of telescope where the spider legs that support the secondary introduce diffraction spikes in the images. There are three baffles, one supported by the secondary support structure, a middle cone baffle that is suspended by cables aligned with the secondary support spiders, and a baffle supported from within the central hole of the primary. The corrector lenses are labeled in order of passage as L1, L2, and L3 which is also the cryostat window. Between L2 and L3 are the filter mechanism and the shutter. The filter mechanism has 3 layers which can store 6 filters.}  
\label{fig:optics}
\end{figure}

The Pan-STARRS1 optical design \citep{2004SPIE.5489..667H, 2004AN....325..636H,2008SPIE.7012E..1KM} has a wide field 
Richey-Chretien configuration with a 1.8~meter diameter $f$/4.44 primary mirror, and 0.9~m secondary. 
The resulting converging beam then passes through two refractive correctors, one of six possible 
interference filters with a clear aperture diameter of 496 mm, and a final refractive corrector that is the cryostat
window. Note that the Pan-STARRS1 as-built optics are described by the Zemax model NOADC-3.0. See Figure\,\ref{fig:optics}. Table\,\ref{table:ps1} has a summary of the Pan-STARRS1 telescope characteristics. 

The optical design has 4 aspheric surfaces; one each on the primary and secondary mirrors, one  on the first corrector lens, L1, and a final aspheric on L3, the last corrector lens in the optical path and which also serves as the cryostat window. The secondary mirror has a conic constant of $-20.43$ and a $6^{\rm th}$ order aspheric term of $4.5 \times 10^{-19}$,
which made it a challenge to fabricate \citep{2008SPIE.7012E..1KM}.

The Secondary Mirror is mounted on a hexapod and can be moved in five axes: $x$, $y$, $z$, tip, and tilt. 
The primary mirror (M1) is on a pneumatic support system and can be commanded in $y$, $z$, tip, and tilt.
M1 can be moved in the $x$ direction as well, but this is not on a powered
actuator and must be done manually. 
Furthermore M1 has a 12 point astigmatic correction system. 
Thus there are 22 independent mirror actuators that can be used to bring the optics into proper
 collimation and alignment with the optical axis as defined by the axis of the instrument rotator. 
 These actuators
allow for modest amounts of M1 deformation to remove trefoil, coma, and astigmatism. 
The procedure for establishing the proper collimation and alignment is described in \cite{2008SPIE.7012E..1KM}. Given the system matrix,
only minor adjustments are required to maintain collimation and alignment. 
PS1 does have significant flexure, so empirical models have been determined
to correct for that. In practice the largest corrections are in the M2 tip and $y$ (tangent to altitude) de-center. 
The M1 figure correction also has
an altitude dependent term. The OTIS software applies these corrections 
for the destination of any commanded slew, corrections are disabled during
exposures and the system tracks quiescently during the short exposures --
generally not more than 2 minutes. 
A focus offset is determined 
from each exposure based on the measured astigmatism, and this offset is applied to the empirically derived focus model. 
The offset is calculated from an analysis of the ellipticity of the PSF across the focal plane calculated by the GPC1 software. 
The calculation of the correction takes approximately one minute, and then cannot be applied until the next pause between exposures while the camera is reading out. 
Thus the 
telescope focus is maintained by the local focus model with an observationally based offset determined within a few minutes of a new exposure. 

After large slews or starting a new pointing, a short exposure (10 seconds) is made to obtain a current focus correction. 
This system maintains the correct M2 focus position to within $\pm 5$ microns of true focus. The collimation and alignment do drift occasionally, especially if there is maintenance performed on the telescope. These drifts are corrected by a procedure of using above and below focus images of stars (donuts) 
\citep{2008SPIE.7012E..1KM} to make a correction. 
The system to maintain the image quality is imperfect, and the results can be seen in some images. Typically the impact is some combination of higher order aberrations that result in a asymmetric PSF. The IPP fits only an elliptical PSF, so there is no systematic measure of this asymmetry or its effect on photometry, albeit it must be small. 
The telescope illuminates a diameter of 3.3 degrees, with low distortion, 
and mild vignetting at the edge of this illuminated region. 
The field of view is approximately 7 square degrees. The 8 meter focal length at $f/4.4$ gives an approximate 10 micron pixel scale of 0.258 arcsec/pixel. 

\begin{table}
\caption{Summary of PS1 Telescope Characteristics}
\begin{center}
\begin{tabular}{ll}
\hline
\hline
{Characteristic} & { Quantity}\\
\hline
Focal Length    & 8000 mm \\
Nominal Field of view& 3.0 degree diameter circle \\
Primary mirror       & 1800 mm diameter \\
M1 coating           & protected aluminum  \\
Secondary mirror     & 947 mm diameter \\
M2 coating           & protected silver  \\
f/number             & f/4.44 \\
Effective aperture   & $ 0.65 \times \pi\, 92^2\, {\rm cm}^2 = 17284\, {\rm cm}^2 $ \\
                     & including diffraction and obscuration \\
Rotator range        & 179 degrees \\
Telescope/Dome wrap  & 420 degrees \\
\hline
\end{tabular}
\end{center}
\label{table:ps1}
\end{table}

\subsection{GPC1 - the Gigapixel Camera \#1 }
\label{subsec:gpc1}

The Gigapixel Camera \#1 (GPC1) uses 
Orthogonal Transfer Array (OTA) devices, a concept developed by 
\cite{1997PASP..109.1154T} and 
their development was key to the Pan-STARRS concept \citep{2000PASP..112..768K}.
The detectors in GPC1 
are CCID58 back-illuminated 
(OTAs), manufactured by Lincoln Laboratory 
\citep{2006amos.confE..47T,2008SPIE.7021E..05T}.
They have a novel pixel structure with 4 parallel phases per pixel
\citep{2008SPIE.7021E..05T} and required the development of a new type of controller \citep{2008SPIE.7014E..0DO}. 
GPC1 is populated with two different kinds of CCID58s, the CCID58a with a three phase serial register, and the CCID58b which has a two phase serial register \citep{2012SPIE.8453E..0KO}.
Table\,\ref{tab:gpc1char} has summary of GPC1 characteristics. 
The intent of the
OTA design was to allow charge to be moved in orthogonal directions providing an on-CCD tip-tilt image correction given a guide signal from a nearby cell being read at video rates, and Tonry's OPTIC camera did this successfully, \citep[e.g.][]{2009ApJS..185..124S}.
However,  with GPC1 when the Orthogonal Transfer mode of the detectors
was turned on, it produced an unacceptable 
level of non-uniform background noise 
\citep{2012SPIE.8453E..0KO}.
The Pan-STARRS1 Surveys did not use the detectors in Orthogonal
Transfer mode. All PS1 Survey data was taken with  
the GPC1 devices operating as ``normal'' CCDs.

\begin{figure}
\includegraphics[width=\columnwidth,angle=0]{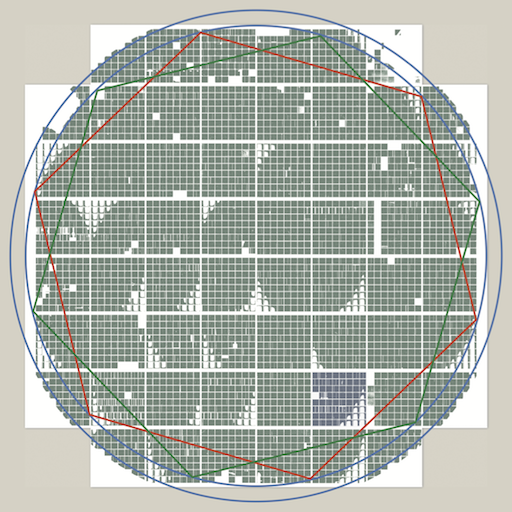}
\caption{Gigapixel Camera 1 focal plane layout and mask. The non-functioning cells are blanked out in white. 
} 
\label{fig:cameramask}
\end{figure}

The focal plane of GPC1 comprises a total of 60  CCID58 OTA devices
\citep{2008SPIE.7021E..05T}.
Each of these devices consists of an $8\times8$ array of individual addressable CCDs called ``cells.'' 
The overall format of a single OTA is a $4846\times4868$ pixel array with a pixel size of 10~$\mu$m which subtends 
0.258~arcsec.
Each OTA device is made up of 64 cells where each cell is $590 \times 598 $
pixels. The cells are separated by a gap between columns, 
that is 18 ``inactive" pixels in size, and a gap between rows that is 12 inactive pixels in size. 
Thus a single OTA device contains a single piece of silicon with 64 cells in an $8\times8$ array 
separated by a grid of $7\times 7$ internal streets.  We will often refer to the OTA devices as ``chips" in 
the data processing discussions. Furthermore,
there are physical gaps between the devices as mounted in GPC1. 
The placement of the devices in the focalplane is shown in Figure\,\ref{fig:cameramask}. 
The relative positions of each device, including rotation, were determined
from a vast number of astrometric measurements on sky. 

The separation between the OTA devices is 1400 microns (approximately 36 arcsec) in the $x$ direction and 2800 microns ( approximately 70 arcsec) in the $y$ direction. In practice the devices are not perfectly spaced and can have some small
rotation with respect to one another. The astrometric solution for each device is solved independently without reference to one another, the only place where the determined relative position is used is telescope pointing and guiding. 
Note there is a slight optical pin-cushion distortion of the 
sky on the focal plane, all of this is removed in the process of the astrometric registration (warping) by the IPP
\citep[see][]{magnier2017a}. 

\begin{table}
\caption{Summary of GPC1 Characteristics }
\begin{center}
\begin{tabular}{ll}
\hline
\hline
{Characteristic} & { Quantity}\\
\hline
Device         & CCID58a; three phase \\
Device         & CCID58b; two phase \\
Read noise     & 8 $e^-$ \\
Dark current    & small, temp dependent \dag \\
Persistence     & moderate \dag \\
Charge transfer & bad regions masked \\
Non-linearity   & begins $~40,000$ DN \dag  \\
Saturation      & median $60,400$ DN \dag \\
Pixel size      & 10 $\mu m $ \\
Pixel size      & 0.258 arcseconds \\
Camera fill factor   & 90\%: \\
Shutter         & opening time 1 sec \\
Shutter         & accuracy 10 msec \\
\hline
Total Overhead  & 10.3 sec \\
Initialization  & 0.3 sec  \\
Exposure start  & 2.0 sec \\
Exposure readout     & 7.0 sec \\
Exposure save/clean  & 1.0 sec \\
\hline
Pixel Mask fractions     &      \\
Good Pixels   & 76\% \\
No Pixel/gap  & 10.1\%  \\
Detector flaws & 10.7\% \\
Poor Charge Transfer Efficiency & 2.2\% \\
Other defect flags       & 1\% \\                  
\hline
\tablefootnote{\dag Device dependent. See \cite{waters2017} for details and how these are treated in the detrending and masking.}
\end{tabular}
\end{center}
\label{tab:gpc1char}
\end{table}

%

The telescope, detector devices, and control electronics each contribute a variety of artifacts to the GPC1 images. Where possible these artifacts are identified and the pixels are masked or modified during processing and flags are set in the database. These include optical ghosts from reflections in the optics, glints from scattered moon light, glints from structure in the camera, regions of poor charge transfer in the devices, persistence or ``sticky charge" from saturation leaving ``burn-trails" that persist for all successive images for tens of minutes, electronic ghosts from cross-talk in the electronics, and correlated read noise from the fiberflex that transmit the signal through the cryostat wall.  These are identified and masked where possible in the detrending procedure as part of the chip processing stage in the IPP. 
The dark current is small, but temperature dependent and therefore a pixel by pixel model of the dark current is subtracted from each image based on the temperatures measured for that exposure. Non-linearity sets in starting at about 40,000 counts and a device dependent model is applied. 
There is a detailed discussion of the detrending and the treatment of defects and how they are individually masked in \cite{waters2017}. These defects are visible, sometimes strikingly so, in the individual warped images, but less evident
in the stacked images made from the multiple images taken over the course of the survey. 

We have used observations from the Pan-STARRS 3π survey described below to further characterize the behavior of the deep-depletion devices used in the Gigapixel Camera. There are systematic spatial variations in the photometric measurements and stellar profiles that are similar in pattern to the so-called “tree rings” identified in the Dark Energy Camera \citep{2014PASP..126..750P}.
Unlike those devices, the photometric and morphological modifications observed in the GPC1 detectors are caused by variations in the vertical charge transportation rate and the resulting charge diffusion variations \cite{2018PASP..130f5002M}.

If we take the sky area covered by the GPC1 footprint to be the area of the inner blue circle in Figure\,\ref{fig:cameramask}
(7 sq degrees) then the dead cells, pixel gaps and masking of defective pixels account for an overall loss of 20\% of the focal 
plane in any one exposure. There is an additional dynamic masking of around 2-3\% per exposure, which mostly  covers the ``burn-trails". Therefore the overall fill factor of the camera is  $76\pm1$\%  per exposure and this is mitigated by the 
dither and stack techniques that were employed in the $3\pi$ and Medium Deep Surveys. 
DR1 has only the stack images and hence the images are mostly 
continuous, although there are areas where a combination of poor devices and fewer than 12 exposures mean 
some small masked regions exist in the final stacked images.

A subset of bright stars (mag$<12$) which fall on the focal plane are selected to be used as guide stars (suitably located across the camera), and a $100 \times 100$ pixel box is defined, centered on the position where these stars are predicted to land based on the commanded telescope position. This set of sub-arrays on different devices are read at video rates. 
The centroid from these video frames are used to send a guide signal to the telescope control system. 

Typically there are 4 to 10 stars chosen, which means these cells are then masked in the science exposures. These additional masked cells
are included in the ``dynamic" mask developed for each exposure that includes 
the masking due to the artifacts of that particular exposure and is added to the "static" mask as seen in Figure 3. 

The shutter, built by the team at Bonn University, is a dual blade
design. The shutter aperture is approximately 40 cm across, and in 
closed position one blade covers the aperture and one is stored to the side. When the shutter is opened, one side of the focal plane is exposed first. 
At the conclusion of the exposure, the second blade traverses the aperture
in the same direction, hence the total exposure time seen by each pixel 
is the same to the precision of the movement, or 10 milliseconds. 
For the subsequent exposure the motion is in the opposite direction. 
Short exposures are possible, where the blades follow each other trailing closely. 

This means that the center time of the exposure is different
by up to 0.5 seconds depending on placement in the focal plane,
and thus the quoted UT of any given detection can be in error by up to 0.5 seconds depending on its position relative to the shutter blade motion for that exposure. 
The metadata exists to calculate the precise time of each detection but this correction has not been made for DR2. Even for moving asteroids this is not a serious limitation. 

The detectors are read out using a StarGrasp CCD controller \citep{2008SPIE.7021E..05T}, 
with a total overhead of 10.3 seconds for a full unbinned image.
The overhead is defined as the time
taken from the request to start an exposure until the point where the
camera is available for the next exposure, minus the requested
exposure time. For bookkeeping purposes, this is broken down into
four parts. The first, initialization, is the time consumed
by parsing the request and checking the readiness of the system. The
second concerns the exposure itself, and includes guide star selection based on the telescope pointing, spreading the guide video pipelines out over the summit cluster of servers that runs the camera, as well as a fixed shutter motion time of around 1 second per exposure.
The readout phase begins once the shutter is closed and the CCDs are
read out; this is the largest contributor to overheads, and is
determined mainly by the clocking pattern used to sample the images
from the CCDs. Some smaller contribution is also involved in managing
the readout threads across the cluster, one for each of the 60 CCDs.
While the bulk of the data transfer from the camera to the cluster is
done in parallel with the readout, there is some small wait at the end for the last pixel data to be transferred after the readout is
completed; this is accounted for in the post-readout phase along with
any other tasks that need to be completed before the camera is
considered ready for the next exposure.

In practice the total overhead time between adjacent exposures, 
including the 1 second shutter movement time, is 10.3 seconds on average, there is some variance.  See Table\ \ref{tab:gpc1char} for a breakdown of the overhead. 

Other performance characteristics of GPC1 are presented in 
\cite{2008SPIE.7021E..05T,2012SPIE.8453E..0KO}.


\subsection{Filter bandpasses and PS1 sensitivity}
\label{subsec:filt}

\begin{figure}
\includegraphics[width=\columnwidth,angle=0]
{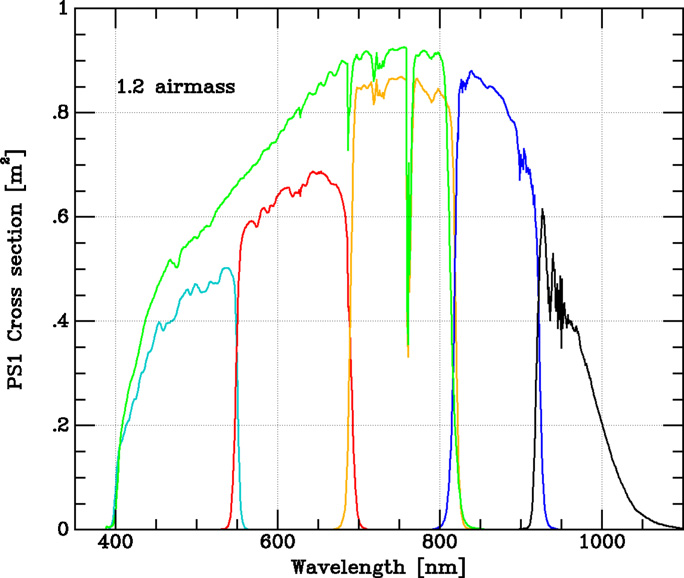}
\caption{This figure is reproduced from \cite{2012ApJ...750...99T} for ease of reference. The PS1 capture cross-section in 
m$^{2}$e$^{-1}$photon$^{-1}$ to produce a detected e$^{-1}$ for an incident photon for the six Pan-STARRS1 bandpasses, 
\grizy\ and \wps\ for a standard airmass of 1.2. } 
\label{fig:tonry}
\end{figure}

The \PS\ observations are obtained through a set of five broadband filters, designated as \gps, \rps, \ips, \zps, and \yps. Under certain circumstances \PS\ observations are obtained with a sixth, ``wide'' filter designated as \wps\ that essentially spans \gps, \rps, and \ips.   
Although the filter system for \PS\ has much in common with that used in previous surveys, such as the Sloan Digital Sky Survey (SDSS,  \citet{2000AJ....120.1579Y}), there are important differences, which is why the filters are labelled specifically with the ``P1'' subscript. 
The \gps\ filter extends 20~nm redward of
$g_{\rm SDSS}$ with the intention of providing greater
sensitivity and lower systematics for photometric redshifts. 
The strong  [\ion{O}{1}] 5577\AA\ sky emission is on the filter edge but only at 1\% transmission. 
The \zps\ filter has a sharply defined cut-off at 922~nm, which  is in contrast to the SDSS $z-$band which has no red cut off and 
the response is defined by the detector response. The \rps and \ips filters are very similar to SDSS and color differences
between the two magnitude systems are small. 
 SDSS has no corresponding \yps\ filter.   The transmission of the Pan-STARRS1 filters, optics and total throughout were 
precisely measured with a calibrated photodiode and a  tuneable laser, without use of celestial standards by 
\cite{2010ApJS..191..376S} and this procedure was repeated in November 2016 (Stubbs et al. in prep). 
The definition of the  photometric system has already been discussed in detail and published in \cite{2012ApJ...750...99T}. 
Tabular data of the overall throughput of the PS1 system is available in the online data of \cite{2012ApJ...750...99T}
and the individual filter throughputs are in \cite{2010ApJS..191..376S}. The PS1 total filter throughputs from \cite{2012ApJ...750...99T}
are reproduced here in Figure\,\ref{fig:tonry}. 

%

Photometry is in the ``natural'' \PS\  system in  ``monochromatic AB magnitudes''\citep{1983ApJ...266..713O} as described in \cite{2012ApJ...750...99T}
\begin{align}
m_{\rm AB}(\nu) &= -2.5\log(f_\nu/3631~\hbox{Jy})\\
 &= -48.600 - 2.5\log(f_\nu[\hbox{erg/sec/cm$^2$/Hz])}
\end{align}
\PS\ magnitudes are interpreted as being at the top of the atmosphere, with  1.2 airmasses of atmospheric attenuation being included in the system response function. No correction for Galactic extinction is applied to the \PS\ magnitudes. 
We stress that, like SDSS, \PS\ uses the AB photometric system and there is no arbitrariness in the definition. Flux representations are limited only by how accurately we know the system response function vs. wavelength. 
See e.g. \cite{1994AJ....108.1476F} for conversions of AB magnitudes to a Vega or Johnson based system. 


The DR1 and DR2 data have been calibrated with the updated values
from \cite{2015ApJ...815..117S}, for details see
\cite{magnier2017c}.



\subsection{IPP - Image Processing Pipeline}
\label{subsec:ipp}

\begin{figure}
\includegraphics[width=\columnwidth,angle=0]
{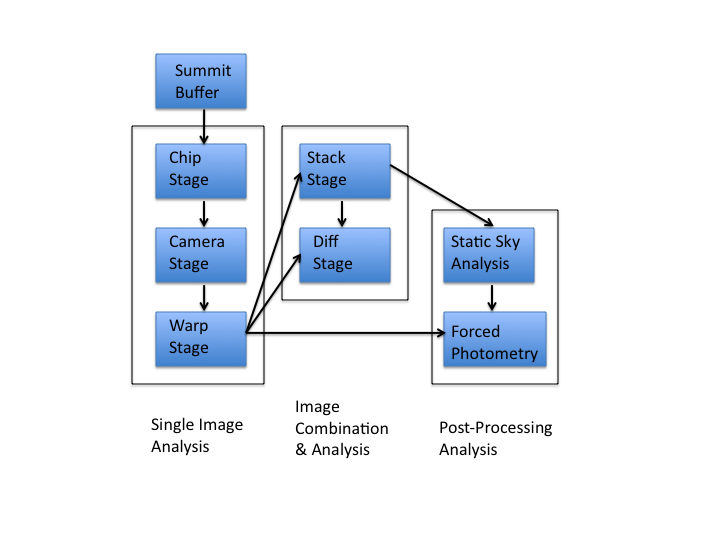}
\caption{Schematic of the image and analysis processing stages of the IPP (Magnier et al 2017b). The images are read from the GPC1 to  buffer storage at the summit. The IPP polls this buffer and retrieves new images whenever they are created. During the night the raw images are retrieved, and are processed individually through the single image analysis. The nightly difference images for moving object or transient detection are created from a warp-stack image combination if the stack exists, or a warp-warp difference from a pair if there is no stack image.  
The post-processing stages work on the stacked images for the Static Sky Analysis and on the individual warp images for the Forced photometry.} 
\label{fig:ippflow}
\end{figure}

All images obtained by the Pan-STARRS1 system are processed through the Image Processing Pipeline (IPP) 
on a computer cluster at the Maui High Performance Computer Center. The pipeline runs the images through a succession of stages, 
including de-trending or removing the instrumental signature, a flux-conserving warping to a sky-based image plane, masking and artifact removal, and object detection and 
photometry. 
The IPP also performs image subtraction to allow for the prompt detection of moving objects, variables and transient phenomena. 
Mask and variance arrays are carried forward at each stage of the IPP processing. 
Photometric and astrometric measurements performed by the IPP system are published in a MySQL relational database.  
Below we give a brief summary of the Pan-STARRS image processing, full details are provided in the companion papers of 
\cite{magnier2017a,magnier2017b,magnier2017c,flewelling2017,waters2017}. 
Figure \ref{fig:ippflow} gives a simplified schematic of the processing stages. 




\subsubsection{Chip Stage}
In the ``Chip Stage" raw exposures are detrended 
\citep[dark subtracted, flattened, masked, etc;][]{waters2017}
and sources in the images are detected and basic instrumental characterization is performed.  A PSF model is generated and all sources fitted with that model.  For sources above a minimal signal-to-noise limit (nominally 20), a simple galaxy model is fitted if the source appears to be extended.  The best model (PSF or galaxy) is subtracted and 
an additional source detection pass is made (down to S/N = 5). 
This provides for some de-blending. Reported values include instrumental positions, fluxes (PSF, seeing-matched aperture, Kron aperture), moments, and various quality flags are recorded for each source in the image.  The output from this stage consists of fits tables of detections and their properties called CMF files and detrended images and their associated variance, and mask pixel images.

\subsubsection{Camera Stage}
In the ``Camera Stage" the instrumental measurements from 
all the chips in one exposure 
are gathered together for astrometric and photometric calibration by comparison with a reference catalog. Initially a synthetic reference catalog was created based on 2MASS, USNO-B, and Tycho. This was used for a photometric calibration as the survey proceeded. 
In the re-processing and re-calibration that produced the data in DR1 and DR2 the reference catalog uses Pan-STARRS itself,
to create a precise and consistent internal calibration \citep{magnier2017b,magnier2017c} based on the ``ubercal"
methods described in \cite{2012ApJ...756..158S} and \cite{2015arXiv151201214F}. 
The primary data product from the Camera Stage is the collection of calibrated detection tables.  



\subsubsection{Warp Stage}
In the ``Warp Stage" the detrended pixel images generated by the chip stage are geometrically transformed to a predefined set of images which tessellate the relevant portion of the sky.  
Specific examples are discussed in Section\ \ref{sec:surveys} below.  
%
%
%
A set of virtual rectilinear images with square pixels of 0.25 arcseconds size,
on a local tangential projection center no bigger than about 4 degrees across
are defined. These virtual images are called ``projection cells" and 
one or more projection centers can be defined for specific areas of interest or arranged in some defined tessellation of the entire celestical sphere. 
%
%
%
%
The total output from the warp stage is the collection of images that describe the signal, variance, and masking for each skycell. 

%

\subsubsection{Stack Stage}

Individual epoch skycell images (from the Warp Stage) are combined together to form deeper stack images of the sky, the details of the algorithms are in \cite{waters2017}. 
In the IPP analysis, stacks of different depths/quality may be made depending on the individual survey goals. This is of particular application to the Medium Deep Survey\ \ref{subsec:mdeep}).
The output from the stack stage consists of the signal, variance, and mask stack images.

\subsubsection{Difference Image Stage}
The primary means for detecting a transient, moving, or variable object is through the process of subtracting a template image of a source from a single image 
to create a ``difference" or ``diff" image. 
The IPP generates Alard-Lupton \citep{1998ApJ...503..325A}
convolved difference images for skycells in various combinations depending on the survey goals. 
The output from the diff stage is a collection of detections from the difference images, including both positive and negative difference detections.

\subsubsection{Static Sky Stage}
The ``Static Sky" refers to a final stacked image. 
The stack images from all filters are processed in a single analysis step to perform the deep source detection and characterization of objects detectable in the stacks.  This analysis step is similar to the source detection and characterization performed at the chip stage, with some important additions:  First, 3 PSF-convolved galaxy models (Sersic, DeVaucouleurs, Exponential) are fitted to all objects with sufficient signal-to-noise and in regions outside the densest portions of the Galactic plane. In addition, sources which are detected in only two of the 5 filters (or just in the \yps band, to allow for the presence of astrophysical objects which are dropouts in the bluer bands) 
are then used to force PSF photometry (and aperture and Kron flux measurements) at that same location in the other 3 (or 4) filters.  Finally, flux is measured for 7 radial aperture annuli, using apertures of the same radii in arc-seconds on the sky as used by SDSS.  These radial aperture fluxes are measured for the raw stack with its natural seeing as well as on a version of the stack convolved to match 1.5 and 2.0 arc-second seeing. 

\subsubsection{Skycal Stage}
The skycal stage performs the photometric calibration of the "Static Sky" outputs relative to the reference catalog in an analogous fashion to the camera stage wherein the photometry is calibrated relative to the reference catalog.  The intial reference catalog was synthetic, generated from model colors based on fits to 
Tycho \citep{2000A&A...355L..27H}, 
USNO \citep{2003AJ....125..984M},
and 2MASS \citep{2006AJ....131.1163S} 
stars as the only available all sky photometry. After May 2012, the photometric reference catalog was generated from an internal re-calibration of the data set to date. The final re-calibration of 
the data is discussed in \cite{magnier2017c}. The pixels are not recalibrated, and hence the authoritative photometry is catalog based, not pixels based. 


\subsubsection{Full Forced Stage}

Image quality variations between different exposures
(and even within a single exposure)
result in a stack PSF which can vary discontinuously on small scales.  
PSF photometry and PSF-convolved galaxy model fitting on the stack cannot follow these variations.  The result is degraded performance in the stack photometry and morphology analysis.  To avoid this problem, we use the outputs from the ``Static Sky" stage analysis as the input to a ``forced" photometry analysis on each of the input warp images.
 
In this analysis, the positions of all objects detected in the stacks are used to measure the PSF photometry of those objects on each of the input warps images, using the appropriate PSF model determined for that position on that warp image (Kron and aperture measurements are also made).  The individual warp measurements are then combined in catalog space (in our photometry databasing system) to determine the mean photometry for each object.  
 
In this step, input measurements with excessive masking are also excluded from 
this mean photometry calculation. 
The result is a reliable photometry measurement for all objects down to the detection limits of the stack, as well as the data to study the variability and transient nature of the faintest sources. 
 
In this stage, we also perform an analysis of the galaxy morphology using the ``static sky" galaxy model measurements as the seed (\cite{magnier2017b}). 
 
 

\subsubsection{Post-Processing and DVO}
\label{sub:dvo}

After the pixel-level processing is performed, the catalogs of measurements extracted from the images are ingested into an instance of the Desktop Virtual Observatory or DVO 
\cite[for more details see][]{magnier2017a}. 
DVO is a set of stand alone tools within the IPP system created to perform calibrations 
and provide further analysis of systematic effects. 


In addition to the ingest into DVO at the IPP, the team of Eddie Schlafly (MPIA, LBL), Doug Finkbeiner (Harvard), 
and Greg Green (Stanford) also ingest the camera-stage data into a separate databasing system 
called LSD \citep{2011AAS...21743319J}.
This system is similar in scope to DVO and allows similar calibration operations.  This team runs the ``ubercal" analysis on the detections from the chip and camera stage to measure zero points for photometric data.  In this analysis, relative photometry of overlapping images is used to constrain the zero points and airmass terms.  A rigid solution is determined by requiring a single zero point and airmass term for each night.  The resulting photometric system is shown to have a precision of 8, 7, 9, 11, 12 millimags for each of 
\grizy\ respectively
\citep[as described in][]{2012ApJ...756..158S}. 

\subsubsection{IPP-to-PSPS}

Given the way the Pan-STARRS System evolved, it  has been necessary to implement a translation layer to collate the catalog products produced by the IPP \citep[so called ``CMF" and ``SMF" fits tables containing measured attributes,][]{magnier2017a} in an optimal manner for ingest into the PSPS. The IPP-to-PSPS produces batches of binary fits files containing
catalog data. There is a different kind of batch for each type of database table (e.g. objects, stacks, detections, difference detections).
Each batch contains data from a localized region of the sky.  
Some units are rationalized in the IPP-to-PSPS, so 
there is some manipulation of data values in this subsystem. 
See \cite{flewelling2017} for a detailed discussion.

\subsection{PSPS - Published Science Products System}
\label{subsec:psps}

The Pan-STARRS Project teamed with the database development group at Johns Hopkins University to undertake the task of providing a hierarchical database for Pan-STARRS \citep{2008AIPC.1082..352H}.
Since the JHU team was the major developer of the SDSS database \citep{2003ASPC..295..217T}, our goal was to reuse as much of the software developed for the SDSS as possible. The Pan-STARRS database is commonly refered to as the ``PSPS". 

The key to moving from the SDSS database to a system capable of dealing with Pan-STARRS data is the design of the Data Storage layers. It was immediately clear that a single monolithic
database design (like SDSS) would not work for the PS1 problem. Our approach has been to use several features available within the Microsoft SQL Server product line to implement a system that would meet our requirements. 
While SQL Server does not
have (at present) a cluster implementation, this can be implemented by hand using a combination of distributed partition views and slices 
\citep{2008AIPC.1082..352H}. 
This allows us to partition data into
smaller databases spread over multiple server machines and still treat the information as a unified table (from the users' perspective). 
Further, by staying with SQL Server we
are able to retain a wealth of software tools developed for SDSS, including the use of Hierarchical Triangular Mesh indexing for efficient spatial searches.

\begin{figure*}
\includegraphics[width=\textwidth,angle=0]{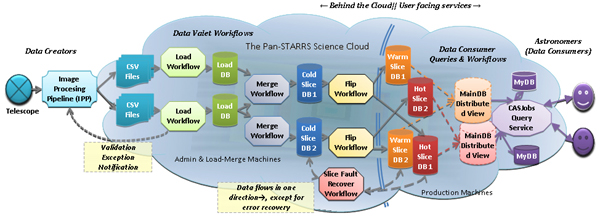}
\caption{Data from the IPP enters the ODM through the DX layer as FITS binary tables. The DX layer converts these into comma separated variable (CSV) files that are then passed to the loading pipeline. The data flow is illustrated in the diagram above. As illustrated in this figure, most of the processing inside the ODM takes place under the hood and is invisible to the users, who only see the data once it is loaded into the data stores that serve the hot and warm processing queues.
The attributes are examined for basic validation (e.g., data in range checks). The loading workflow process the input from the loading through to the merging of the new data records with the information already contained in the cold database.
When a sufficient quantity of new input has been merged into the cold database we execute the copy-flip workflow. In this stage the hot (fast queue) database is allowed to drain its query queue and is then taken off line. The cold database is then copied in total to the hot database storage. While the hot copy is taking place the warm (slow queue) is kept running to drain any remaining queries that have been staged for processing. Once the hot copy has been completed we pause the slow queue, flip the active queue back to the fast queue and resume processing there, and execute the warm copy.
} 
\label{fig:psps}
\end{figure*}

An overview of the PSPS system is shown in Figure\ \ref{fig:psps}.

\subsubsection{Object Data Manager}
The Object Data Manager is a collection of systems that are responsible for publishing data attributes measured by the IPP or other Pan-STARRS Science Servers to the end user (scientist). 
The ODM manages the ingest of data products from the IPP (or other sources), integrates the new products with existing information in its data stores, and then makes the information available to the users in relational databases.
 
Catalog data from the IPP as prepared by the IPP-to-PSPS layer is contained in batches of binary fits tables. These fits tables are read by a Data Transformation
(or DX) Layer where data are grouped by declination zone and throttled in Right Ascension by the IPP-to-PSPS layer. Then the data is loaded
into `cold' or load slice machines by the DLP or data loading pipeline.  The slices are variable bands in declination, established to have nearly constant data density. Once data are loaded on all declination slices through a given RA range, the
data are merged, wherein they are stored and indexed on the slice machines so that data that are nearby in the sky are similarly nearby on disks and grouped by machine. Once the data are successfully merged across the whole sky, the database is copied from the load/merge machines to the data storage machines where the user can access the database through the Query Manager (QM)
and web-based Pan-STARRS Science Interface (PSI).

\subsubsection{The Data Retrieval Layer}

The Data Retrieval Layer or DRL is the unseen hub of the PSPS system. It sits between software that provides user access and the underlying data stores themselves. The DRL provides the access to users and the databases through web browsers. Only those users who want to write their own access clients will interact with the DRL directly. A simple application programming interface (API) has been developed to allow one to develop such applications. 

The DRL also provides the internal mechanisms for routing result sets from the PSPS databases back to the user.

The DRL API allows the PSPS to expand to incorporate the addition of new databases that can make science products created by PS1SC science servers available to the user community. The API has been demonstrated to work with Microsoft SQL Server, MySQL, and PostgreSQL databases.

The DRL Layer is accessible through the CasJobs interface at the Pan-STARRS1 
Archive at MAST. 

\subsubsection{PSI Interface}
The Pan-STARRS Science Interface (PSI) is a web application that has been developed by the PSPS development team. It is designed provide users with easy access to the PSPS through a web browser. PSI has tools to simplify the construction of querys and flags and a variety of useful features. PSI is built on a improved version of CASJOBS, but it is not immediately backwards compatible with the version of CASJOBS  at STScI. 
Access to the Pan-STARRS1 archive at MAST at STScI is through the standard CASJOBS  (\cite{casjobs,10.1109/MCSE.2008.6}) interface.

\subsection{Science Servers}
\label{subsec:servers}

The PS1 Science Servers were a project concept to add science value to the basic data products of 
object, position and flux. The three projects that evolved to provide working code and data products are 
breifly described here.

\subsubsection{MOPS - Moving Object Processing System}
\label{subsec:mops}

The Pan-STARRS Moving Object Processing System 
(MOPS; \cite{2013PASP..125..357D})
is a modern software package that produces automatic asteroid discoveries and identifications from catalogs of transient detections from Pan-STARRS or any
next-generation astronomical survey telescope. 

As implemented as a subsytem
in the Pan-STARRS System, it obtains difference detections from the IPP, 
performs linkages between detections, and makes initial orbit determinations.
Potential moving objects are evaluated by a human inspection system, and 
candidates are passed to the Minor Planet Center of the IAU.

Funded by the Pan-STARRS Project prior to the formation of the PS1SC, MOPS was the first integrated asteroid detector system able capable of automatically producing high-quality orbits from individual per-exposure transient catalogs. MOPS is also able to search its own historical data for orphaned one-night detections after an orbit is generated. 

As implemented as a subsytem
in the Pan-STARRS System, it obtains difference detections from the IPP, 
performs linkages between detections, and makes initial orbit determinations.
Potential moving objects are evaluated by a human inspection system, and 
candiates are passed to the Minor Planet Center of the IAU.

MOPS has additional value as a research tool in survey design, able to simulate years of observations and detections given a catalog of synthetic asteroids and a hypothetical observation schedule.  The synthetic solar system model \citep[S3M;][]{Grav2011}, containing $10^7$ objects representing populations of all major solar system bodies, remains the standard synthetic population for evaluating survey performance.


\subsubsection{TSS  - Transient Science Server }
\label{subsec:tss}
The vast majority detections in difference images requires a system
for classifying real vs artifacts to manually select the most promising candidates. 
The Queen's University Belfast group developed the
Transient Science Server to systematically process difference detections from
stationary transients from the IPP stream and apply machine learning techniques to classify them.
(\cite{2015MNRAS.449..451W}). This system continues to process transient events
from Pan-STARRS and post discoveries on the IAU Transient Name Server.
In parallel the team at CfA, Harvard developed a custom version of the 
{\it photpipe} image subtraction and analysis pipeline and analyse the 
MDS data in real time \citep{2012ApJ...755L..29B,2014ApJ...795...44R}
The two teams cross-correlated transient discoveries 
and photometric measurements from both streams to improve efficiency and 
measurement precision of the IPP products. Both were successful in different ways, 
and the QUB based TSS was the only one currently in operation for the 
3$\pi$ based searches and the ongoing Pan-STARRS Survey for Transients \citep{2015IAUGA..2258303H,2016MNRAS.462.4094S}.

\subsubsection{PCS  - Photometric Classification Server }

{The Photometric Classification Server (\cite{2012ApJ...746..128S})
is a set of software tools and hardware set up to 
compute photometric, color-based star/QSO/galaxy classification
and best-fitting spectral energy distribution (SED) and photometric
redshifts (photo-z) with errors for (reddish) galaxies.
The system can establish an interface to the PSPS database 
and results can be ingested back into the PSPS. 
Results from the Photometric Classification Server will not
be available in DR2. }



\subsection{Pan-STARRS Operations}
\label{subsec:ops}

The observatories are operated remotely from the Pan-STARRS Remote Operations Center in the Institute for Astronomy (IfA)  Advanced Technology Research Center (ATRC) in Pukalani, Maui.  There is no one at the summit at night or on weekends except in urgent or emergency situations. The Observatory is approximately 45 minute drive from the ATRC. A Pan-STARRS observer on a swing shift schedules the night's operations based on the overall science goals, state of the survey, and expected conditions. The night observer executes the plan prepared by the swing shift observer and modifies it in real time as circumstances demand. The observing staff rotate  through the swing and night shift and support the day crew at the summit. The Staff at the ATRC also provides support for the telescope, scheduling software,  and system administration for the IPP cluster in Kihei.    

The IPP is a linux cluster that has evolved continually as the survey progressed. At the time of the DR1 reprocessing it had about 3100 cores and 5.5 Petabytes of storage. The IPP cluster was located at the Maui Research and Technology Center in Kihei, Maui. The computing facility (power, cooling, network connectivity to the outside world) was administered by the Maui High Performance Computing Center. 
Additional computing resources were required for the PS1 Surveys including the Mustang Cluster (30,000 cores) at Los Alamos National Laboratory and the Cray cluster at the University of Hawaii (3600 cores). The DR2 processing took place after the cluster was moved to the University of Hawaii Information Technology Center (ITC) on the Manoa Campus on Oahu and at the time of the DR2 release has about 5000 cores and 15 Petabytes and processes the nightly data from both PS1 and PS2. 

Operationally the IPP and the PSPS are run remotely by the IPP team from IfA Manoa.  
During night time operations, the raw exposures are immediately downloaded to the IPP cluster.  Nightly data processing occurs automatically for exposures as they are obtained, with the analysis emphasis on the discovery of transient events, as well as data characterization for future re-processing. The reprocessing versions and status are discussed in detail below and in the companion papers. These data products have been loaded and merged in the PSPS database and transferred to STScI for distribution through the MAST archive.

\section{The PS1 Surveys}
\label{sec:surveys}

\subsection{The PS1 Science Goals}
\label{subsec:goals}


The primary science design drivers for PS1 were originally put forth in the PS1 Science Goals Statement (\citep{chamberskc2007199832}). The top level goals were:

\begin{itemize}

\item{} Precision photometric and astrometric survey of stars
        in the Milky Way and the Local Group;
\item{} Surveying our Solar System, including
        searching for Potentially Hazardous Objects
        amongst Near Earth Asteroids;
\item{} New constraints on Dark Energy and Dark Matter;
\item{} Exploration and categorization of the astrophysical time domain;
        including, but not limited to,
        explosive transients, microlensing events in M31,
        and a transit search for exo-planets.
\item{} Providing a development platform for prototyping PS4
        components, subsystems, and survey strategy.
 
\end{itemize}

These goals drove the initial design and engineering requirements, and shaped real time development decisions. 
On the last point, while the PS4 system has not yet been funded, PS1 did serve in this capacity for the 
development of PS2 \citep{2012SPIE.8444E..0HM}. 
The above outline goals do  not  begin  to cover 
the vast array of solar system, Galactic, extragalactic, and cosmological studies that can be done with the PS1 data products. 
To refine this, the project and the PS1 Science Consortium Science Council generated the PS1 Mission Concept Statement
\citep{chamberskc2006199843} 
with a set of surveys as follows:
(1) A 3$\pi$ Steradian Survey; of 60 epochs in five passbands (\grizy) of the entire sky north of 
declination $\delta=-30$ degrees, 
(2) A Medium Deep Survey with data in all of \grizy\ of ten PS1 footprints on well studied fields totaling 70 square degrees at
high Galactic latitudes spaced around the sky, 
(3) A solar system ecliptic plane survey in the wide \wps\ passband with
cadencing optimized for the discovery of Near Earth Objects and Kuiper Belt Objects, 
(4) a Stellar Transit Survey of 50  square degrees in the Galactic bulge; 
and (5) a Deep Survey of M31 with an observing cadence designed to detect micro-lensing events and other transients.
In addition a special series of observations of spectro-photometric standards was carried out for
calibration, and the Celestial North Pole was observed nightly for the last two years of the survey to track performance and measure atmospheric properties. 
Table\,\ref{tab:surveys} summarises these surveys and the approximate percentage time spent on each of the total operational science time. 

The operational plan for execution of these surveys 
was articulated in the 
PS1 Design Reference Mission (\cite{chamberskc2008199860}) 
or DRM,
that served as a benchmark as the system transitioned from commissioning to operations. 
This survey strategy evolved into a Modified Design Reference Mission as lessons learned were incorporated as the surveys progressed.


\begin{table}
\caption{The Pan-STARRS1 Surveys}
\begin{center}
\begin{tabular}{llll}
\hline
\hline
{Surveys} & {Filters} & {Percent} &{Dates}\\
          &           & (time)   &       \\
\hline
3 $\pi$ Steradian Survey         & \grizy   & 56 & 2009-14\\  

Medium Deep Survey               &  \grizy &25   & 2009-14 \\
Solar System Survey              & \wps  &  5 $\rightarrow$ 11& 2012-14\\
Pan-Planets Transit Surv       & \ips                 & 4 & 2010-12\\
PAndromeda Surv of M31         & \rps,\ips               &  2 & 2010-12\\
\hline
Calibration: & & \\
\hline
Spectro-photometric stds     & \grizy,\wps & 1  & 2010-14  \\ 
Small Area Survey 2          & \grizy   & 1  & 2010 \\
Celestial North Pole         & \grizy  &  1  &2012-14 \\
\hline
\end{tabular}
\end{center}
\label{tab:surveys}
\end{table}

\begin{figure*}
\begin{center}
\includegraphics[width=3.2in,angle=0]
{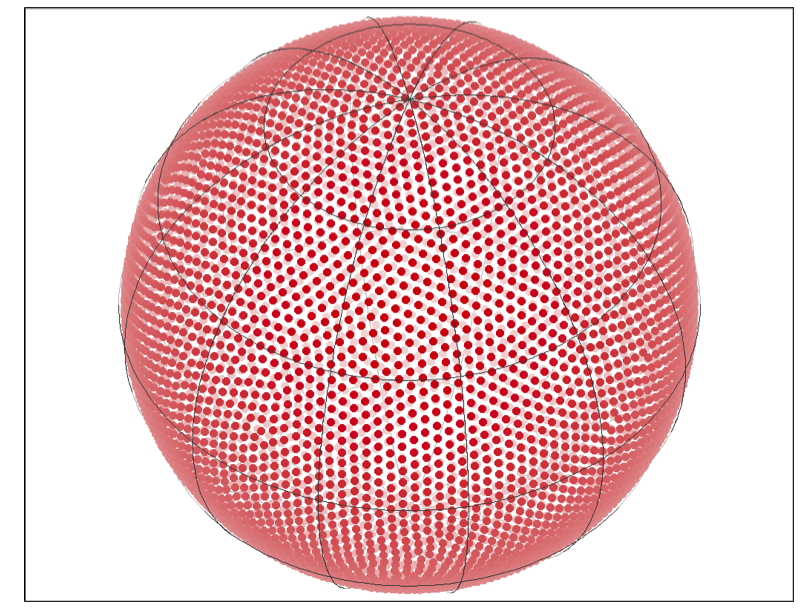}
\includegraphics[width=3.0in,angle=0]
{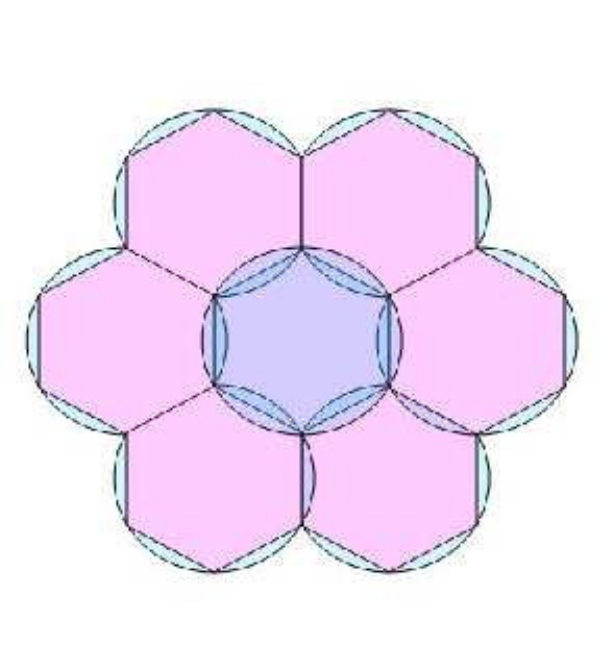}
\includegraphics[width=3.0in,angle=0]
{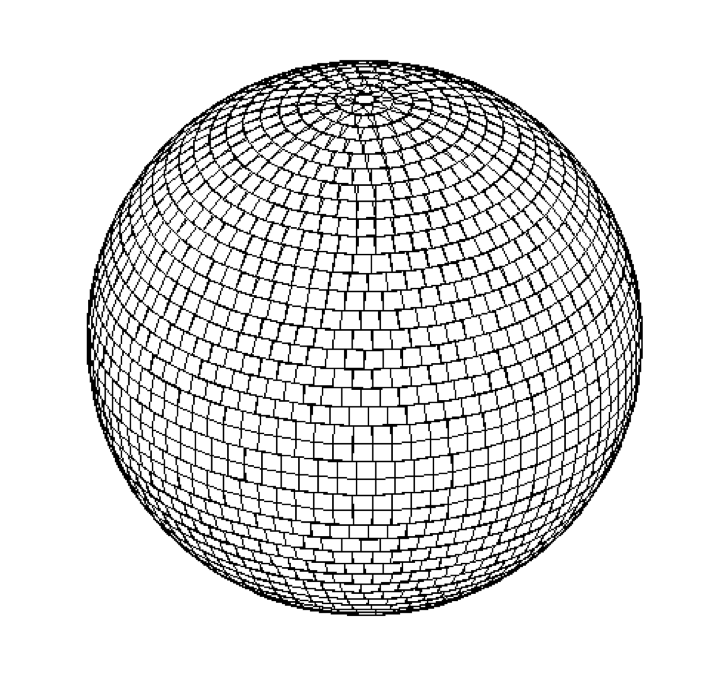}
\includegraphics[width=3.0in, angle=0]
{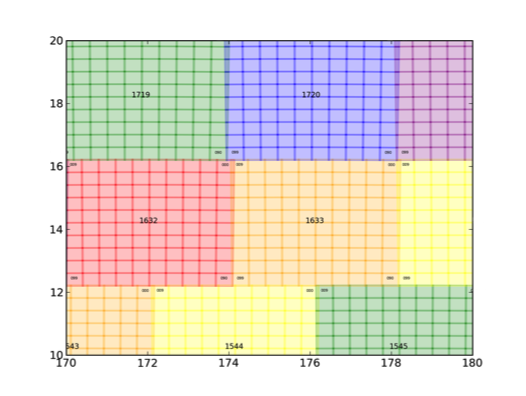}
\end{center}
\caption{Top Left: One realization of the boresight tessellation. Top Right: Schematic of seven field-of-views laid out in a flat hexagon pattern. Most of the vertices of the boresight tessellation have six nearest neighbors, a few have five. The figure shows the hexagon that is inscribed upon the field of view in Figure 3. This shows the nominal overlap of adjacent field-of-views. For a given exposure the (x,y) axes of the camera can have any orientation with respect to North, as the telescope is alt-az.  Lower Left: the RINGSV.3 tessellation of the sky with virtual rectangular images on tangential projection centers. For the 3$\pi$ Survey with a boresight southern declination limit of Dec $> -3$ degrees, the RINGSV.3 has 2009 tangential projection centers, and a nominal 200,900 sky cells which extend to Dec$=-31.81$ degrees. However the southern edge of the set of images is ragged from the footprint extending down from the southern most possible boresight of Dec$=-30$. The number of skycells containing populated imaging data is then 200,684, see \cite{flewelling2017} for more details. 
Furthermore there is a special tessellation for the north pole (\cite{magnier2017a}).
Bottom Right: a zoom showing several projection cells, each in a different color, and each divided up into overlapping sky cells. This shows the overlap of sky cells and the overlap of projection cells. Nearly all analysis of Pan-STARRS1 images is done on a sky cell basis.} 
\label{fig:obstess}
\end{figure*}

\subsection{The 3$\pi$ Steradian Survey }
\label{subsec:3pi}

The $3\pi$ Steradian Survey covers the sky north of Dec $= -30$ degrees in
five filters (\grizy) and includes data taken between
2009-06-02 and 2014-03-31.
This means that for a given sky tessellation, a field center was included in the survey only if it was above declination $\delta=-30$ degrees. 
For pointings with field centers that are close to $\delta=-30^{\circ}$,  
close to half the field (up to 1.5 degrees) extended below the limit. This means there is a ragged edge and an uneven 
declination limit to the survey between  $-31.5\degree < \delta < -30\degree$. 

The survey pattern and scheduling followed
two different strategies over the course of the 3$\pi$ survey:
the initial pattern laid out in the Design Reference Mission (DRM) 
\cite{chamberskc2008199860} followed by the Modified Design Reference Mission (MDRM). We switched to MDRM on 
2012-01-14. 
All exposures in the DRM were taken in pairs, with each exposure separated by a Transient Time Interval or TTI of 12 to 24 minutes, 
for the purpose of detecting moving objects within the Solar System. These were referred to as ``TTI pairs".
The original plan was then to take 2 TTI pairs over an observing season with \gps, \rps\ and \ips\ taken within the same lunation and separated by 
days to weeks. The \zps\ and \yps\ were to be taken approximately 6 months apart to optimise stellar parallax and proper motion measurements (for low mass 
stars). Over 3.5 years this would give (allowing for weather interruptions) 12 exposures in each band or 60 in total over
all 5 filters. 

In the MDRM, a series of 4 exposures, ``quads", all separated by approximately 15 minutes (therefore completed within about 1hr), were implemented for about half of the \gps\rps\ips exposures with the express purpose of increasing the recovery of Near Earth Objects (NEOs). The relative exposure times in  each were also chosen to make an asteroid of mean solar color (taken to be $(\gps-\rps) = 0.44, (\rps-\ips)=0.14$) to have approximately the same signal-to-noise. 



%

\subsubsection{The reduced image tessellation }
\label{sec:tessell}
During data processing, the  ``Warp Stage"  takes the detrended pixel images generated by the ``Chip stage"
and geometrically transforms (warps) and 
re-samples them onto a predefined set of images which tessellate the relevant portion of the sky (these processing stages are discussed in  Section \ref{subsec:ipp}). 
For the $3\pi$ survey, PS1 uses a modification (RINGS.V3) of the Budavari rings tessellation with tangential projection centers spaced $~4$ degrees apart.  A set of virtual images called  ``projection cells" are defined to cover the sky about these projection centers without gaps.  These virtual projection cells are subdivided along cartesian pixel boundaries into ``skycells", the image regions onto which the native device pixels are warped.  All skycells have a pixel scale of 0.25 arcsec per pixel and are roughly 20 arcminutes on a side, which is comparable in size to the native device images (these chip images are the $4846\times4868$ pixel arrays at 0.258 arcsec per pixel). 
The main output from this stage is the collection of three separate pixel images
each representing the signal, variance, and masking for the skycells.
The MD and similar surveys use special local projection cells centered on the fields of interest.

%

\subsubsection{Primary object resolution on the sky}
\label{sec:primary}

The skycells and projection cells are defined to have an overlap of 60 arcseconds
(~232 pixels) on each edge 
in order to avoid objects being split between adjacent skycells. 
Note that it is the same data which goes into the overlap regions - there is no new data involved here (although there may be slight variations in the way the data are stacked). The problem of identifying a unique area, and thus assigning an object to a particular skycell, is called the primary resolution problem. This is important, as data analysis is performed on each skycell independently, so an object near a boundary will have duplicate measurements. IPP produces a tessellation tree file which contains RA and DEC limits for each projection cell, which can be used to define unique areas. Objects landing within these limits are classed as primary objects and have the primary flag set in PSPS. This flag should always be used to define a unique sample of objects on the sky.
However, it is possible in some cases for objects not to have a detection classed as primary. This can occur, for example, where the  particular area of sky is only masked in the primary skycell, or where an object very close to the detection limit is only detected on the non-primary skycell.

\begin{table}[htbp]
\caption{Properties of images in each filter. The solar elongation indicates when twilight for that filter effectively starts.  }
\begin{center}
\begin{tabular}{lll} \hline\hline
Filter & Solar elongation  \\
          & (degrees) \\
\hline
\gps  &  16 \\ 
\rps  &  15.5  \\
\ips  &  15 \\
\zps  &  13 \\
\yps  &  10    \\
\wps  &  16    \\
\hline
\end{tabular}
\end{center}
\label{table:twilight}
\end{table}

\subsubsection{Scheduling of PS1 Surveys} 
\label{sec:scheduling}

The primary reason for a discussion of the scheduling of the PS1 Surveys is to explain why the time domain of the 3$\pi$ Survey has the detailed structure that it has. 
Prior to the formal start of the PS1 Mission on May 13, 2010, we used a contemporaneous version of the LSST scheduler to model the PS1 Mission as defined by the
Design Reference Mission \citep{chamberskc2008199860} 
and smaller in summer in accordance with the length of night.  
We further tweaked the size of the $3\pi$ slices to accommodate time for the smaller PAndromedra and Pan-Planets surveys, and assumed that the MD surveys, which are fairly evenly distributed in RA, could be fit into a constant nightly time allocation.

\begin{figure*}
\begin{center}
\includegraphics[width=\textwidth,angle=0]{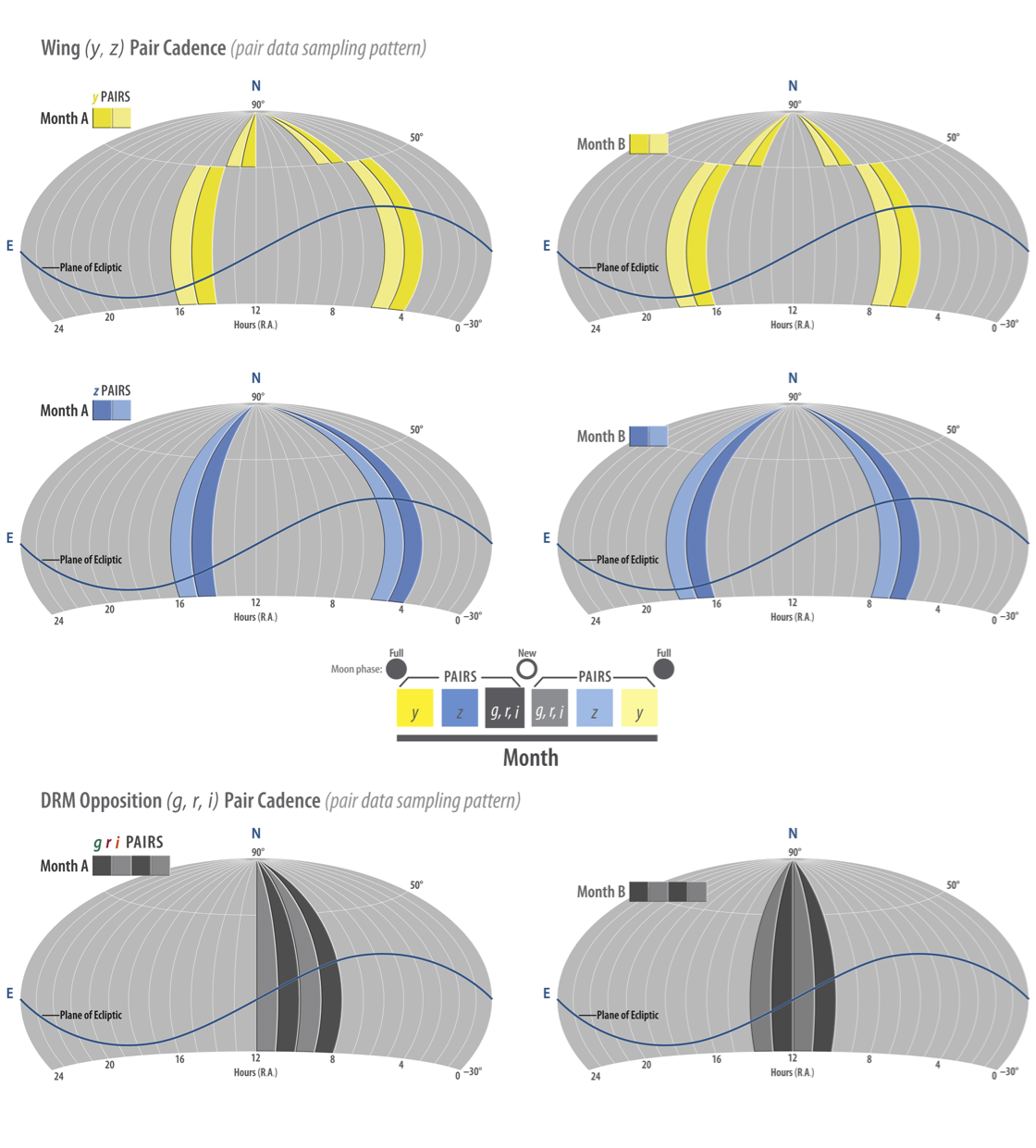}
\end{center}
\caption{DRM twilight and opposition scheduling cadence}
\label{fig:drm}
\end{figure*}

The observing pattern from the DRM  (applicable from 2010-05-10 - 2012-01-14) is schematically shown in
Figure\,\ref{fig:drm}. 
An Observing Cycle (OC) is defined as one lunation. The sky areas and filter coverage observed in an OC example are illustrated
in this figure. 
Clearly one needs to
observe, on average,  1/12.37  of the sky per Observing Cycle per filter 
(12.37 is the number of lunations in one year). This corresponds to a slice of sky from the pole to $\delta=-30$ which is roughly 4\,hrs in 
right ascension. This mean value was expanded or compressed {\it a-priori} for the length of night and to adjust for the non-uniform  impact of the smaller surveys in their RA distribution. 
One aspect of the PS1 3$\pi$ Survey is that the $z,y$ bands are observed out of phase with $g,r,i$ by months, whereas
$g,r,i$ might be taken in the same night or be out of phase by days.  
We defined two distinct kinds of slices, the Opposition slices, where the sky within about 2 hours RA of opposition was observed in $g,r$ and $i$ bands, and ``Wing" slices which were near the meridian at twilight. 

There were several reasons for this ``strategic" approach: 
(i) because twilight (defined as the moment when the night sky reaches a constant sky brightness) occurs  at increasing solar elongation as one proceeds through the filter set from
red to blue $y,z,i,r,g$, there is a period of time when the sky is as dark as it is going to get in $y$ band, but it is still in twilight in $z,i,r,g$ bands. Thus it is most advantageous to use this time in $y$ band. Once the sky becomes dark in $z$ band, the same is true. The time differences in the other bands are more modest. 
To illustrate this quantitatively Table\,\ref{table:twilight} gives details of the solar elongation angle at which 
the sky reaches its constant dark level (i.e. end of twilight) in each filter.

(ii) We desire to measure as many stars with measurable parallaxes as possible. These are the closest stars and are thus most likely to be brown dwarfs. PS1 is a red sensitive instrument, and is already delivering on its goal of finding new populations of L and T dwarfs
\citep{2011AJ....142...77D,2013ApJ...777L..20L}. 
It is therefore desirable to observe in $z,y$ bands at maximum parallax, i.e. with a cadence of nearly six months.
As illustrated in Figure\,\ref{fig:drm}, the Wing slices here in the $z,y$ bands are separated by nearly six months: 
as the pattern marches to the left from  ``Month A" to ``Month B". This shows that 
in about six months time the same region of the sky will be observed again in the same filters. 
(iii). This approach also ensures that the sky areas surveyed in the $z,y$ bands are observed near the meridian, or close to optimum airmass. 
During operations there typically was not 
quite enough time to get all of the $y$ and $z$ band observations in the twilight time of $g, r, i$. However near full moon, the 
sky is bright even in $i-$band, and some $y$ band fields were observed closer to the middle of the night. This required that the polar regions, which were beyond the 30 degree moon avoidance region, be shifted slightly closer to opposition and yet could still be observed at reasonable airmass. 

During a night's observing the pattern from the above  strategy was to observe ``chunks" where a chunk is simply a contiguous region of sky 
of approximately $4 \times 4$ GPC1 footprints, with a pair of visits separated by a TTI in $y$ band, and then if possible the same chunk in $z$ band as the sky got darker. Then the available Medium Deep Survey fields were inserted between the Wing Slices and the Opposition Slices. Depending on the sky brightness (lunar illumination and distance) a chunk in $g,r$ or $i$ band would be observed near opposition. Once opposition passed it would be back to the other programs and the morning Wing  slice at twilight. The prioritization of the chunks in declination was by image quality and transparency, so generally by airmass, followed by wind direction or partial clouds.

%

\begin{figure*}
\begin{center}
\includegraphics[width=\textwidth,angle=0]{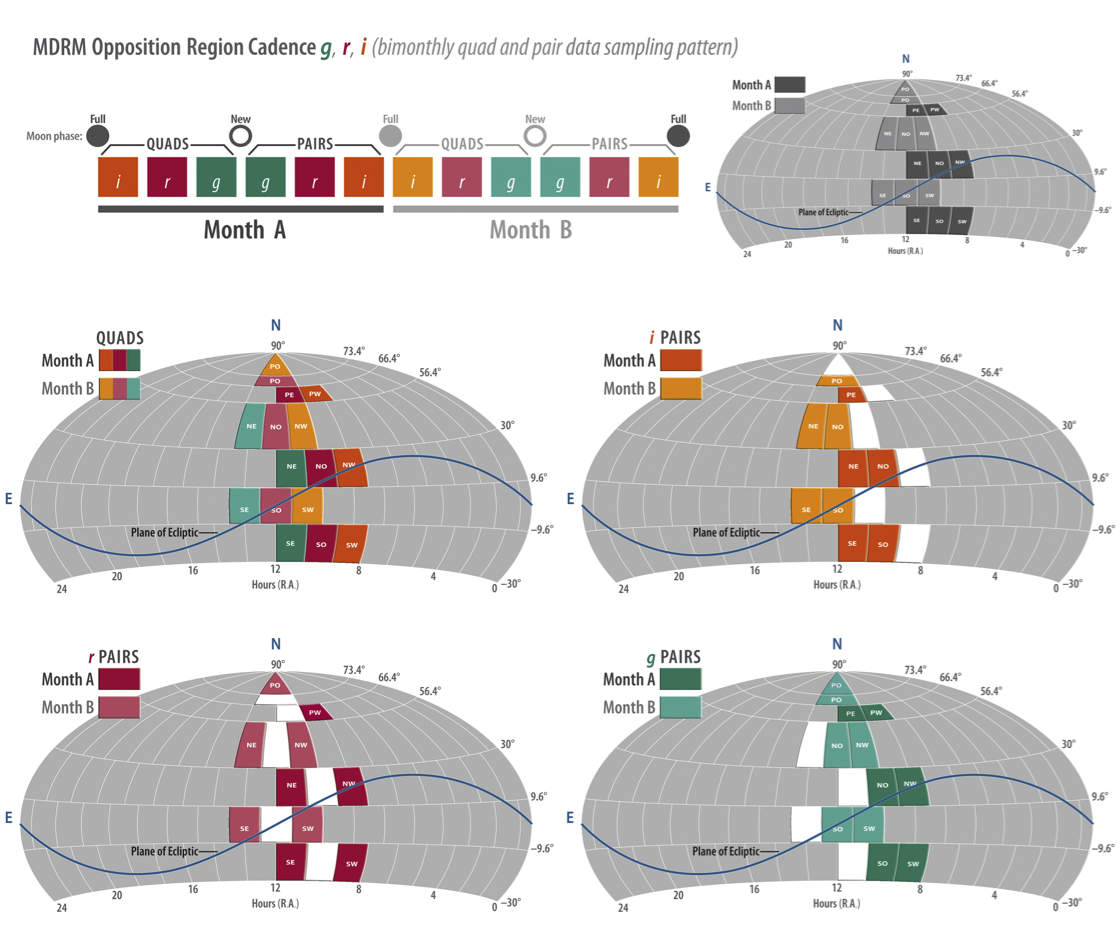}
\end{center}
\caption{MDRM Opposition Cadence. For the Modified Design Reference Mission the \gps, \rps, and \ips\ observing cadence and pattern were changed, but the Wing cadence pairs of the DRM for \zps\ and \yps\ remained the same as in Figure\,\ref{fig:drm} } 
\label{fig:mdrm}
\end{figure*}

\begin{figure}
\begin{center}
\includegraphics[width=\columnwidth,angle=0]
{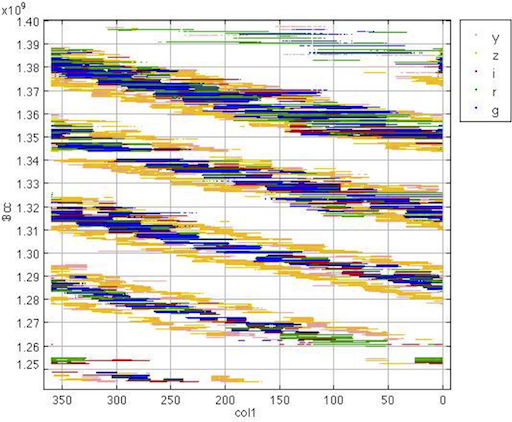}
\end{center}
\caption{Right Ascension in degrees vs Time in Unix seconds. 
} 
\label{fig:timeRA}
\end{figure}

It was eventually realized that a modification to the DRM was necessary. 
The DRM was done entirely in pairs with the assumption that observations of NEOs in pairs separated by one or two nights could be linked. As a matter of experience that turned out not to be the case. However simply switching
to quads, or four exposures per night separated by TTIs would put the years worth of exposures for a given field in a given filter all into one night. This would have endangered the photometric survey, reducing the number of opportunities to have a photometric night, critical for ubercalibration \citep[see][]{magnier2017b}.

The solution, called the Modified Design Reference Mission or MDRM, which was settled upon is shown in Figure\,\ref{fig:mdrm}. 
The Wings pairs in $z,y$ remained the same in the MDRM as the DRM. The sensitivity to discovering asteroids was low in $z,y$ although a few have been found by a pair in $y$ and a pair in $z$. 
In this compromise solution, one third of the data in a given filter is taken in 
a quad, while the remainder is taken in pairs on different nights. The total number of exposures per field per year is still 4, but the cadence will be different in different parts of the sky. This is roughly smoothed out with data over different years where the pattern is shifted by the position of opposition at new moon. Furthermore the pattern is altered by the scheduling around weather and wind. 

One way of visualizing the time domain is shown in Figure\,\ref{fig:timeRA} where the date 
(in Unix seconds \footnote{Unix time is defined as the number of seconds that have elapsed since 00:00:00 Coordinated Universal Time (UTC), Thursday, 1 January 1970}) of every exposure over the 3$\pi$ Survey is plotted vs the right ascension. The slanting bands are the yearly revisiting of the RA in opposition in $g,r,i$ bounded by two visits in $z,y$ per year. For a given
day, one can look along a row of constant time and see the range in RA ascension covered in a night. This banding is by design in the ``strategic" approach, but because the constraints of airmass and sky brightness are generic to an all sky survey, one imagines that the results of the LSST scheduler should show the same pattern if the tension between the parallax cadence (six months) and the sky brightness (twilight) is balanced. 
Compare Figure\,\ref{fig:timeRA} with Figure 1 of \citep{chamberskc2006199843} in advance of the survey. 

\begin{figure*}
\begin{center}
\includegraphics[width=3.0in,angle=0]{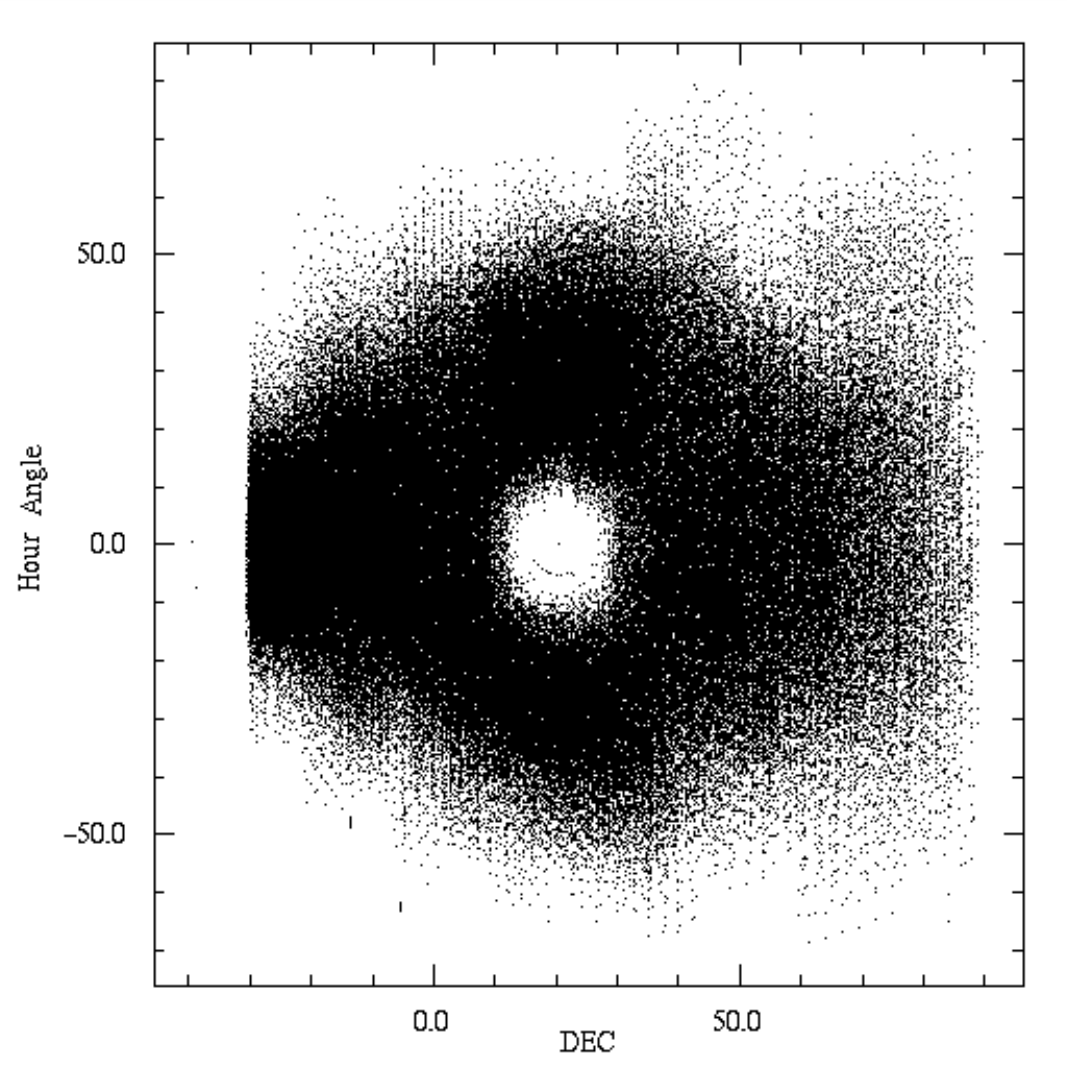}
\includegraphics[width=3.0in,angle=0]{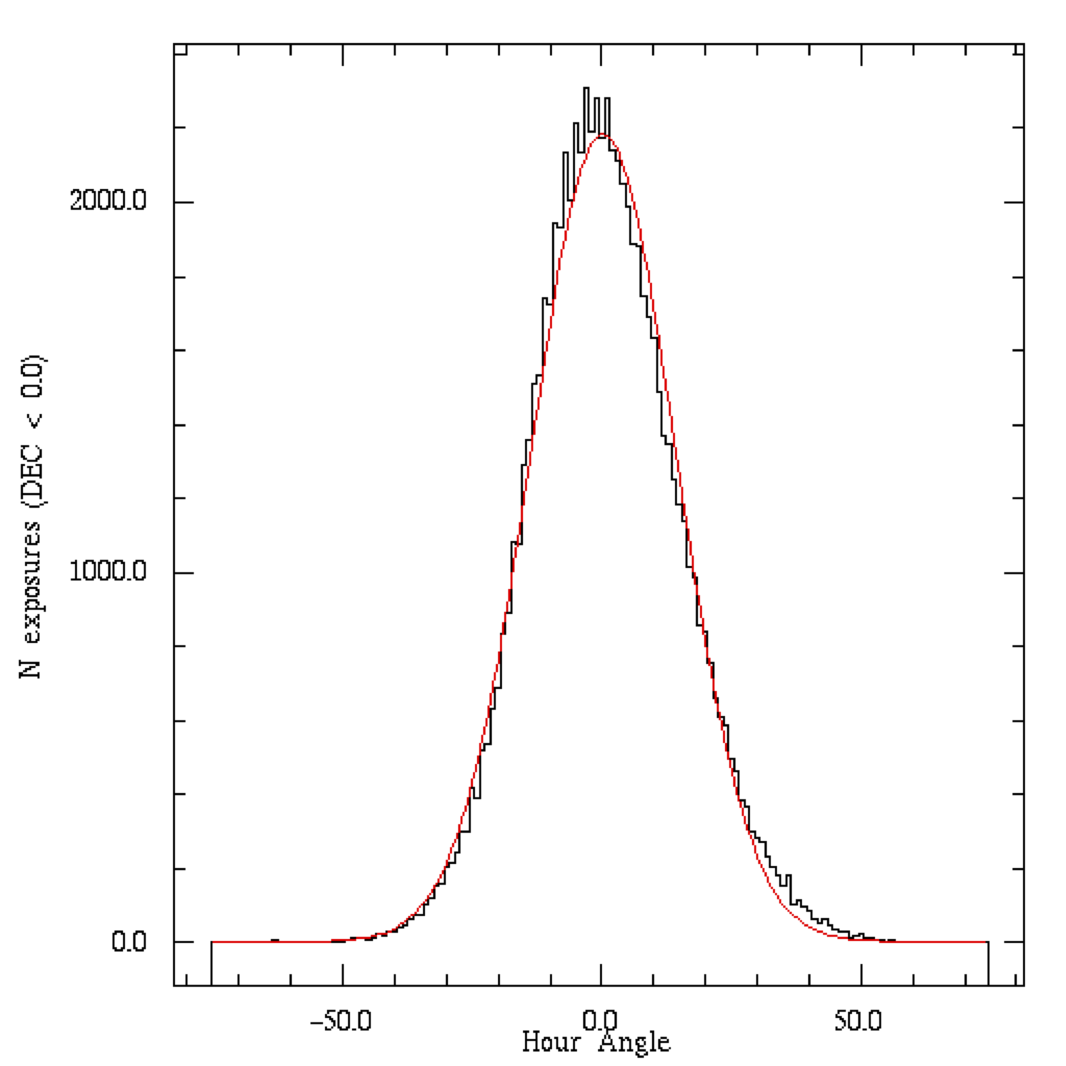}
\end{center}
\caption{Distribution of telescope pointings for the $3\pi$ Survey in Hour Angle vs Declination. The hole in the middle is the keyhole of the Alt-Az PS1 telescope at latitude = XX. The width of the HA distribution is XX.}
\label{fig:HADec}
\end{figure*}

The effectiveness of the strategic scheduling approach and the patient efforts of the PS1 observers who used the scheduling tools to solve the travelling salesman problem in multiple dimensions and responded interactively to the nightly conditions (clouds, wind speed, cadence, sky brightness, survey completeness) is demonstrated in Figure\,\ref{fig:HADec}. This shows the actual distribution of pointings in Dec vs Hour Angle for the entire $3\pi$ Survey. The hole in the middle is the keyhole characteristic of an Alt-Az telescope.  
The hour angle distribution 
shows 
that  65\% of the data is taken within $\sim$1.5 hours of the meridian. 

\subsection{The Medium Deep Survey}
\label{subsec:mdeep}
The Medium Deep Survey consisted of 10 single PS1 footprints on well studied fields spaced approximately uniformly 
around the sky in Right Ascension. The pointing centers of these 10 fields are listed in Table\,\ref{table:mdfields}. The
table includes two additional fields of M31 which can be considered an MDS like field (see Section\,\ref{sec:m31}) and
a field at the north ecliptic pole (NEP). The latter was not observed as extensively as the 10 main fields 
and  was only observed over the period 2010-09-20 to 2011-06-17.

\begin{figure*}
\begin{center}
\includegraphics[width=3.0in,angle=0]
{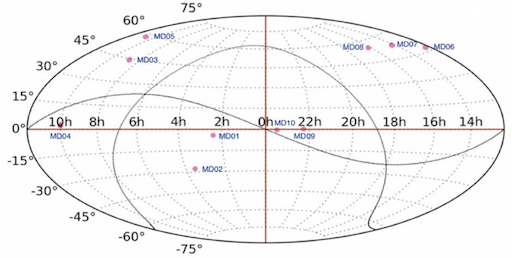}
\includegraphics[width=3.0in,angle=0]
{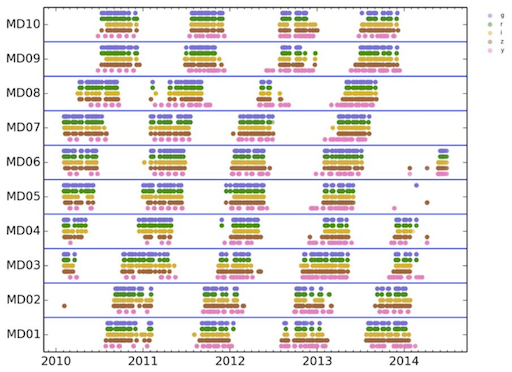}
\end{center}
\caption{Medium Deep Field Survey: on the left the sky positions of the MDS fields are shown. On the right, the cadence of the observing is illustrated. With the \grizy filters labelled in blue, green yellow, brown and green respectively } 
\label{fig:mdsky}
\end{figure*}

The Medium Deep Survey (MDS) component of the program regularly visited these 10
fields (~7 sq. deg. each). Each field was picked to have significant multi-wavelength overlap from previous and concurrent surveys
by other teams and facilities 
(e.g. DEEP2, ELIAS-N, CDFS, COSMOS, GALEX). In total, 25\% of the PS1 time was allocated to the MDS.  
The cadence was generally composed of \gps\ and \rps\ together on one night, followed by \ips\ on the second 
and \zps\ on the third. This pattern was repeated continually on a 3 day cycle over the 6-8 month observing season 
for the field, interrupted only by weather and the moon. Around full moon, the \yps\ filter was primarily used and hence
it does not have the same time cadence as the other 4. Figure\ \ref{fig:mdsky} 
illustrates the cadence and observing seasons 
while Table\,\ref{tab:MDFcadence} \citep[originally presented in][]{2014ApJ...795...44R}
summarises the individual exposure times per filter, which were considerably longer than those for the 3$\pi$ survey.

On each night the 8 separate exposures were dithered and the field was rotated. The
images were then combined into nightly stacks of 904 sec (\gps and \rps) and 1902 sec (\ips, \zps\ and \yps). Roughly once a year these stacks are further combined to produce so-called reference stacks, which are then used as templates for difference imaging. Finally, all the data are combined to produce very deep stacks, which contain several tens of hours worth of exposure.

The description of the MDS was initially presented in \cite{2012ApJ...745...42T} and many papers on transients have
already given an overview of the data products and survey
\citep[e.g.][]{2011ApJ...743..114C,2012Natur.485..217G,2015MNRAS.448.1206M,2015ApJ...799..208S,2016ApJ...831..144L}. 
Estimates of the  typical 5$\sigma$ depths of the MDS nightly stacks were given in \cite{2014ApJ...795...44R}
and are also listed in Table\,\ref{tab:MDFcadence} here. 
Development work continued to improve the single exposure
processing though to deep stacks during the transient event discovery and other science consortium
programs over the course of the survey, the culmination of those improvements being applied in a more
uniformly reprocessed dataset used for the public data release. 
A full discussion of the Medium Deep Fields,  including improved estimates of depths and their special processing will 
be presented in Huber et al. 2019 (in preparation - Paper VII). 
No Medium Deep data will be released in DR2. 

\begin{table}
\begin{center}
\caption{The MDS cadence and exposure times. This table was originally presented in \cite{2014ApJ...795...44R} and is reproduced here in identical form. }
\begin{tabular}{lccc}\hline\hline 
Night  & Filter & Exposure Time & $5\sigma$ Depth \\
       &         & (seconds) & (AB mag) \\ \hline
1 & \gps, \rps & 8$\times$113 each & 23.1, 23.3 \\
2 & \ips & 8$\times$240  & 23.2 \\
3 & \zps & 8$\times$240  & 22.8 \\
repeats \\
Full Moon$\pm$3 & \yps & 8$\times$240  & 21.9 \\
\hline
\end{tabular}
\label{tab:MDFcadence}
\end{center}
\end{table}



 

\begin{table}[htbp]
\caption{Pan-STARRS1 Medium-Deep Fields }
\begin{center}
\begin{tabular}{lrrr}
\hline
\hline
{Field} & {RA J2000} & {Dec J2000} & Overlaps \\
\hline
MD00  &  10.675 & $ 41.267$ & M31\\
MD01  &  35.875 & $ -4.250$ & XMM-LSS-DXS/VVDS-02h\\
MD02  &  53.100 & $-27.800$ & CDGS/GOODS/GEMS \\
MD03  & 130.592 & $ 44.317$ & IFA/Lynx \\
MD04  & 150.000 & $  2.200$ & COSMOS\\
MD05  & 161.917 & $ 58.083$ & Lockman-DXS \\
MD06  & 185.000 & $ 47.117$ & NGC4258 \\
MD07  & 213.704 & $ 53.083$ & DEEP2/Groth Strip\\
MD08  & 242.787 & $ 54.950$ & Elias N1- DXS\\
MD09  & 334.188 & $  0.283$ & SA22-DXS/VVDS-22h\\
MD10  & 352.312 & $ -0.433$ & DEEP2-Field 3\\
MD11  & 270.000 & $ 66.561$ & North Ecliptic Pole\\
\hline
\end{tabular}
\end{center}
\label{table:mdfields}
\end{table}

\subsection{Solar System Survey}
\label{subsec:sol}



\subsection{Pan-Planets stellar transit survey}

For Pan-Planets, seven slightly overlapping fields with overall 40 sq. deg. were observed with PS1, making up about 4\% of the total survey time (see Table \ref{table:ppfields}). Data were collected between 2009 and 2012 in the \ips-band. Depending on seeing, exposure times were either 30 sec or 15 sec. In the first two years of the survey, three fields were observed. From 2011 on, four additional fields were added to the survey area, meaning that the previous three fields have a higher number of visits. On each survey night, the exposures were cycled through the seven fields to minimize saturation effects. We obtained at least 2000 exposures for each point in our FOV and up to 6000 in the overlapping areas between the fields.
The main goal of Pan-Planets is the search for transits from extrasolar planets, mainly hot gas giants close to their star with a special focus on M-dwarfs \citep{2007ASPC..366..326A}.
There are up to 60000 M-dwarfs in the FOV with magnitudes between 13mag and 18mag in the i-band, which makes the survey one of the most comprehensive transit searches for M-dwarf exoplanets. A description of the scientific results
and analysis can be found in \cite{2016A&A...587A..49O}. 
The Pan-Planets stellar transit data is not included in DR1 or DR2. 

\begin{table}[htbp]
\caption{Pan-Planets Stellar Transit Survey Fields }
\begin{center}
\begin{tabular}{lrrr}
\hline
\hline
{Field} & {RA J2000} & {Dec J2000} & Overlaps \\
\hline
PP1 &  298.286 &  19.677& \\
PP2 &  295.937 &  19.100& \\                       
PP3 &  300.124 &  17.638& \\
PP4 &  297.700 &  17.060& \\
PP5 &  295.271 &  16.527& \\
PP6 &  299.462 &  14.994& \\                    
PP7 &  297.033 &  14.450& \\   
\hline
\end{tabular}
\end{center}
\label{table:ppfields}
\end{table}

\subsection{PAndromeda, the M31 transient survey}
\label{sec:m31}
PS1 had a special monitoring survey for M31 for 2\% of the original PS1 survey time. Data were taken from 2010 to 2012 (3 seasons), during the second half of each year when M31 was easily visible. M31 was also covered in the regular 3$\pi$ Steradian Survey. 
As part of the separate survey, 
M31 was visited up to two times per night in the \rps\ and \ips\ filters. Depending on the weather conditions, we obtained up to fourteen 60-second exposures in \rps\ and ten 60-second exposures in \ips. 
These exposures were spread across the two visits per night to give some intra-night time resolution.
The survey strategy was optimized to detect short-term M31 microlensing events, but to also allow one to identify and analyze the variable star content in M31.
Observations were taken much more sparsely in the remaining filters (\gps, \zps\ and \yps) in order to give multicolour maps of M31 in the full PS1 filter complement. The first results and demonstration of data quality from the first 90 nights in 2010 were presented in 
\cite{2012AJ....143...89L}. 
M31 data will not be released in DR1 or DR2. 

\subsection{Calibration observations, CNP, SAS2}
\label{subsec:cal}


\subsubsection{Spectro-photometric and Calspec Standard Stars}

The AB magnitude system calibration of the Pan-STARRS1 photometric system by \cite{2001AJ....122.2118B}
used data from a single photometric night.
and special observations of the HST Calspec sample \citep{2001AJ....122.2118B}.
 All standard stars were
placed on OTA 34 and cell 33, so their integration was on the
same silicon and used the same amplifier for read-out.
However, this position was very close to the center of
the focal plane, where it has been noted that there is a strong
gradient in the behavior of the chip (Rest et al. 2014),
and thus these observations were not included in the 
subsequent study by \cite{2014ApJ...795...45S,2015ApJ...815..117S}.
They analyzed a sample of faint Calspec standards observed over
the course of the $3\pi$ survey and re-determined the AB offsets for the
\gps, \rps, \ips, \zps bands of the PS1 system. 
The super-cal  \cite{2015ApJ...815..117S} AB offsets were used 
in the calibration of all the DR1 and DR2 data \cite{magnier2017c}.
However \cite{2015ApJ...815..117S} note that primary difference 
in the update arises from changes in the Calspec standards. 

\subsubsection{The Celestial North Pole}
The 3$\pi$ Steradian Survey extends to the North Pole. It was soon realized that a dedicated nightly pointing near the Celestial North Pole would provide continuous time coverage that
could monitor the performance of the system as well as be of scientific interest for the 
unique cadence. So a set of \grizy exposures of 
30 seconds was obtained each night
on the meridian and at a declination of 89.5 degrees. Observations were obtained every clear night
between 2010-10-13 and 2014-02-13. The net result is about
a 4 square degree area with regular observations for 3.3 years. These data are not included in DR1 or DR2. 

\subsubsection{Small Area Survey 2}
In July 2011 a test area of the 3$\pi$ survey, consisting of about 70 deg$^2$ centered on $(\alpha,\delta) = (334\degree, 0\degree)$ (J2000), was observed to the expected final depth of the survey in \grizy . These data are described in depth in \cite{2013MNRAS.435.1825M} where general issues of the data and the PS1 reduction software is subject to a rigorous investigation, with emphasis on the depth of the stacked survey. A further paper \citep{2014MNRAS.437..748F} demonstrates how galaxy number
counts and the angular two-point galaxy correlation function, w($\theta$),can be reliably measured. These data are not included in DR1 or DR2. 



\begin{table*}
\caption{Image Processing Pipeline stages and data products}
\begin{center}
\begin{tabular}{lllrrlrll}
\hline
\hline
processing & image      & file  & ID    & Avg No.      &No.& No. IDs  &location  & Release \\
stage      & class      & type  & type & components&filters&in $3\pi$&            &        \\
           &            &       &       & per ID    &per   & Survey  &            &        \\
           &            &       &       &           &ID    &         &            &        \\
\hline
raw   & raw image       &fits   & exp   &60.0000&         1& 374k    & UH$^{a}$& \nodata \\
         
chip  & signal image    &fits   & exp   &59.9984&         1& 374k    & \nodata$^{b}$& \nodata \\
      & variance image  &fits   & exp   &59.9984&         1& 374k    & \nodata    & \nodata  \\
      & mask image      &fits   & exp   &59.9984&         1& 374k    & \nodata    & \nodata  \\
      & detections      &cmf    & exp   &59.9984&         1& 374k    &  both$^{c}$& \nodata \\  
      
camera&detection table  &smf    & exp   &1      &         1&374k    & both    & \nodata \\
           
warp  &  signal image    &fits  &skycell&72.6060&         1&1050k     & MAST    & DR2  \\
      &  variance image  &fits  &skycell&72.6060&         1&1050k     & MAST    & DR2  \\
      &  mask image      &fits  &skycell&72.6060&         1&1050k    & MAST    & DR2 \\
      & detections       &cmf   &skycell&72.6060&         1&1050k     & both    & \nodata \\
          
stack & signal image    &fits   &skycell&1      &         1&1050k     & both    & DR1 \\ 
      & variance image  &fits   &skycell&1      &         1&1050k        & both    & DR1 \\ 
      & mask image      &fits   &skycell&1      &         1&1050k        & both    & DR1 \\
      & number image    &fits   &skycell&1      &         1&1050k        & both    & DR1  \\
      & exp image       &fits   &skycell&1      &         1&1050k        & both    & DR1  \\
      & expwt image     &fits   &skycell&1      &         1&1050k        & both    & DR1  \\
& convolved 6\&8 signal &fits   &skycell&1      &         1&1050k        & \nodata & \nodata \\
& convolved 6\&8 variance&fits  &skycell&1      &         1&1050k        & \nodata & \nodata\\
& convolved 6\&8 mask   &fits   &skycell&1      &         1&1050k        & \nodata & \nodata \\
& convolved 6\&8 number &fits   &skycell&1      &         1&1050k        & \nodata & \nodata \\
& convolved 6\&8 exp    &fits   &skycell&1      &         1&1050k        & \nodata & \nodata \\
& convolved 6\&8 expwt  &fits   &skycell&1      &         1&1050k        & \nodata & \nodata \\
      & detections      &cmf    &skycell&1      &         1&1050k        & both    & \nodata \\ 
      
static sky & detections &cmf    &skycell&1      &         5&201k    & both    & \nodata       \\ 

sky cal & detections    &cmf    &skycell&1      &         1&1050k      & both    & \nodata \\ 

forced & detections     &cmf    &forced &19.1184&         1&19m     &  both    & \nodata \\

 diff & signal image    &fits   &skycell&51.6105&         1&19m     &  \nodata & \nodata \\        
      & variance image  &fits   &skycell&51.6105&         1&19m     &  \nodata & \nodata \\
      & mask image      &fits   &skycell&51.6105&         1&19m     &  \nodata & \nodata \\
      & diff detections &cmf    &skycell&51.6105&         1&19m     &  both    & DR2     \\

\hline
\end{tabular}
\end{center}
\label{tab:products}
\footnotetext{UH means the data is stored in two geographically different locations of the University of Hawaii; Maui (ATRC)  and Oahu (ITC).}
\footnotetext{`\nodata' for location means the source image was not saved after all processing for that region of the sky was completed. They can only be regenerated from the raw (for chip) or stacked and warp images (for the difference images) by re-processing at UH.}
\footnotetext{`both' indicates copies are kept at both MAST and UH.}
\footnotetext{`\nodata' for Release means the cmf and smf files are not
part of the Data Releases, but they are archived at both MAST and UH.}
\end{table*}

\begin{table*}
\caption{Fundamental PSPS database tables}
\begin{center}
\begin{tabular}{lll}
\hline
 Table Class & PSPS Table Name & Release \\
\hline
\hline
Detection & Detection         & DR2 \\ 
Object & ObjectThin           & DR1 \\ 
       & MeanObject           & DR1 \\
       & GaiaFrameCoordinate  & DR1 \\ 
       
Stack & StackObjectThin       & DR1 \\
      & StackObjectAttributes & DR1 \\
      & StackApFlx            & DR1 \\
      & StackApFlxExGalUnc    & DR1 \\
      & StackApFlxExGalCon6   & DR1 \\
      & StackApFlxExGalCon8   & DR1 \\
      & StackPetrosian        & DR1 \\
      & StackModelFitExp      & DR1 \\
      & StackModelFitDeV      & DR1 \\
      & StackModelFitSer      & DR1 \\
Difference& DiffDetection     & DR2 \\
      & DiffDetObject         & DR2 \\ 
      
Forced & ForcedMeanObject     & DR1 \\ 
       & ForcedWarpMeasurement& DR2  \\
       & ForcedMeanLensing    & DR2  \\
       & ForcedWarpLensing    & DR2 \\
       & ForcedGalaxyShape    & DR2  \\
       & ForcedWarpExtended   & DR2\\ 
       & ForcedWarpMasked     & DR2 \\

\hline
\end{tabular}
\end{center}
\label{tab:PSPS}
\end{table*}


\section{Overview of PS1 Data Products}
\label{sec:products}

The PS1 Data Products consist of images of various kinds, catalogs of attributes measured from the images organized in a hierarchical relational database, and potentially derived data products such as proper motions and photometric redshifts, 
and metadata for linking and tracking all of the above. 
Here we provide a brief overview, see \cite{flewelling2017} for details. 
We refer below to types of images and data files listed in 
Tables \ref{tab:products} and \ref{tab:PSPS}. 



The proper convention when reporting Pan-STARRS1 magnitudes 
is to use the nomenclature {\grizy} (see Section\,\ref{subsec:filt}) and the convention for IAU names is 
$$ {\bf PSO JRRR.rrrr+DD.dddd} $$ 
where the PSO identifier stands for Pan-STARRS Object, and the coordinates are in decimal degrees. 
Another  point of interest is the use of magnitudes and fluxes. There are advantages and disadvantages to the use of these, 
and we employ both magnitude and fluxes where useful. 
Luptitudes \citep{1999AJ....118.1406L} also have 
some advantages, but we have made the decision not to use them in the PS1 data products. 
One noteable advantage of fluxes, for example, is that the magnitude of an aperture flux measurement
can correctly be negative when measured on a sky subtracted image and, when the mean of a series of such measurements is computed, the result is well behaved. 
All Pan-STARRS magnitudes are in AB magnitudes \citep{2012ApJ...750...99T} and the fluxes are reported in the corresponding Janskys, where the absolute calibration is discussed in \cite{2014ApJ...795...45S}. 
See also Section\,\ref{subsec:filt}. 


\begin{figure*}
\includegraphics[width=3.6in,angle=0]
{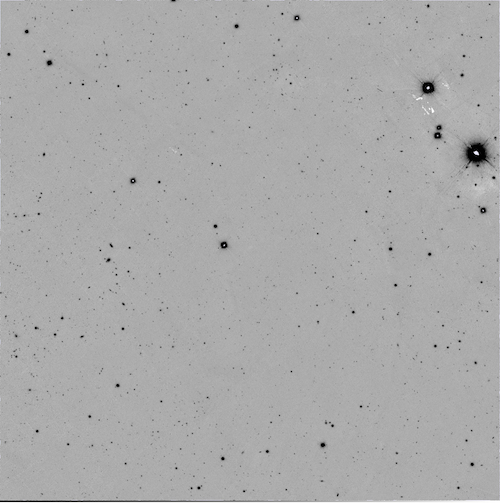}
\includegraphics[width=3.6in,angle=0]
{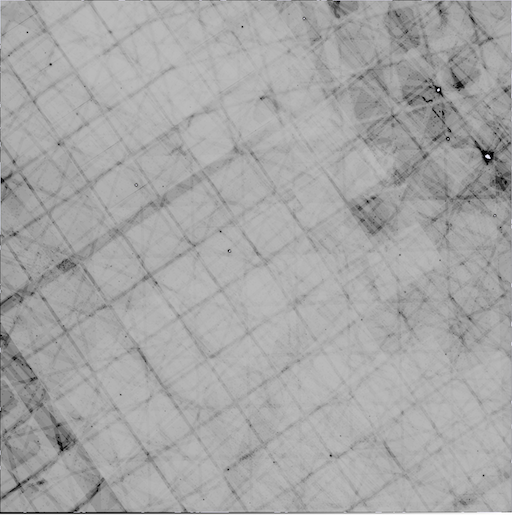}
\includegraphics[width=3.6in,angle=0]
{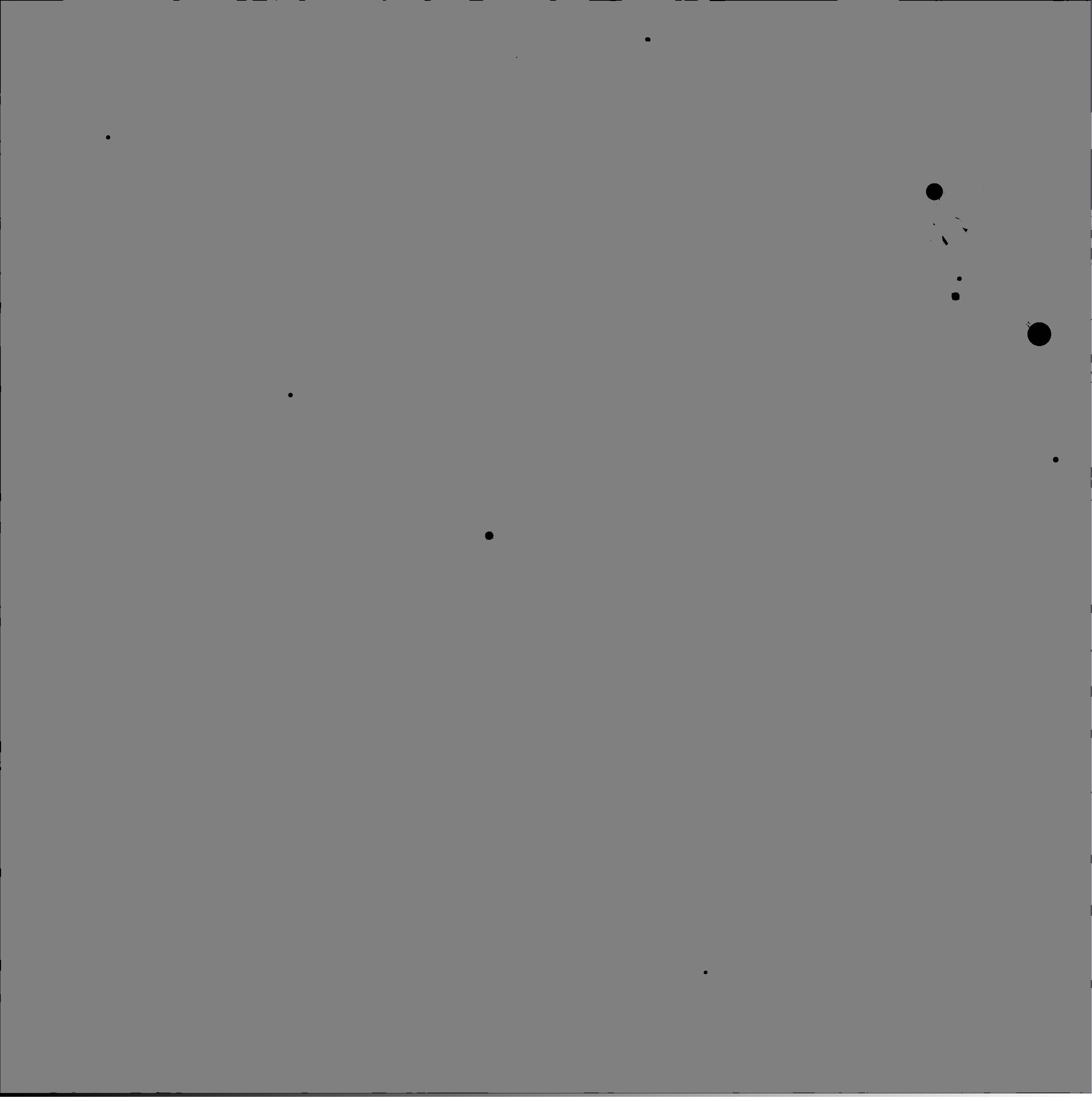}
\includegraphics[width=3.6in,angle=0]
{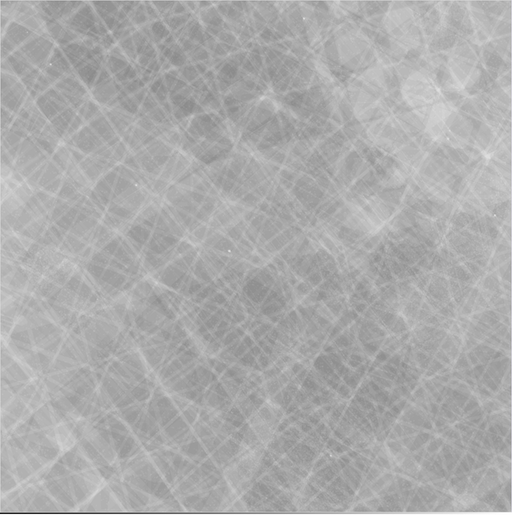}
\caption{Example of image types from a sky cell. Upper Left:signal image. Upper right: variance image. Lower left: mask image. Lower right: number image.} 
\label{fig:images}
\end{figure*}

\subsection{Image Data Products}
\label{subsec:products}

In this section we discuss the specific kinds of images, their properties and location as produced by
the IPP processing stages discussed above in Section\,\ref{subsec:ipp}.
A summary of the kinds of data files that exist including images is provided in Table\,\ref{tab:products}.

The raw pixel data are 
archived in two geographically separate locations; one archival copy is retained on storage machines at the IfA's ATRC on Maui, and another is stored in the IPP cluster, which has moved from its inital location at the Maui High Performance Computing Center (MHPCC) to the Maui Research and Technology Center (MRTC-B) to its permanent location in the UH Information Technology Center on the Manoa campus. Within the IPP all other files have at least two instances on separate 
Raid 10
machines. 

Each Pan-STARRS image (an ``exposure" or  ``frame") creates  60 FITS image files, one for each device in the camera, and each FITS file has 64 extensions, where each extension is 
the pixel data from one OTA cell (see Section\,\ref{subsec:gpc1})
Table \ref{tab:products} lists the various image 
and binary fits table files produced by the IPP by each of its stages. 
Some images are intermediate products and are not saved permanently, although they can be reproduced from the raw data. 

The ``chip" images are the detrended images. The signal image is now a float, and a matching mask and variance image are also produced. 
The detection of objects and measurements of their positions 
and attributes (in detrended pixels) are stored in binary fits tables 
(internally called CMF files). 
These measurements are therefore in $(x,y)$ pixel coordinates and can have any orientation on the sky. Together with the astrometric calibration from 
the ``camera stage" these measurements and their position in $(ra, dec)$  are the basis of the ``Detection Table" \citep{flewelling2017}, 
These are also binary fits tables, (internally called SMF files). 


The ``warp" images are astrometrically registered by a geometric transformation onto rectilinear North-South pixels in a tangential projection using the nearest projection center as defined by the RINGS3 tessellation (Section \ref{sec:primary}).
The warp stage also produces warped mask and variance images, see
\cite{waters2017} for complete discussion.

The ``stack" images are additions of accumulated warp images which should be precisely registered. 
Variance and mask images for the stack are also created as well as a 
number image that shows the number of warps that contributed to the stack at any pixel.
Note, the different warps likely have different PSFs having been taken at different times and at different places in the focal plane, leading to what are essentially intractable problems in PSF measurements performed on the stacks. This is the motivation for the ``Forced Photometry" stage \citep{magnier2017a}.
The results of the analysis of the stacked images are stored in binary fits tables 
(again, internally labelled as CMF files)
and are available in the Stack tables \citep{flewelling2017}. 

Two convolved versions of the stack images are created by convolving with gaussians of width 
6 or 8 pixels 
(precisely 1.5 and 2.0 arcseconds in 0.25 arcsecond sky cell pixels).
These are then used for aperture measurements \citep{magnier2017a}. 
The convolved images are intermediate products and are not saved. The 
aperture measurements are stored in binary fits CMF files and are available 
in the Stack Aperture tables.

Difference images and their associated variance and mask images are 
created by the ``Diff Stage" and measured, with the results going in the
Difference Tables. The Difference images are not retained, but could in principle be regenerated from the stacks and warps. 
 
\begin{figure*}
\begin{center}
\includegraphics[width=\textwidth,angle=0]
{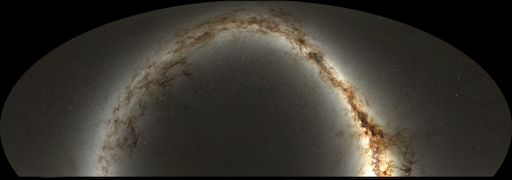}
\end{center}
\caption{Color image constructed from 16x16 binned \gps, \rps, and \ips versions of the 3$\pi$ stack images by Daniel Farrow. The binned images were converted into HEALpix \citep{2005ApJ...622..759G} pixels and from this a color image was created using the software presented in \cite{2012ASPC..461..263B}.} 
\label{fig:threepi}
\end{figure*}
               


\subsection{Data Products}
\label{subsec:fund}




The Pan-STARRS1 database schema (\citep{flewelling2017}) is 
organized into four sections:
\begin{enumerate}
\item Fundamental Data Products. These are attributes that are calculated from either detrended  but untransformed pixels or warped pixels. The instrumental fluxes or magnitudes have been re-calibrated, as have their positions. Because of these calibrations, the catalog values are to be preferred to making a new measurement from the images. See Table \ref{tab:PSPS} for a list of PS1 catalogs.
\item Derived Data Products. These are higher order science products that have been calculated from the Fundamental data products, such as proper motions, photometric redshifts, associations of detections of moving objects by MOPS. 
\item Observational Metadata. These are metadata that provides detailed information about the individual exposures (e.g. PSF model fit) or which exposures went into an image combination (stacks and diffs) of exposures, 
as well as information such as detection efficiencies. 
\item System Metadata. These tables have fixed information about the system and the database itself, including descriptions of various flags. 
\end{enumerate}
Various database ``Views" or logical combinations of Tables are also constructed as an aid for common kinds of queries. Note in the PSPS architecture, large tables (almost all except the ``head node" ObjectThin, MeanObject, and GaiaFrameCoordinate) are actually ``views" joining subsections (slices) of the
data across different file partitions, but this structure is hidden from the casual user.

The classes of tables in the Fundamental Data Products include 
Detection, Object, Stack, Difference, and Forced (Table\ \ref{tab:PSPS}). We now consider each of these in turn.



\subsubsection{Detection Table}
At the most basic level, an individual ``Detection" is a feature, 
likely a star or galaxy or artifact, detected above the noise in an
individual exposure. There are likely, but not always, multiple ``detections" of the same astronomical object from subsequent exposures

The majority of single detections at the faint end are not real, but arise from systematic noise, primarily correlated read noise, in the GPC1 \citep{waters2017}. A wealth of flags are provided 
to help distinguish between real detections and artifacts \citep{flewelling2017}. Nonetheless
there are detections that arise from systematic noise that are,
by themselves, indistinguishable from real features.



For each single epoch detection, the Detection Table contains PSF magnitudes,
total aperture-based magnitudes,  Kron magnitudes \citep{1980ApJS...43..305K},
assorted radial moments and combinations of moments,
and circular radial aperture magnitudes in SDSS radii R3 through R11 \cite{2002AJ....123..485S}.
See \cite{magnier2017b} for details. 


\subsubsection{Object Tables}
\label{subsec:objecttab}
Individual ``detections" are associated into
``Objects" by virtue of being approximately at the same location in
$(ra, dec)$. The IPP makes this association of detections into objects in the ``static sky stage" if they are within 
1.0 arcseconds
%
%
and there are various complications for blended objects or single objects that become resolved in a subsequent, 
higher quality image \citep{magnier2017b}.
It is possible that an astronomical object is only measured once even with multiple exposures with good pixels at the same location - for example a moving object, or a transient object, 
or an object that only rises above the noise in one image. 
Systematic noise, especially the correlated read noise in the GPC1 detector \citep{waters2017}, 
can also contribute a faint artifact that is interpreted as a single detection and becomes an Object. 
Such single instances must also be elevated into ``Objects" because at the time there is no independent way of knowing. Thus the association  of detections into objects is one-or-more to one. 
Thus one-time-only false detections from artifacts 
are also promoted to Objects. 
In the Pan-STARRS1 dataset, as a consequence of these features produced by the GPC1, such artifacts dominate the Detection and Object Tables. 
One simple way to exclude them is to require an Object to have two or more detections; an occurrence which is decreasingly likely to happen if the feature isn't real. This could also obviously exclude real moving and transient objects. Sample queries to produce robust catalogs from the Pan-STARRS1 database are provided in \cite{flewelling2017}, but one should always be aware of this aspect of Pan-STARRS1 data. 


The Object Tables described in \cite{flewelling2017} include the ObjectThin Table, which contains the most minimal information set about an object, primarily its position and various indexes linking it to other tables. There are two (ra,dec) positions provided, a ``Mean" position and a ``Stack" position. Mean positions are the most accurate if available, as they come from a mean of all the individual epoch measurements, each of which have been calibrated on the Gaia \citep{2016A&A...595A...4L} reference frame. Objects that are only detected in the stack are fainter and their positions in DR1 have not been re-calibrated on the Gaia frame. 
This is because their uncertainties are intrinsically larger and hence this is only an issue for the most demanding astrometry. 

The MeanObject Table contains the mean photometric information for objects based on the single epoch data, calculated as described in 
\cite{magnier2017b}.
To be included in this table, an object must be bright enough to have been detected at least once in an individual exposure. PSF, Kron \cite{1980ApJS...43..305K} and aperture magnitudes and statistics are provided for all filters. 

The GaiaFrameCoordinate Table contains the re-calibration  of the astrometric positions of all MeanObjects on the Gaia reference frame \citep{magnier2017c}.


\begin{figure*}
\includegraphics[width=3.5in,angle=0]
{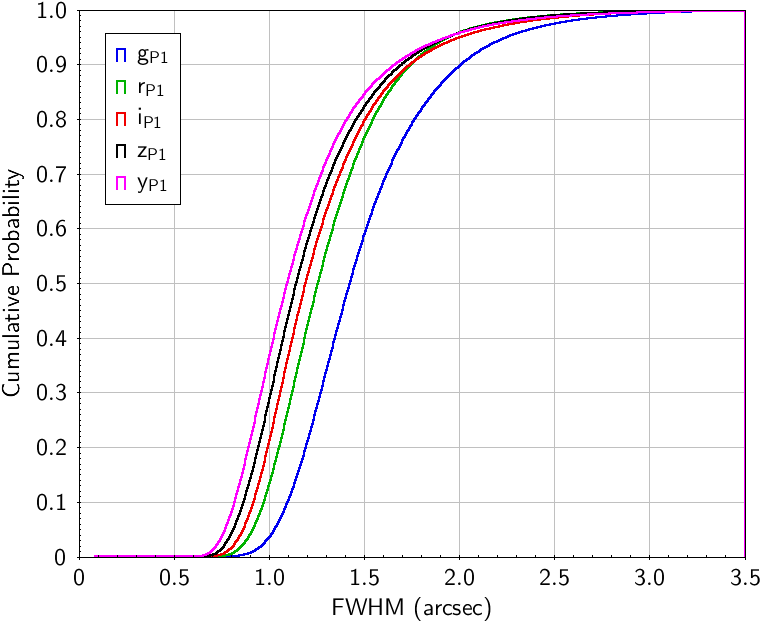}
\includegraphics[width=3.5in,angle=0]
{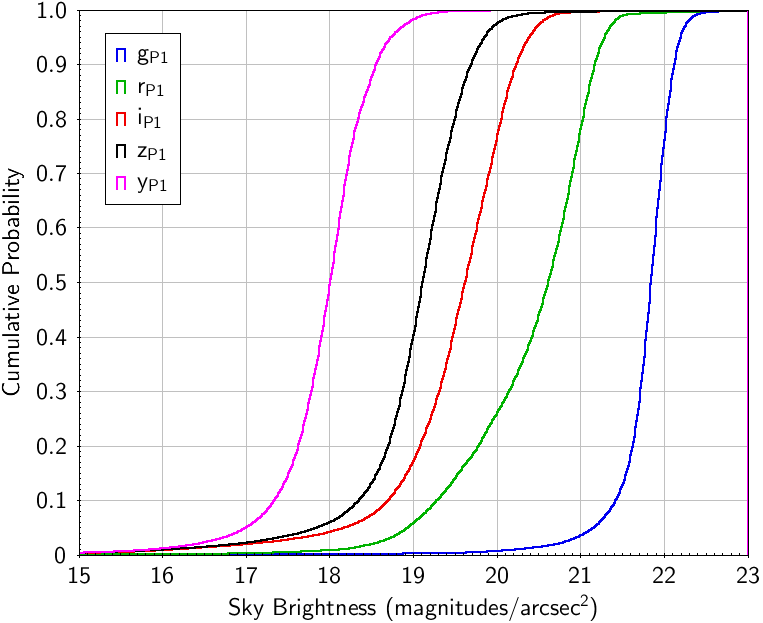}
\caption{Left: FWHM cumulative probability distribution for all the observations in the $3\pi$ Survey. Note that warp images where the FWHM exceeds 2.5 arcseconds are excluded from the stack images. Right: Cumulative probability distribution for the sky brightness in mag/arcsec$^2$ for all the observations in the $3\pi$ Survey.
} 
\label{fig:fwhm}
\end{figure*}

\begin{figure}
\includegraphics[width=\columnwidth,angle=0]
{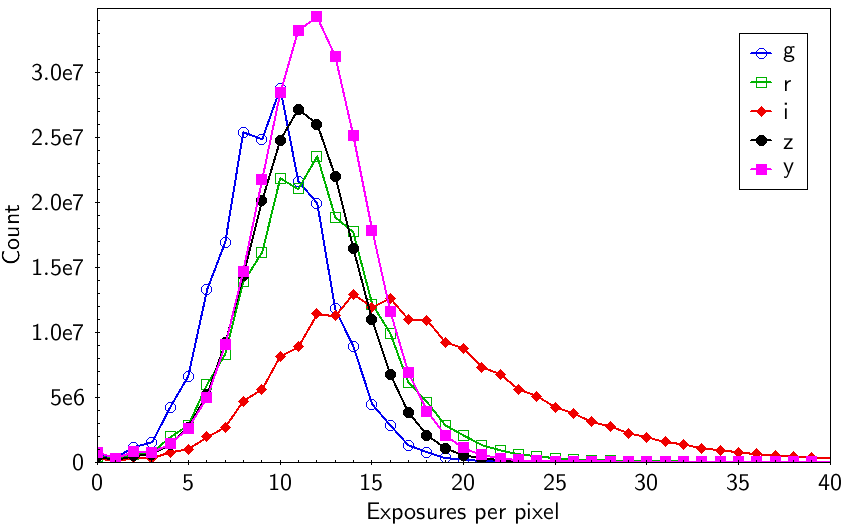}
\caption{The number of exposures contributing to each 4\arcsec\ binned pixel of the stacked 3$\pi$ survey, for \grizy. }
\label{fig:coverage}
\end{figure}

\subsubsection{Stack Tables}
\label{subsec:stack}
Attributes measured on the stacked images are reported in the Stack Tables.
The StackObjectThin table contains the most minimal positional {\it and} photometric information for point-source photometry of stack detections. The information for all filters are joined into a single row, with metadata indicating if this stack object represents the primary detection. See Section\ \ref{sec:primary}.

The StackObjectAttributes table is analogous to the Detection table for
single epoch images and it contains the PSF, \cite{1980ApJS...43..305K}, and aperture fluxes for all filters in a single row, 
along with assorted radial moments and combinations of moments.
 
The StackApFlx Table contains the unconvolved and convolved (to a target PSF of 1.5 and 2 arcseconds) fluxes within the SDSS 
R5, R6, R7 apertures \citep{2002AJ....123..485S} for all Stack Objects. 
The StackApFlxExGal Unc, Conv6, Conv8,  Tables
 contain the unconvolved and convolved fluxes within the SDSS R3 through R11 
 apertures \citep{2002AJ....123..485S} for objects in the extragalactic sky, i.e., they are not provided for objects in the Galactic plane because they are not useful in crowded areas. 
 For each aperture we report: flux (janskys), flux error, flux standard deviation (from the individual measurements), and the fill factor of the aperture (masked pixels could reduce this from 1.0). 
 
The StackPetrosian and StackModelFit Exp, DeV, and Ser tables report the results of fits of extended sources to model PSF convolved surface brightness profiles. The measurements include Petrosian magnitudes and radii \citep{1976ApJ...209L...1P} , exponential, de Vaucouleurs \citep{1948AnAp...11..247D}, and Sersic \citep{1963BAAA....6...41S} magnitudes and radii and elliptical aperture magnitudes and errors for a signal-to-noise ratio 
and galactic latitude limited sample.
See \cite{magnier2017b} for details.

 
\subsubsection{Difference Tables}
The IPP generates Alard-Lupton convolved difference images for skycells in various combinations depending on the survey goals. 
See Price et al. 2019 for details on the implementation of the Alard-Lupton algorithm in the IPP. 
For the 3$\pi$ Survey a difference image: $ diff = (warp - stack )$
is created for each epoch. 

This difference image is then analyzed in the same fashion as an
individual warp. The DiffDetection table is analogous to the Detection
table and has the same measurements. If possible, DiffDetections are associated into DiffObjects, e.g. different points on a light curve 
are associated into a Difference Object. 

No attempt is made to associate Diff Objects with Objects. 
While this might make sense for a variable object, a transient source,
e.g., a supernovae in a galaxy, could be undetectable in either an individual warp or stack, and yet be clear in the difference image. In this case
the closest Object would be the host galaxy, but that association would 
be incorrect. Hence Difference Objects are a unique class and not contained
in the Object Table. On the other hand, a ``good" match between Objects and 
DiffObjects would provide a candidate for a variable object. 
The Difference Tables are not included in DR1 or DR2, they will be included in DR3. 

\subsubsection{Forced Photometry Tables}

Forced photometry is carried out on the warps at the positions of all significant objects 
found in at least two bands in the stacks. This requirement keeps the number of forced objects to a practical number. 
Single band detections, especially $z$-band dropouts, i.e. objects found only in $y$-band, are a non-trivial subject of active research. 

The forced detection measurements made on individual warp images
are reported in the ForcedWarpMeasurement table. 
Where the field-of-view of the exposure contains the position of the 
object, but its properties can not measured because the data happens to be masked at the position, 
the missing object's ObjectID is stored in the ForcedWarpMasked Table to indicate that it was in the field of view of the telescope at the time of the exposure, but could not be measured. It is important to distinguish this case from the case
where a non-detection arises from a lack of signal to noise, but there are active pixels at the position and an upper limit can be obtained from
the properties of the warp image.

The ForcedWarpExtended table contains the single epoch forced photometry fluxes within the SDSS R5, R6, and R7 
apertures \citep{2002AJ....123..485S}.
ForcedWarpLensing contains the 
contains the mean lensing parameters \citep{1995ApJ...449..460K}
of objects detected in stacked images measured 
on the individual single epoch data. The individual epoch measurements
are not reported, only their mean. 
ForcedMeanObject has the mean properties of the individual forced measurements, including PSF, Kron, and aperture magnitudes, and
R5, R6, and R7 apertures. See \cite{magnier2017b}.
The Sersic and DeVaucouleur extended source model fits are also forced and reported in the ForcedGalaxyShape table.  In the forced galaxy shape analysis, the stack model fits are used as a prior for the fits to the galaxy on each warp image.  The positions and aspect ratio, as well as the Sersic index (as appropriate) are fixed to the stack value.  The major and minor axes of the model are allowed to vary in a grid about the stack values and the model realizations are convovled with the PSF for each warp image to determine the normalizations and $\chi^2$ values.  The separate warp measurement are combined to yield a best-fit galaxy model for the full set of observations. 




\begin{figure*}
\includegraphics[width=3.5in,angle=0]
{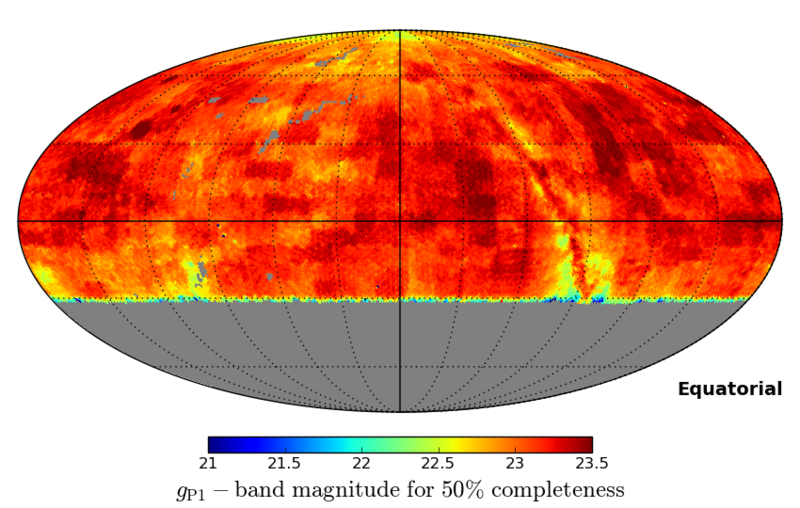}
\includegraphics[width=3.5in,angle=0]
{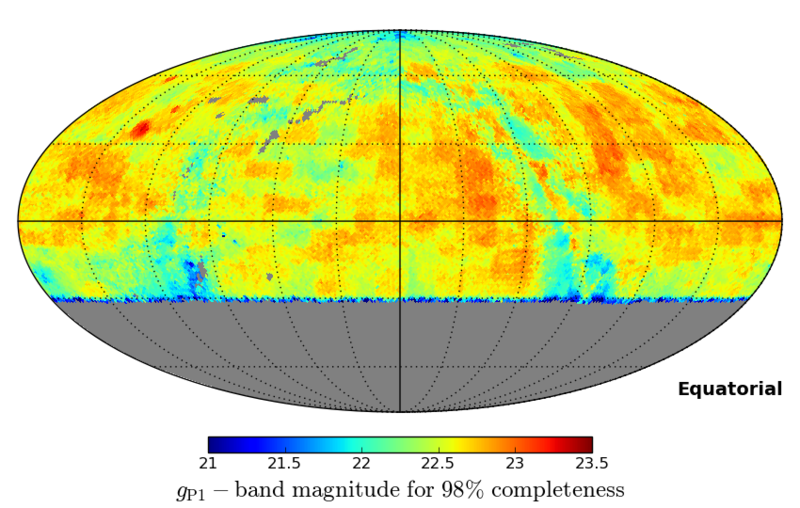}
\includegraphics[width=3.5in,angle=0]
{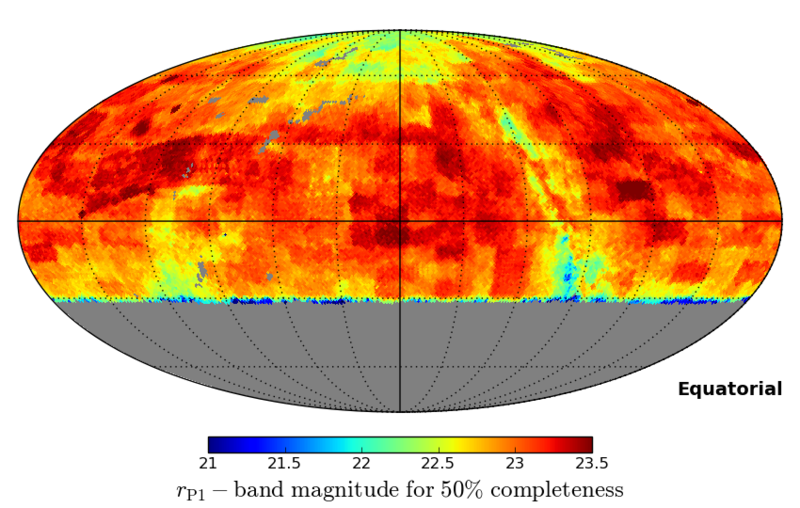}
\includegraphics[width=3.5in,angle=0]
{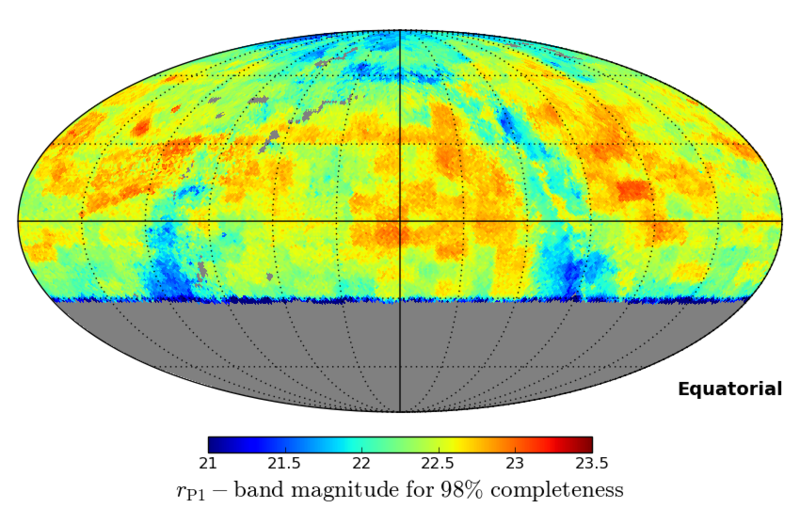}
\includegraphics[width=3.5in,angle=0]
{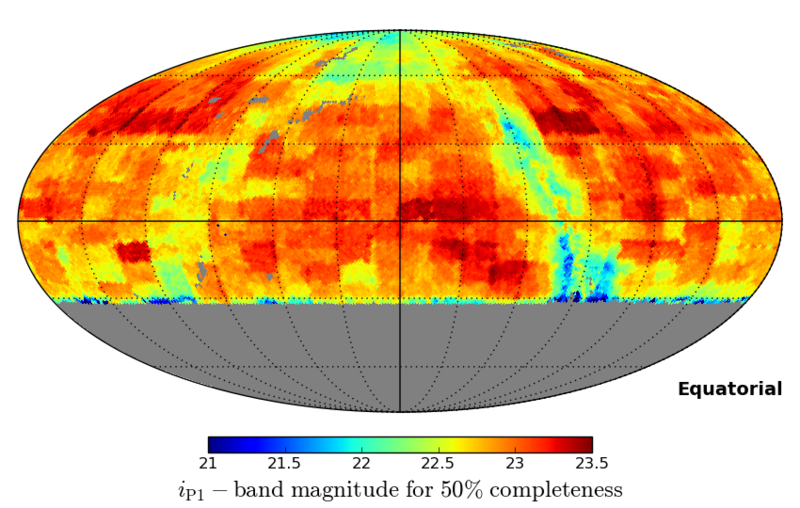}
\includegraphics[width=3.5in,angle=0]
{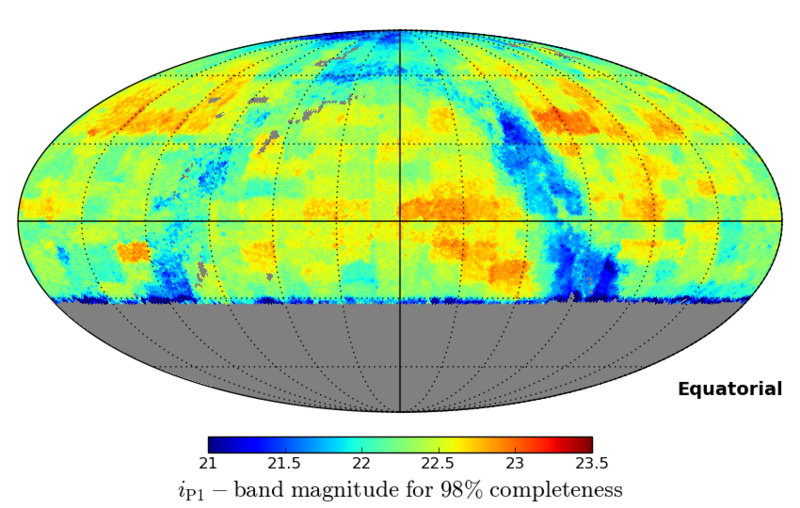}

\caption{left: The all-sky distribution of magnitude limits for 50\% completeness on the 3$\pi$ stacked data in the \gps \rps \ips bands, based on the recovery of injected fake point-sources;
         right: 98\% completeness for the same sample.} 
\label{fig:fakes1}
\end{figure*}

\begin{figure*}
\includegraphics[width=3.5in,angle=0]
{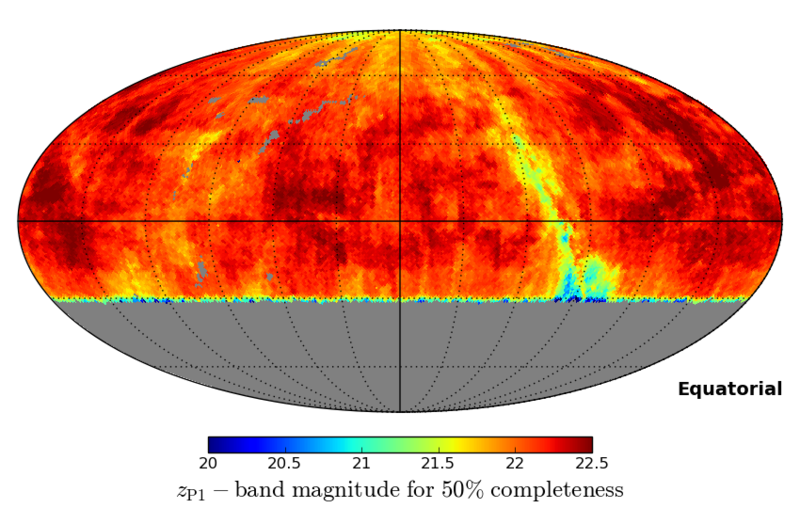}
\includegraphics[width=3.5in,angle=0]
{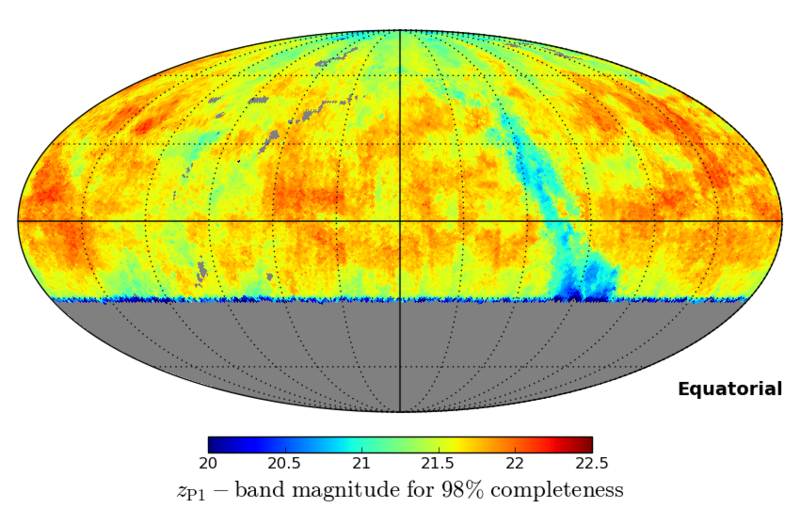}
\includegraphics[width=3.5in,angle=0]
{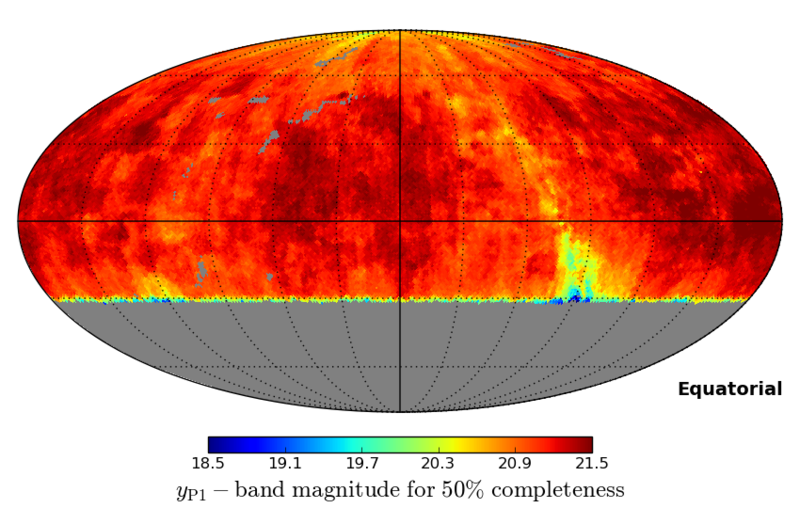}
\includegraphics[width=3.5in,angle=0]
{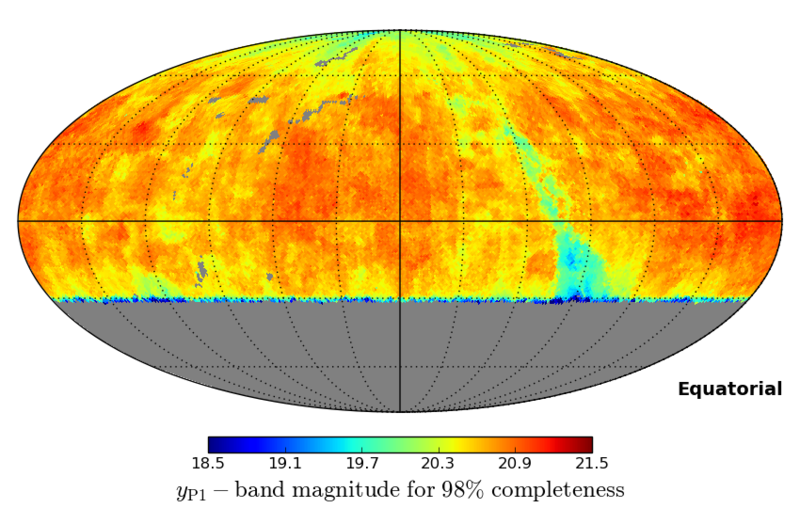}
\caption{As Figure \ref{fig:fakes1} but now showing the completeness for the \zps and \yps bands.}  
\label{fig:fakes2}
\end{figure*}



\subsection{Derived Data Products}
\label{subsec:derived}

Derived data products are results that can not be traced directly back to the pixels but are the result of systematic analysis of the Fundamental Data Products discussed  above. These include  
(i) measurements of proper motion and parallax
(Magnier et al. in prepation) made from an analyis of the minute changes
in the positions of objects; 
(ii) Photometric redshifts deduced from aperture
magnitudes using a variety of machine learning techniques 
,(iii) the extinction and stellar parameters deduced from stellar photometry; 
(iv) associating detections in the database with known or discovered 
moving objects in the MOPS database.  We have the tools to ingest these
derived data products back into the PSPS and make them widely available 
to the community. Our intention is to include these 
derived data products in future Data Releases.


\section{Accuracy and precision of the PS1 data set}
\label{sec:calib}
\subsection{Astrometric}
%
%

       
The Pan-STARRS1 astrometry has been re-calibrated \citep{magnier2017c} 
using Gaia \citep{2016A&A...595A...1G}
The Gaia DR1 \citep{2016A&A...595A...2G} catalog \citep{2016A&A...595A...4L} was used as the input reference catalog. 
After recalibrating all individual epoch measurements to the Gaia Frame, and then re-constructing the mean PS1 positions
we can estimate the astrometric error of the resulting PS1 postions. 
The systematic uncertainty of the astrometric calibration using the Gaia frame comes from a comparison of the results with Gaia: 
the standard deviation of the mean and median residuals ($ \Delta ra, \Delta dec $ )
are (2.3, 1.7) milliarcsec, and (3.1, 4.8) milliarcsec.
The latter is a measurement of the bright end errors for average positions 
while the former is a measurement of the consistency of the PS1 and Gaia systems (\cite{magnier2017c}).





\begin{table*}
\caption{3$\pi$ Steradian Survey Characteristics }
\begin{center}
\begin{tabular}{lrllllllll}
\hline
\hline
{Filter}&No.&Nominal&{5$\sigma$}&{Bright}& Mode   & Median & Mode         &Median & 5$\sigma$\\
   & validated &exposure&single & star   & of PSF &of  PSF & of sky       &of sky & stack \\
   & exposures &secs    &epoch   &limit   &distribution  &distribution  &brightness&brightness &limit\\
   &           &        & mag    & mag    &arcsec  &arcsec  &mag/arcsec$^2$&mag/arcsec$^2$&mag\\
\hline
\gps  & 60528 &    43& 22.0 & 14.5  & 1.31  & 1.47 & 21.89 &  21.86  & 23.3           \\
\rps  & 70918 &    40& 21.8 & 15.0  & 1.15  & 1.31 & 21.04 &  20.62  & 23.2           \\
\ips  &104414 &    45& 21.5 & 15.0  & 1.05  & 1.19 & 19.86 &  19.61  & 23.1           \\
\zps  & 67604 &    30& 20.9 & 14.0  & 1.00  & 1.14 & 19.27 &  19.10  & 22.3           \\
\yps  & 70982 &    30&19.7 & 13.0  & 0.95  & 1.09 & 18.09 &  17.99  & 21.4           \\
\hline
\end{tabular}
\end{center}
\label{tab:stats}
\end{table*}

\subsection{Photometric}
%
%
        
The photometric accuracy of the PS1 data products has been demonstrated in the ubercal analysis 
\citep{2012ApJ...756..158S}
and
relative photometric analysis
\citep{2013ApJS..205...20M}. 
Zero points for photometric data are determined with a reliability of 7-12
millimags. Individual detections in the 3$\pi$ survey have photometric accuracy limited at the bright end
to $\sim$12 millimags per
epoch. The current limits on the photometric precision are driven by our ability to model the 2D variations in the shape of the
PSF. The PSF shape in a given exposure changes on a variety of spatial scales due to 3 major effects: the atmosphere, the
optics, and the detector. To the extent that the PSF model is unable to follow the PSF variations, the PSF photometry is
biased either high or low as the model PSF under or
over predicts
the size of the PSF. The optics introduce image quality
variations due to ripples in the focal surface. These variations occur on spatial scales of ~10 arcminutes and are relatively
stable between exposures, introducing photometry errors of a few millimags. The atmosphere introduces stochastic
variations due to uncorrelated seeing across the focal plane, with a similar level of impact. The detectors introduce PSF
changes due to variable diffusion resulting from variations in the doping characteristics with spatial scales down to 10s of
arcseconds. With the density of PSF stars available in a typical PS1 exposure at high Galactic latitude, we are able to
model the PSF variations on spatial scales of ~3 arcmin, placing a limit on the accuracy of the PSF model on small scales.
See \cite{magnier2017c} for details.

\begin{table*}
\caption{3$\pi$ Steradian Survey Photometry Scope }
\begin{center}
\begin{tabular}{ll}
\hline
\hline
Measurement&Scope\\
\hline
PSFMag  & All objects\\
ApMag  & All objects\\
KronMag & All objects\\
Fixed apertures R5-R7& All objects\\
Fixed apertures R3-R4, R8-R11& Objects outside Galactic plane\\
Petrosian & Objects outside Galactic plane with S/N$>20$\\
Sersic & Objects outside Galactic plane with S/N$>20$ \\
Exponential & Objects outside Galactic plane with S/N$>20$ \\
de Vaucouleurs & Objects outside Galactic plane with S/N$>20$ \\
\hline
\end{tabular}
\end{center}
\label{tab:sscope}
\end{table*}

\section{3$\pi$ Survey Characteristics}
\label{sec:3pi}

The $3\pi$ Steradian Survey is comprised of 374,446 validated images
taken between 2009-06-03 and 2015-02-25. This includes some commissioning data taken before the start of the Mission and a modest number of images taken after the formal end of the survey, primarily in \zps \yps during twilight to smooth out the spatial distribution.

\subsection{Summary of performance metrics}
\label{subsec:metrics}

The image quality in the Pan-STARRS1 Surveys varies significantly,
this is one reason for the forced photometry measurements.
The site, Haleakala Observatories (HO), is a well characterized site, 
and the lower limit to the seeing distribution is equivalent to
Mauna Kea, but the distribution is broader and the median seeing 
as recorded by the HO Differential Image Motion Monitor (DIMM) 
is 0.84 arcsec, with a mode of 0.66 arcsec. 
However, PS1 has a floor to its image quality, arising primarily from the wide field optics, so even the best images do not have a 
FWHM $< 0\farcs6$. The image quality also depends on the filter, with the reddest bands displaying the best. 
Figure\ \ref{fig:fwhm} shows the cumulative distribution of the image quality as characterized by a FWHM for each filter for the PS1 Surveys.  
Haleakala is known for very low atmospheric scattering, that is why it is preferred to Mauna Kea for the Solar Telescope, the sky is even darker than Mauna Kea. Solar astronomers assert this is due to the fact that the summit of Haleakala is  primarily rock, whereas the summit of Mauna Kea is primarily cinder, and that summit of Mauna Kea is constantly surrounded by a halo of microscopic volcanic cinders.  
Figure\ \ref{fig:fwhm} also shows the cumulative distribution of the sky brightness in each filter for the PS1 Surveys. 

Table\ \ref{tab:stats} provides a summary of the characteristics of the 3$\pi$ Survey. Table \ref{tab:sscope} lists the coverage on the sky of the various different kinds of photometric measurements.


%

\subsection{Variation of $3\pi$ Steradian Survey Depth}
Although by the design of the survey each pixel on the sky notionally has 12 visits, in practice the coverage can be much more variable than this. Figure \ref{fig:coverage} shows the distribution of the number of exposures which contribute to each 16x16 binned pixel (4x4\arcsec) over the whole of the 3$\pi$ stacked survey. The result of this is that the depth of the stacked survey varies significantly on quite small scales.
To estimate the depth, in the reduction of a skycell, artificial point-sources are added in magnitude bins and run through the process of being detected. The numbers of these fake sources recovered and inserted as a function of magnitude is stored, for the stacked data, in the StackDetEffMeta table. Maps of depth can be produced by finding at what magnitude a particular percentage of fake point-sources is
recovered for each skycell, using linear interpolation between different
magnitude bins when necessary. To visualise the results across the whole
survey, it is convenient to take the mean of these magnitudes for each
skycell landing in a particular HEALpix\footnote{http://healpix.sourceforge.net} \citep{2005ApJ...622..759G} pixel. Figures \ref{fig:fakes1} and \ref{fig:fakes2} show the results of this procedure for recovery rates of 50\% and 98\%. Not all the variation in limiting magnitude seen is due to the coverage. For instance, in the Galactic plane crowding can significantly reduce the number of recovered fakes. It should also be noted that these limits are for point sources - \cite{2013MNRAS.435.1825M} showed that the limits for extended sources are roughly $0.5$ mag brighter, although this is, of course, depends on the profile of the source.

Tests of how well these fake sources reproduce the true
point-source recovery fractions, as well as a method of producing even
higher resolution maps of depth will be presented in Farrow et al (in
preparation).  

\subsection{Simple star/galaxy separation}
\begin{figure}
\includegraphics[width=\columnwidth,angle=0]
{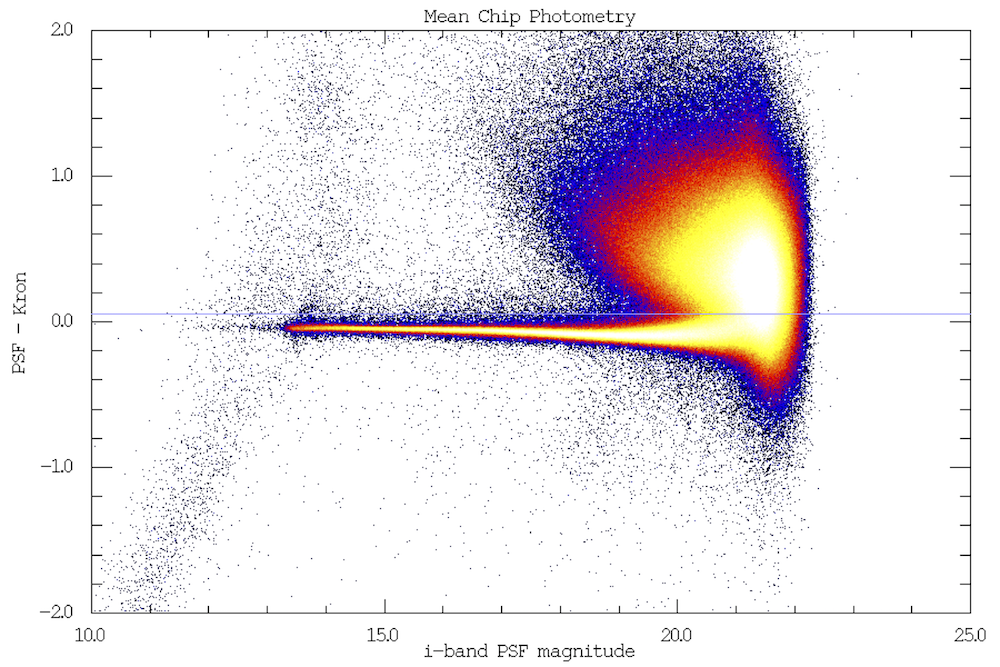}
\caption{A demonstration of simple star galaxy separation using (PSF-Kron) magnitudes for a sample of \ips-band chip detections around the galactic pole. } 
\label{fig:sgsep}
\end{figure}

\begin{figure}
\includegraphics[width=\columnwidth,angle=0]
{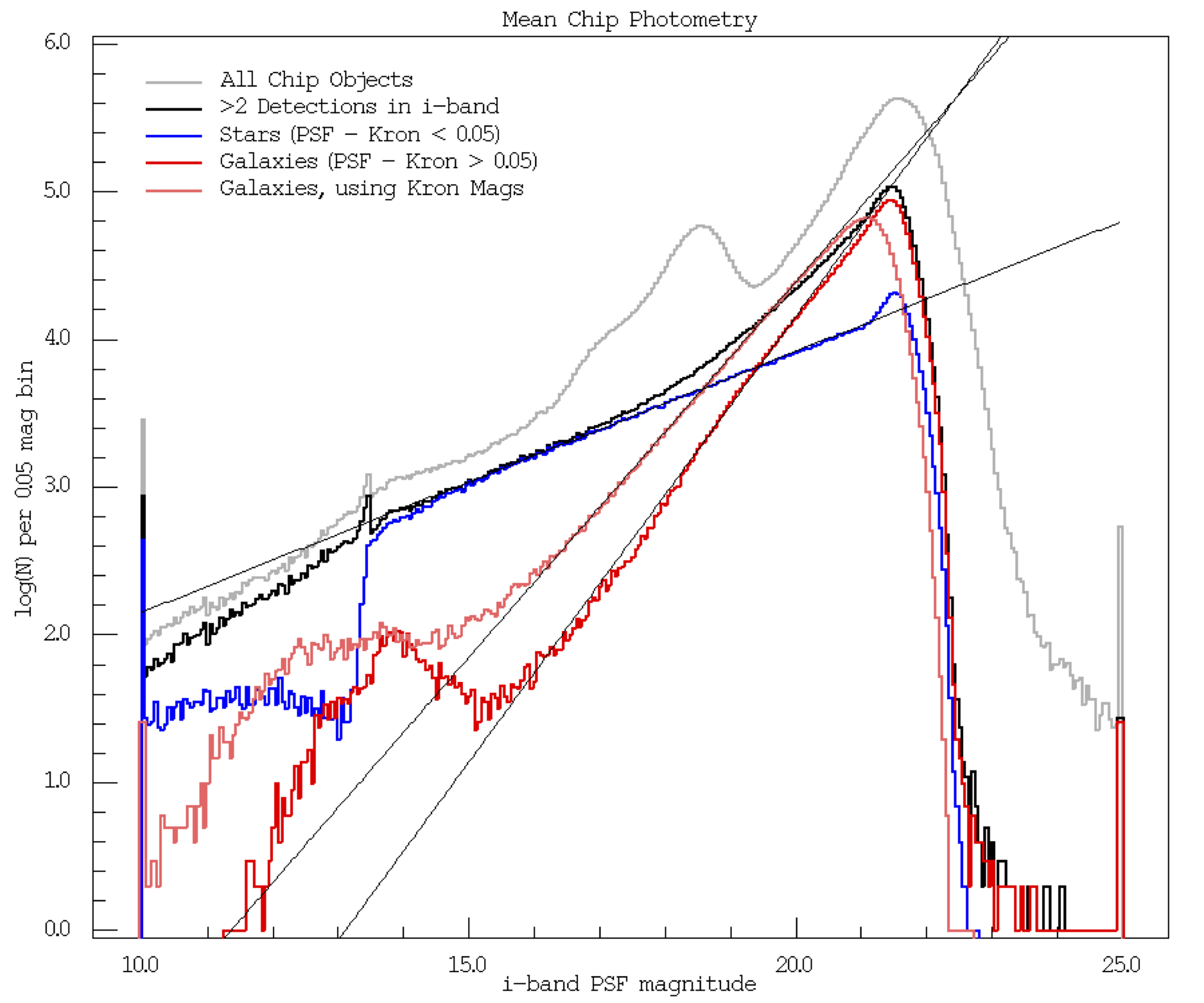}
\includegraphics[width=\columnwidth,angle=0]
{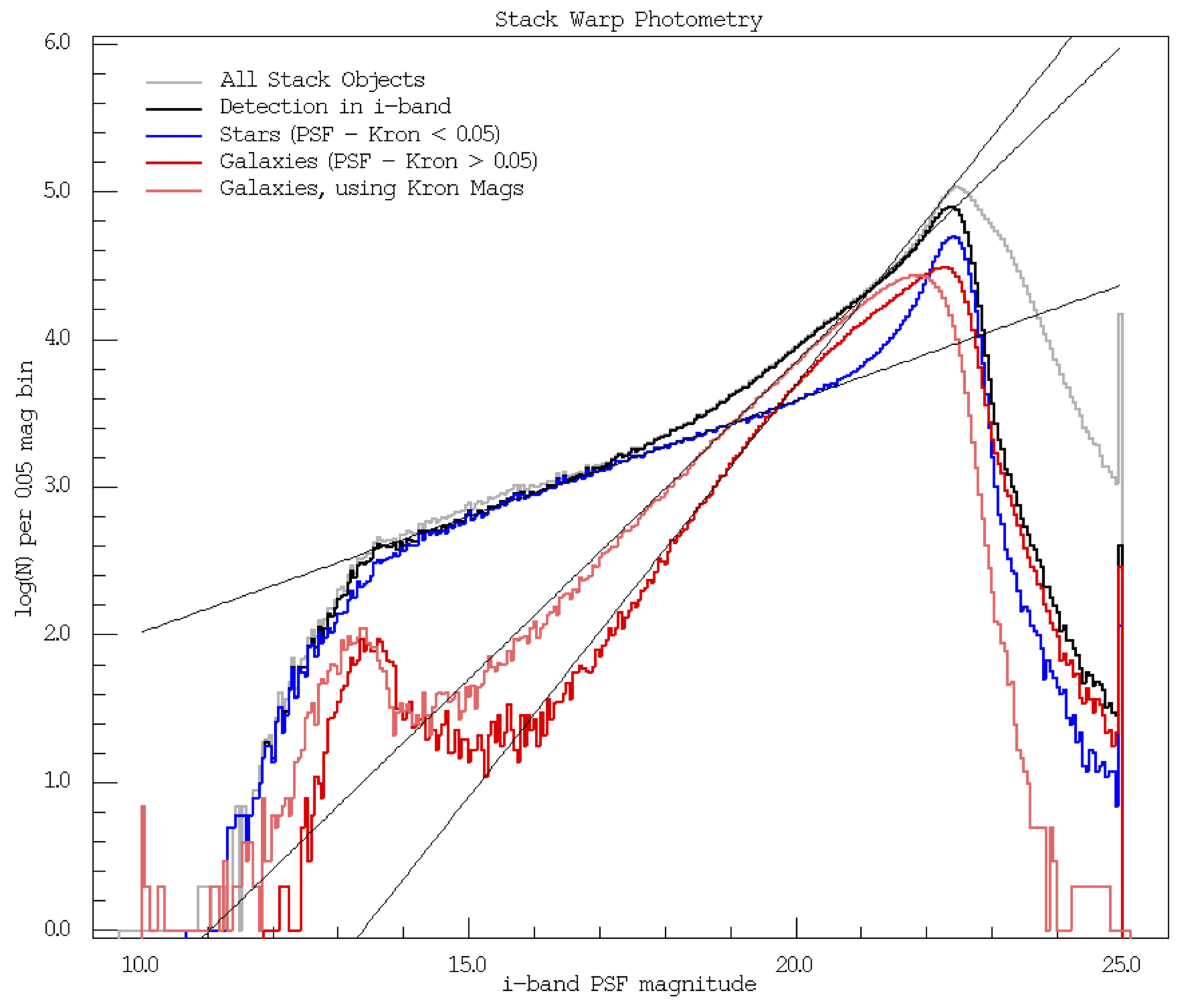}
\caption{Number counts of stars and galaxies from mean chip photometry (top) and stack photometry (bottom) for a region around the North Galactic pole. A simple constant cut in (PSF-Kron) was used to separate stars from galaxies.} 
\label{fig:chip}
\end{figure}

\begin{figure}
\includegraphics[width=\columnwidth,angle=0]
{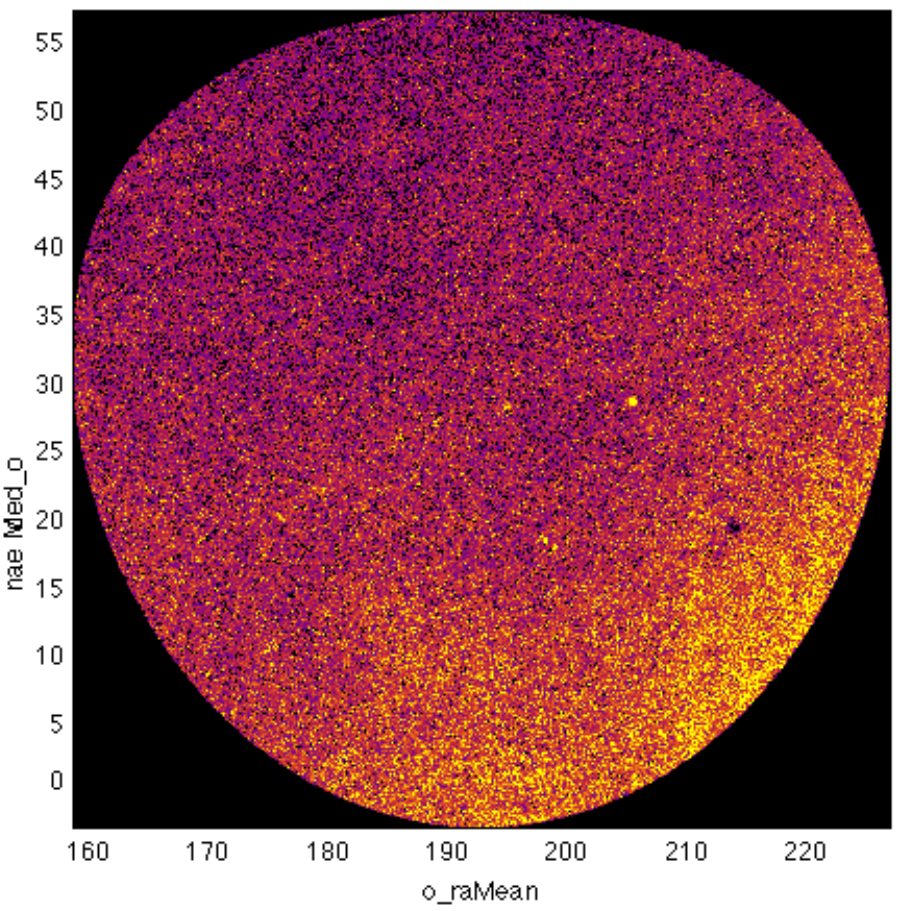}
\includegraphics[width=\columnwidth,angle=0]
{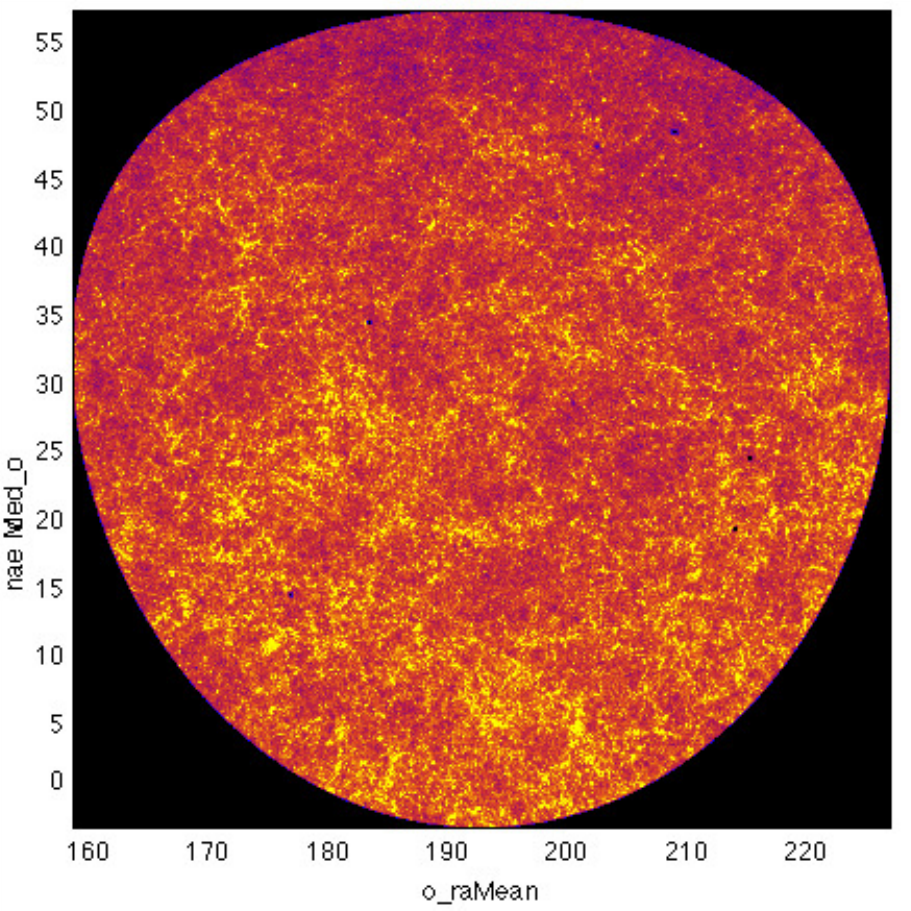}
\caption{The spatial distribution of (top) stars and (bottom) galaxies in the North Galactic pole region  ($|b| > 60$), selected using a simple cut in (PSF-Kron). 
} 
\label{fig:galcap}
\end{figure}

For the DR1 and DR2 releases we recommend using a simple cut in (PSF - Kron) magnitude space to separate stars from galaxies. Figure \ref{fig:sgsep} shows \ips v $i_{PSF}$-$i_{Kron}$ for \ips-band data around the galactic pole region. Unresolved objects form the tight sequence around PSF-Kron$ = 0.0$. A cut of (PSF-Kron)$<0.05$ does a reasonable job of selecting stars down to \ips$\sim21$. Figure \ref{fig:chip} shows the star and galaxy counts resulting from such a cut. Faintward of \ips$\sim21$ the number of stars is over-predicted by a simple linear cut like this. Also, at the brightest magnitudes, saturated stars tend to get classified as extended by this technique, resulting in a peak in the galaxy counts at \ips$\sim13.5$. The use of the IPP flags or a more sophisticated non-linear cut can relieve this problem to some extent. An example distribution of stars and galaxies on the sky is visualised in Figure \ref{fig:galcap}. 

A more detailed discussion of this technique applied to PS1 data, including the behaviour of synthetic stars and galaxies, can be found in \cite{2014MNRAS.437..748F}.


\begin{figure}
\includegraphics[width=\columnwidth,angle=0]
{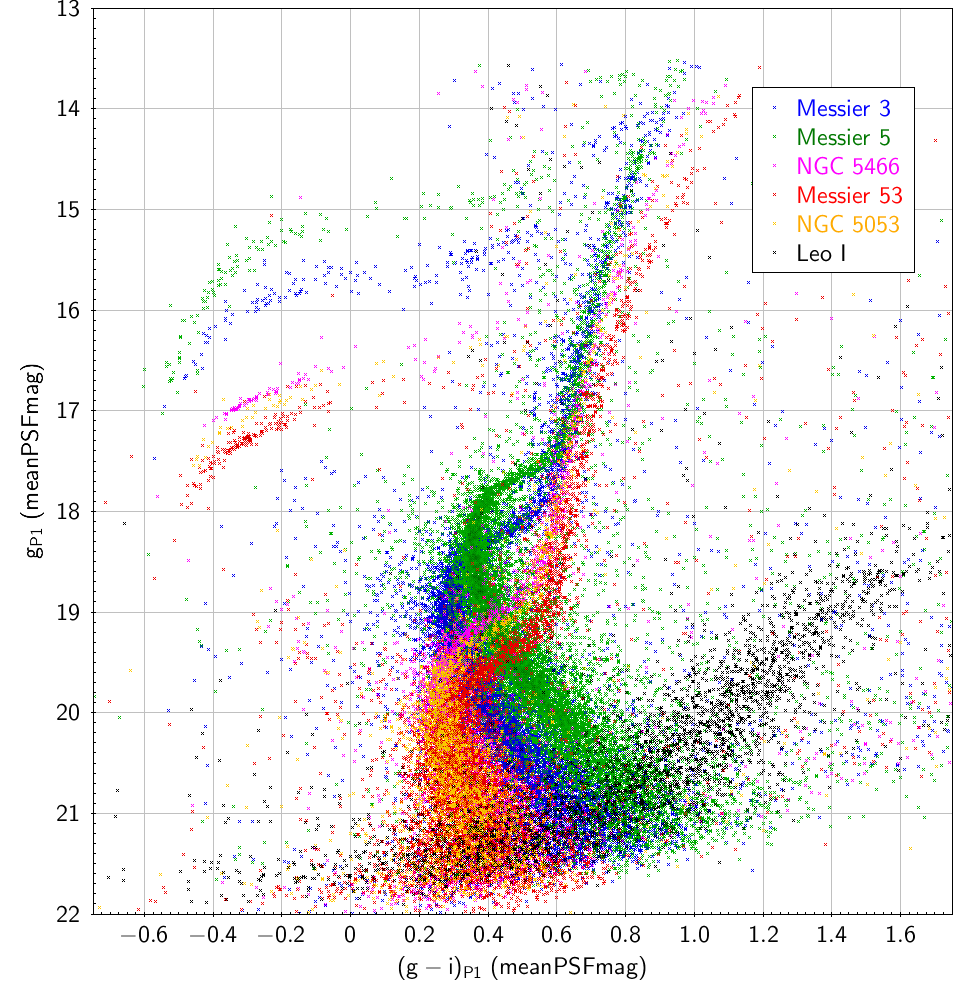}
\caption{\ips v \gps-\ips\ color-magnitude diagrams for a variety of galactic star clusters and for the Local Group dwarf galaxy Leo I. These data are taken from the meanObject table - see Section \ref{subsec:objecttab}.} 
\label{fig:cluster_cm}
\end{figure}

\begin{figure}
\includegraphics[width=\columnwidth,angle=0]
{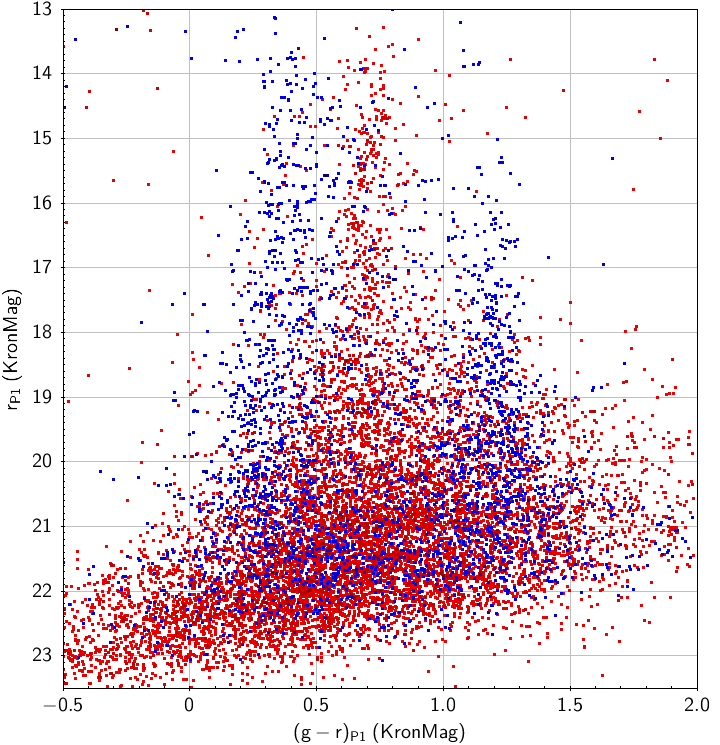}
\caption{Example of a \rps v \gps-\rps\ color-magnitude diagram for a 1 degree square region around the Coma galaxy cluster (Abell 1656). Galaxies are indicated in red, stars in blue. This plot uses Kron magnitudes taken from the StackObjectThin table - see Section \ref{subsec:stack}.} 
\label{fig:coma_cm}
\end{figure}

\begin{figure*}
\begin{center}
\includegraphics[width=3.0in,angle=0]
{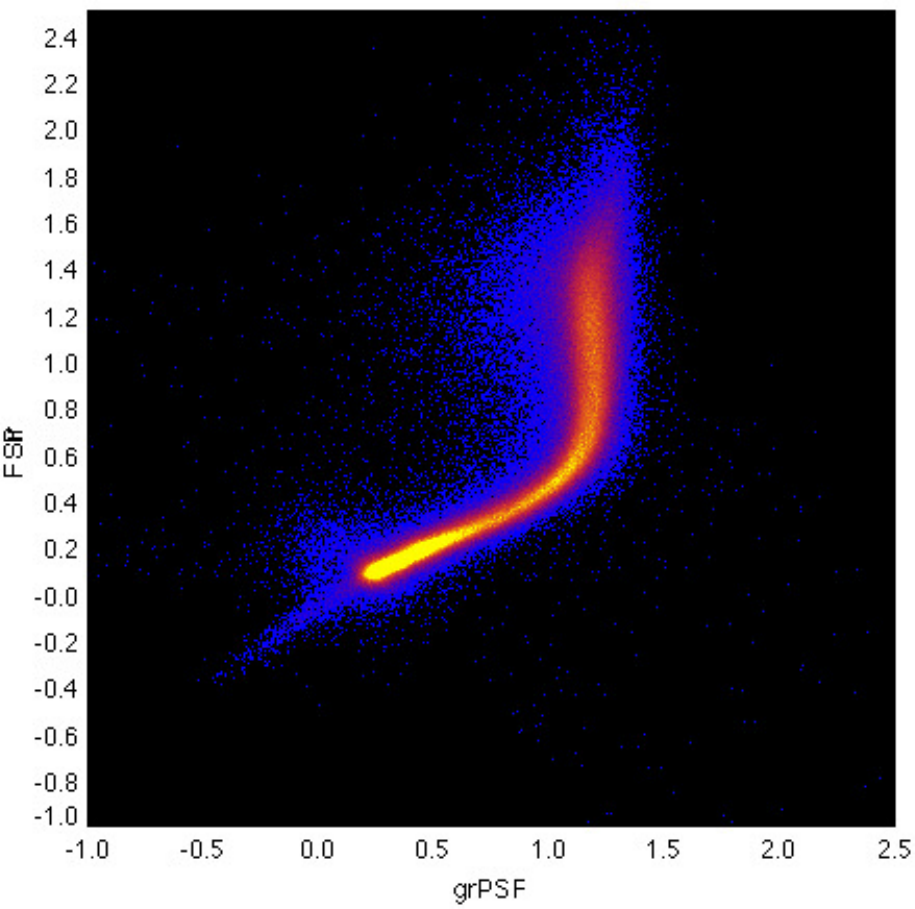}
\includegraphics[width=3.0in,angle=0]
{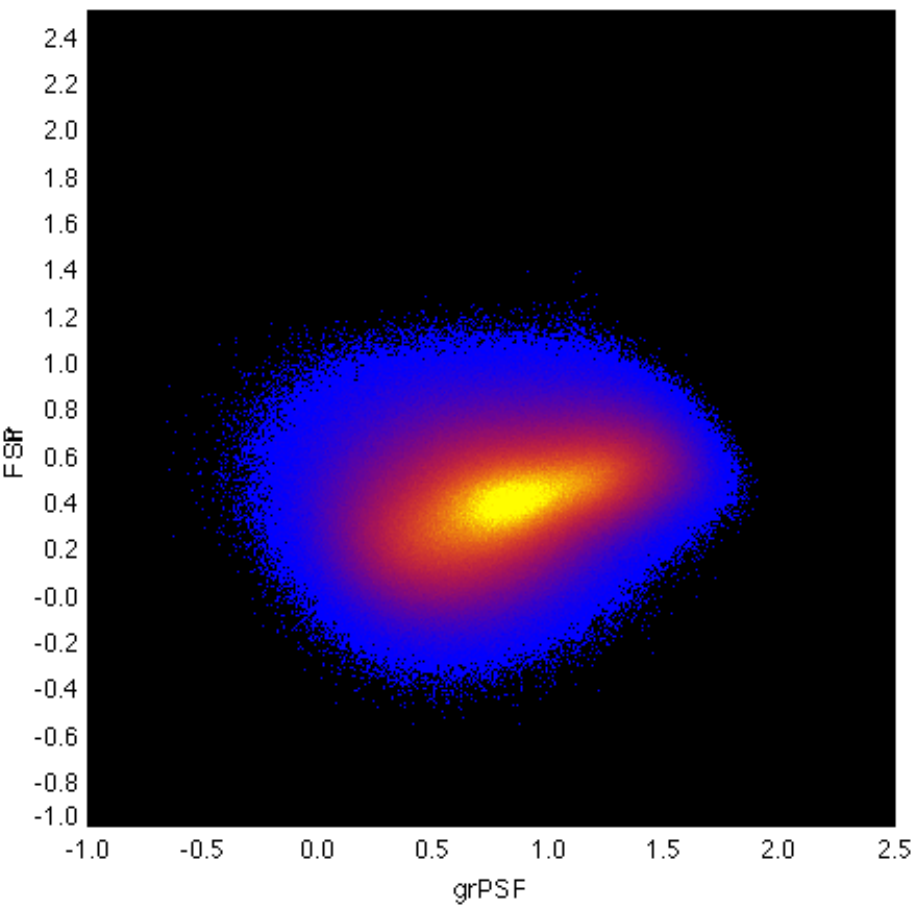}
\includegraphics[width=3.0in,angle=0]
{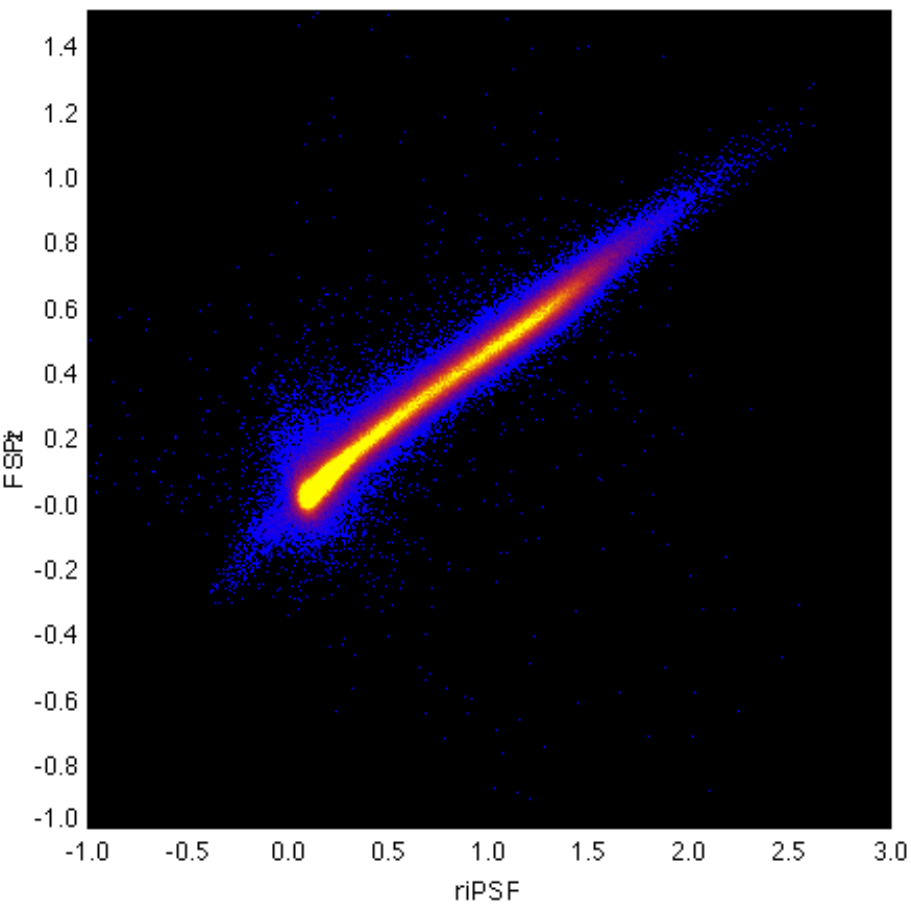}
\includegraphics[width=3.0in,angle=0]
{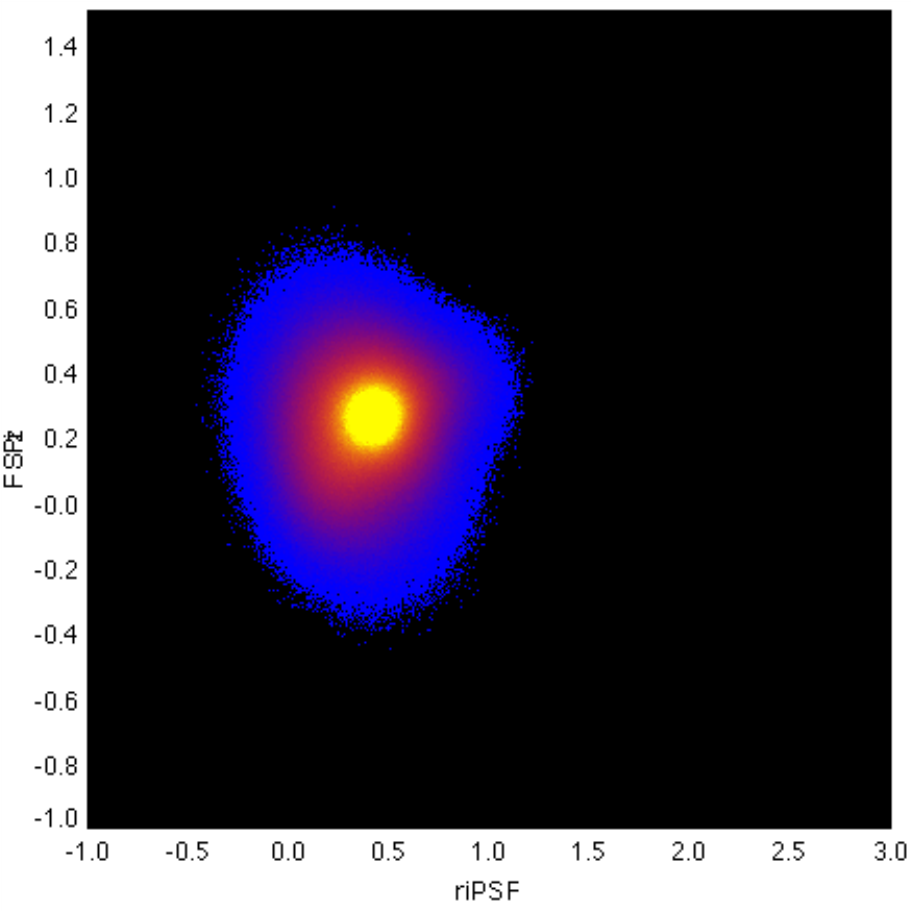}
\includegraphics[width=3.0in,angle=0]
{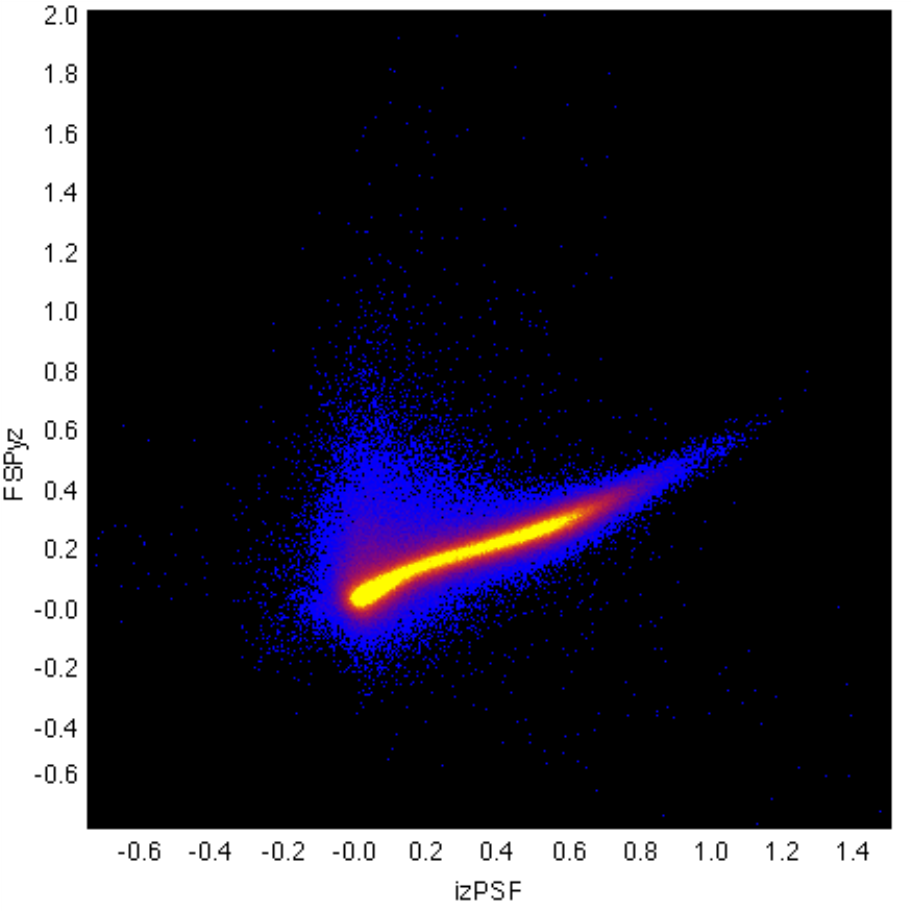}
\includegraphics[width=3.0in,angle=0]
{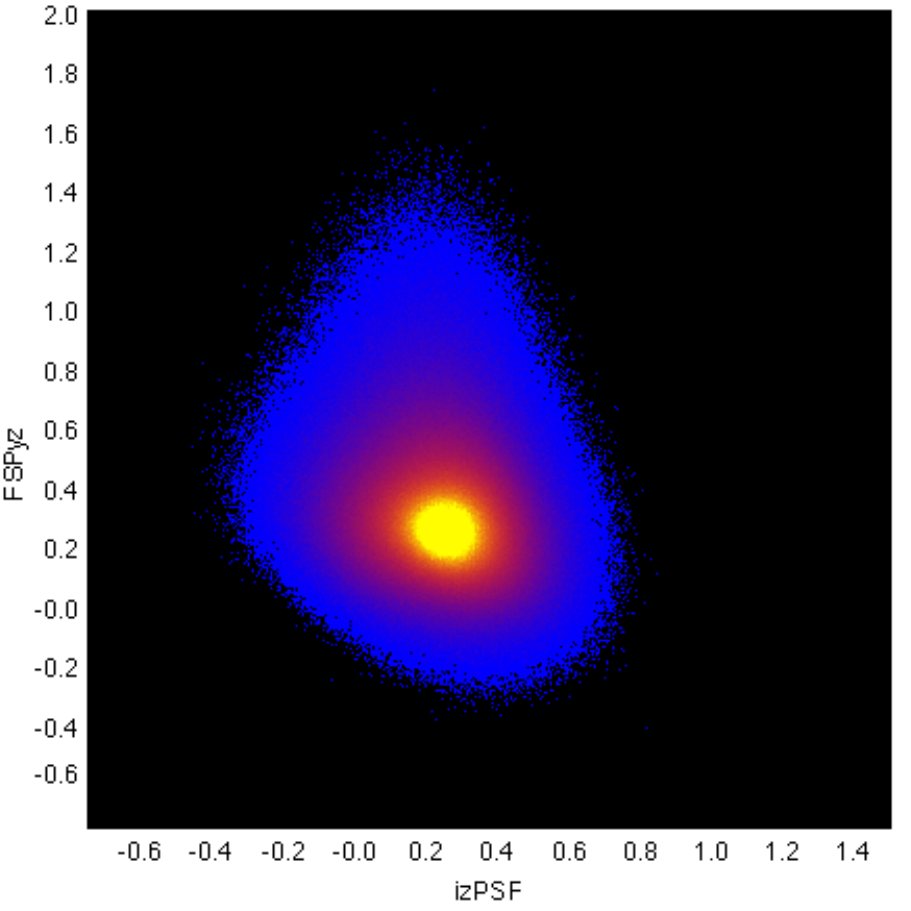}
\end{center}
\caption{Color-Color plot for stars (left column) and galaxies (right column) with b $>$ 60. } 
\label{fig:grristar}
\end{figure*}

\begin{figure*}
\includegraphics[width=\textwidth,angle=0]
{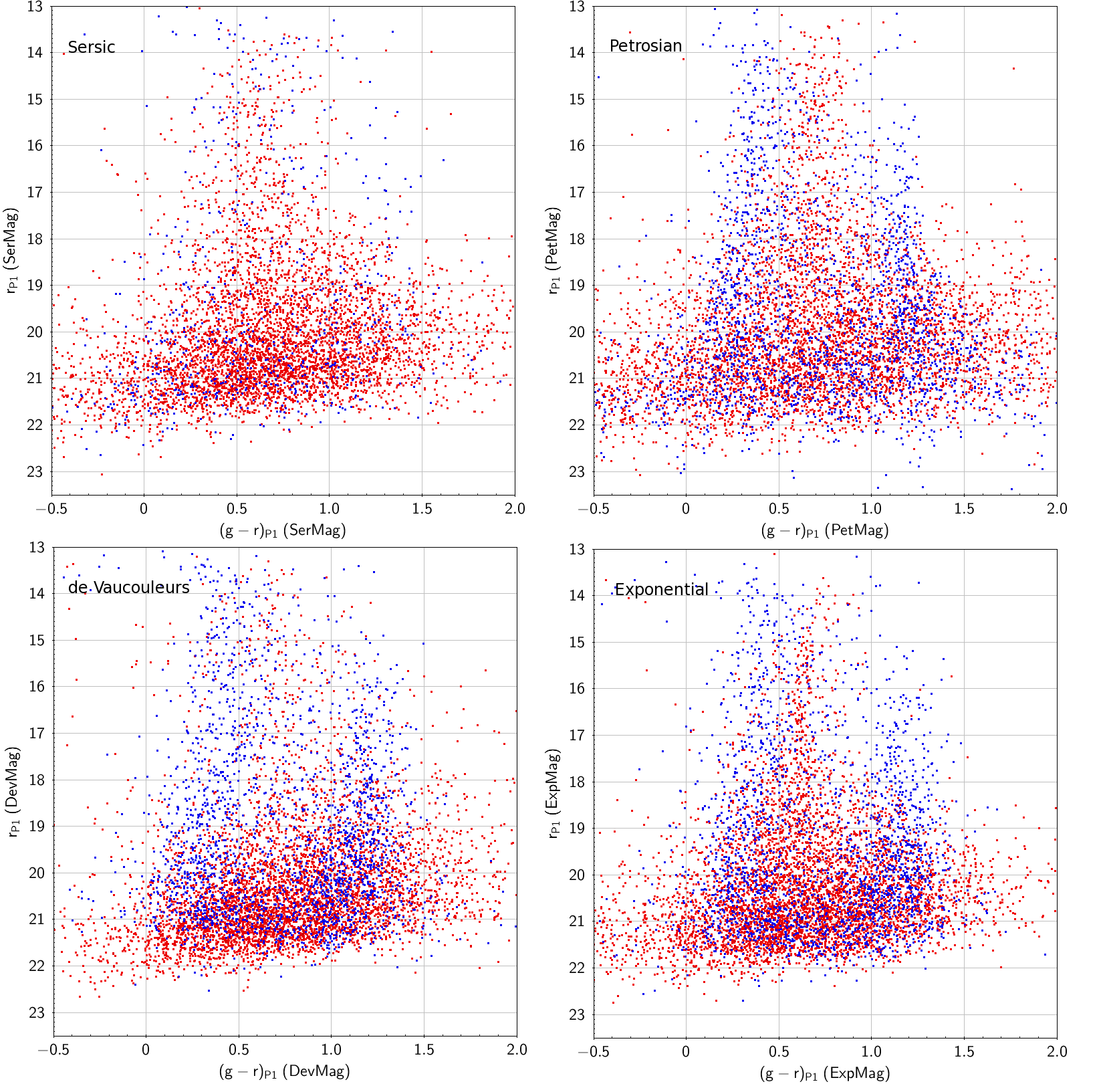}
\caption{Example \rps v \gps-\rps\ color-magnitude diagram for a 1 degree square region around the Coma galaxy cluster (Abell 1656). Galaxies are indicated in red, stars in blue, using the same criteria as in \ref{fig:coma_cm}. This plot shows four different extended source model magnitudes - Sersic, Petrosian, de Vaucouleurs and Exponential. }
\label{fig:extended}
\end{figure*}

\begin{figure}
\includegraphics[width=\columnwidth,angle=0]
{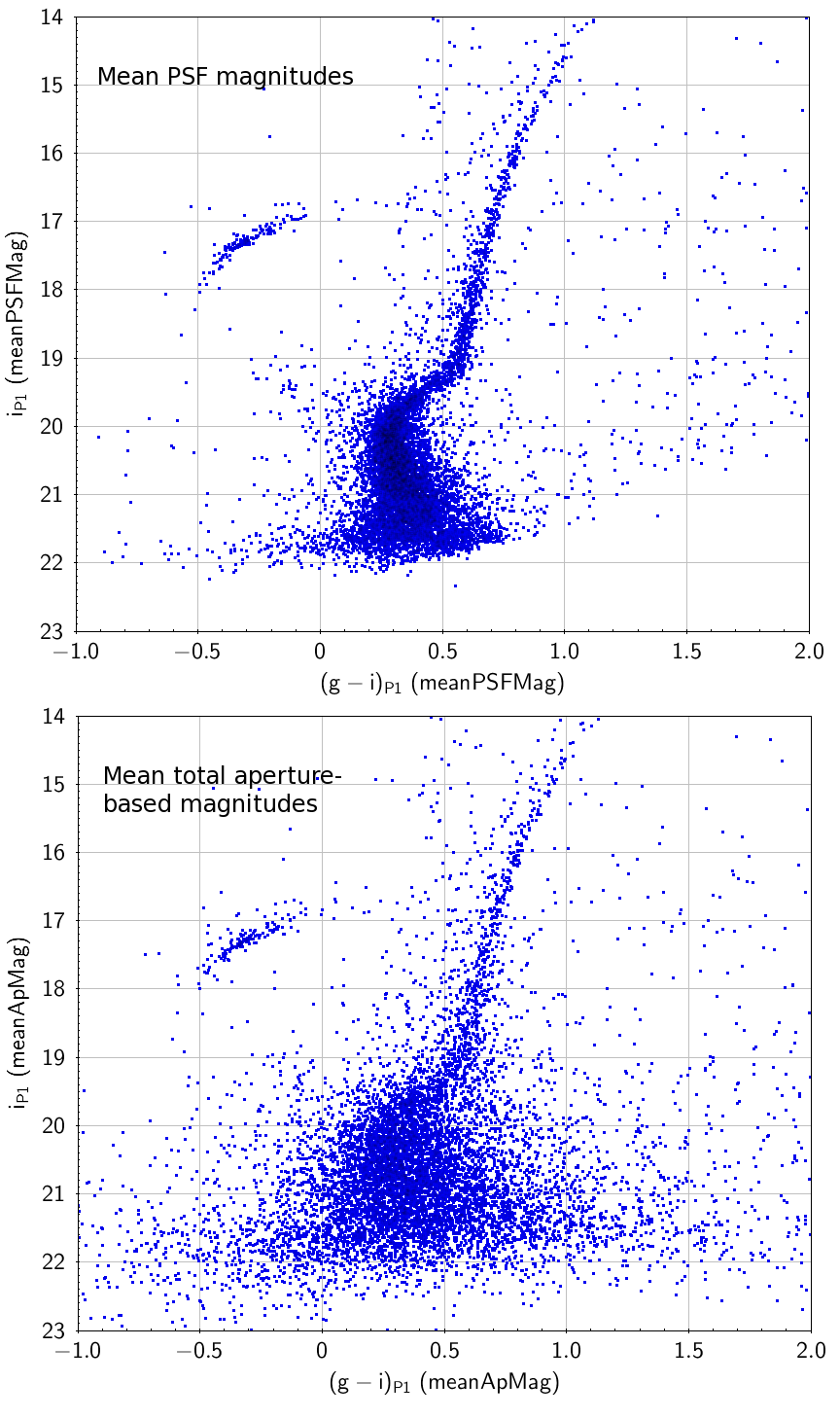}
\caption{Example of a \ips v \gps-\ips\ color-magnitude diagram for the globular cluster Messier 53 degree, showing mean PSF magnitudes (top) and  mean aperture-based total magnitudes (bottom) taken from the meanObject table. }
\label{fig:m53mean}
\end{figure}

\begin{figure}
\includegraphics[width=\columnwidth,angle=0]
{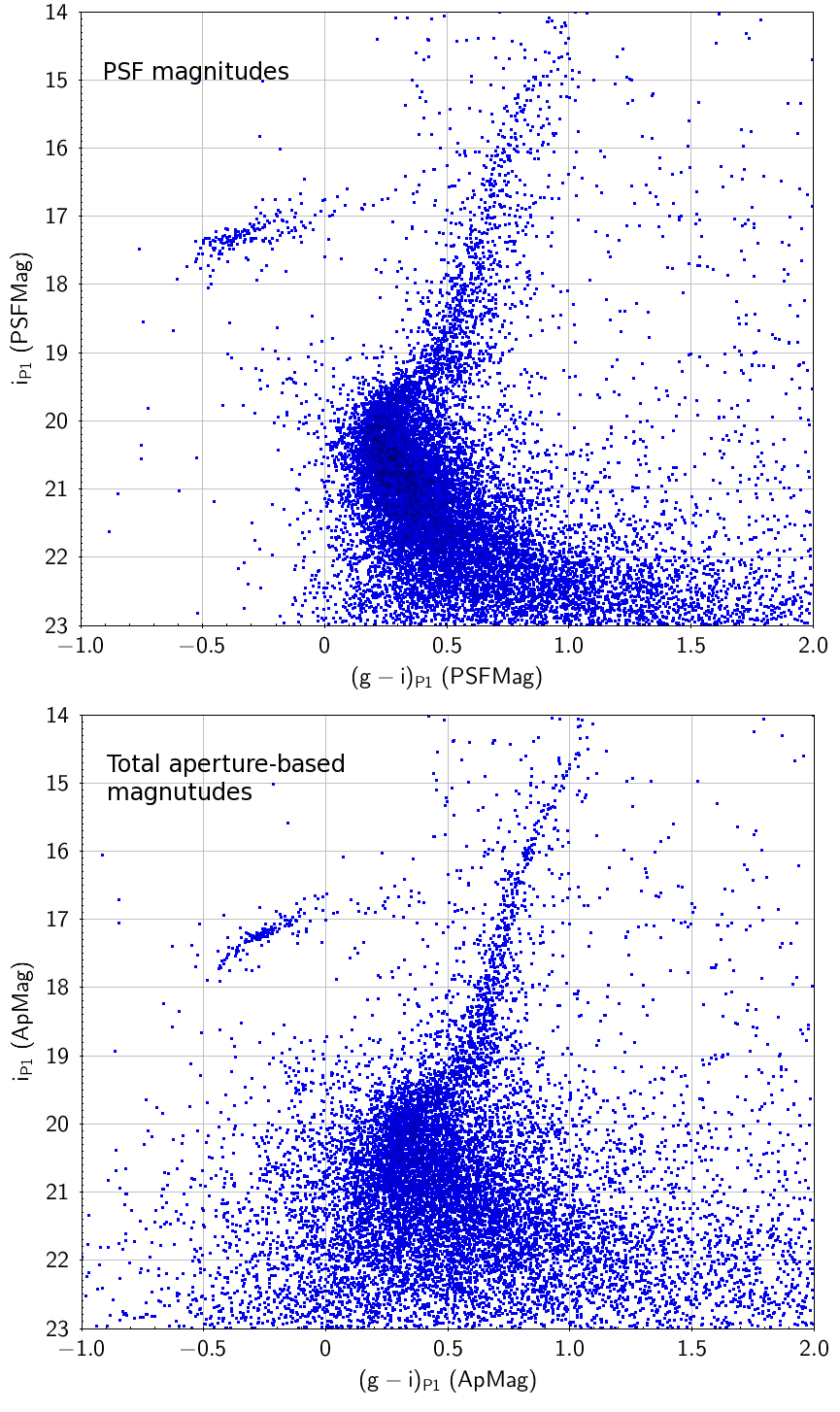}
\caption{Example \ips v \gps-\ips\ color-magnitude diagram for the globular cluster Messier 53 degree, showing stack PSF magnitudes (top) and stack aperture-based total magnitudes (bottom) taken from the StackObject table. }
\label{fig:m53stack}
\end{figure}




\subsection{Examples of Pan-STARRS1 photometry}

\subsubsection{Stellar and galactic loci}
In Figure \ref{fig:cluster_cm} we show examples of stellar \gps $v$ \gps-\ips color-magnitude diagrams for a variety of well-known Galactic globular star clusters, as well as the Local Group dwarf galaxy Leo I. These PSF magnitude data were taken from the MeanObject table, and hence represent the mean of the measurements on individual exposures (and so go no deeper than a single exposure). Despite the crowded nature of these fields, the stellar sequences are still quite tightly defined.

Figure \ref{fig:coma_cm} demonstrates the use of Kron magnitudes for galaxies, this time taken from the stacked data. Here we show \rps $v$ \gps-\rps for the Coma galaxy cluster, with its prominent sequence of early-type galaxies at \gps-\rps $\sim0.7$. 

Finally, in Figure \ref{fig:grristar} we display the \gps-\rps v \rps-\ips, \rps-\ips v \ips-\zps and \ips-\zps v \yps-\zps color-color locii for stars and galaxies around the Galactic pole region. 

In producing all these plots, the simple star/galaxy separation technique described in the previous Section has been adopted.





\begin{figure}
\includegraphics[width=\columnwidth,angle=0]
{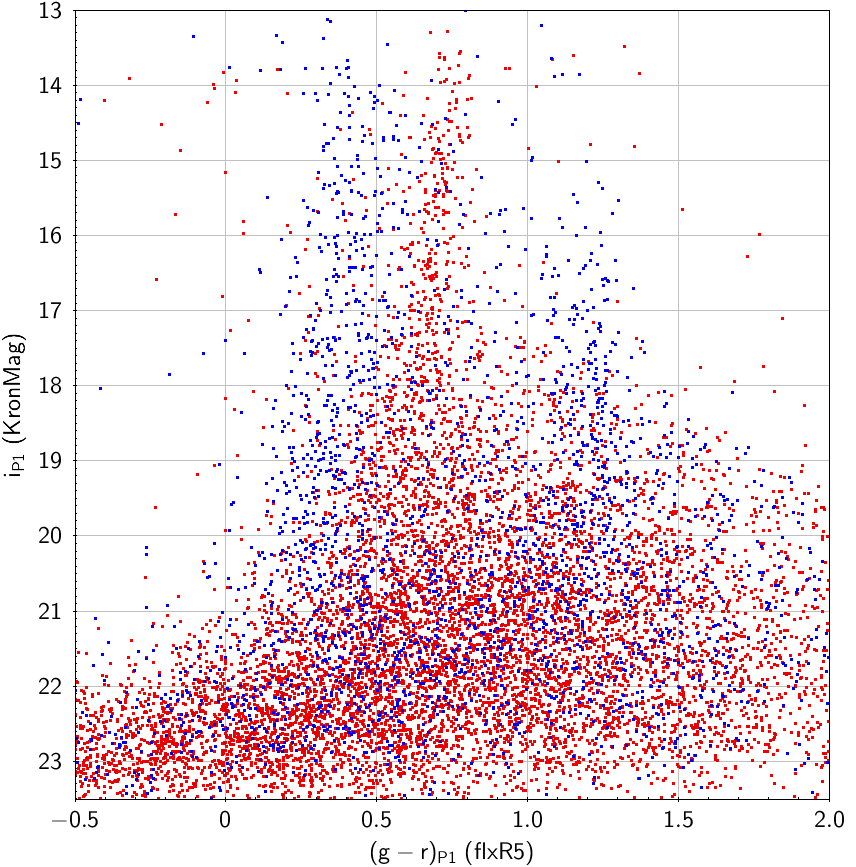}
\caption{Example \rps v \gps-\rps\ color-magnitude diagram for the Coma galaxy cluster, this time using the R5 fixed aperture magnitudes (3 arcsecond radius) to determine the \gps-\rps color. }
\label{fig:coma_ap}
\end{figure}

\subsubsection{Galaxy extended source models}

Both DR1 and DR2 releases provide extended source fits to a subset of objects detected in the stacked data. Exponential, De Vaucouleurs and Sersic profiles are fit to objects (including stars) outside the galactic plane (see Paper IV, Section 4.8 for the definition of this area).  Objects were fitted if the magnitude in one of the filters was brighter than the following limits: (\grizy) = (21.5, 21.5, 21.5, 20.5, 19.5). Petrosian magnitudes are also determined for objects in the same region, without restriction by brightness.
In Figure \ref{fig:extended} we reshow the Coma cluster color-magnitude diagram (Fig. \ref{fig:coma_cm}) using the various alternative magnitudes available. We still include the stars, although, apart from Petrosian, these models are not expected to give meaningful fits to the PSF. Indeed most stars have failed to to be fit at all by the Sersic model and so are missing from the plot. The de Vaucouleurs and Sersic fits show a disappointing large scatter. The cause of this is not known, and we recommend caution if using these measurements. Generally speaking, in DR1 and DR2, the Kron magnitudes appear to give more robust results than the extended model fits.



\subsubsection{Aperture photometry}
Pan-STARRS1 produces a total aperture-based magnitude, accurate for stellar objects only. Figure \ref{fig:m53mean} shows a comparison of these with PSF magnitudes for the globular cluster M53. This plot is based on data from the meanObject table and shows that the PSF magnitudes provide a tighter main-sequence. The situation for the stacked data (Figure \ref{fig:m53stack}), however, is less clear. Here the total aperture-based magnitudes show less scatter for \ips$<20$, but the PSF magnitude is clearly better faintward of this. This is probably due to the discontinuities in the PSF across the stacked data, which cannot be followed accurately by the fitted PSF model, and hence result in systematic errors in the PSF magnitudes. At bright magnitudes these errors dominate over the statistical errors, whereas at fainter magnitudes the reverse is true, and, as with the mean magnitudes, the PSF fitting produces the more accurate result.

Not to be confused with the total aperture-based magnitudes are the fixed, circular aperture magnitudes measured within a series of radii on the stacks. Figure \ref{fig:coma_ap} shows our Coma cluster color-magnitude diagram again, but now using the R5 aperture (3 arcsecond radius) measurements to form the \gps-\rps colors. As the aperture measurements do not measure total flux, we plot choose to plot against the Kron \rps magnitudes, as used in Figure \ref{fig:coma_cm}.

\subsubsection{Light curves with DR2}

\begin{figure}
\includegraphics[width=\columnwidth,angle=0]
{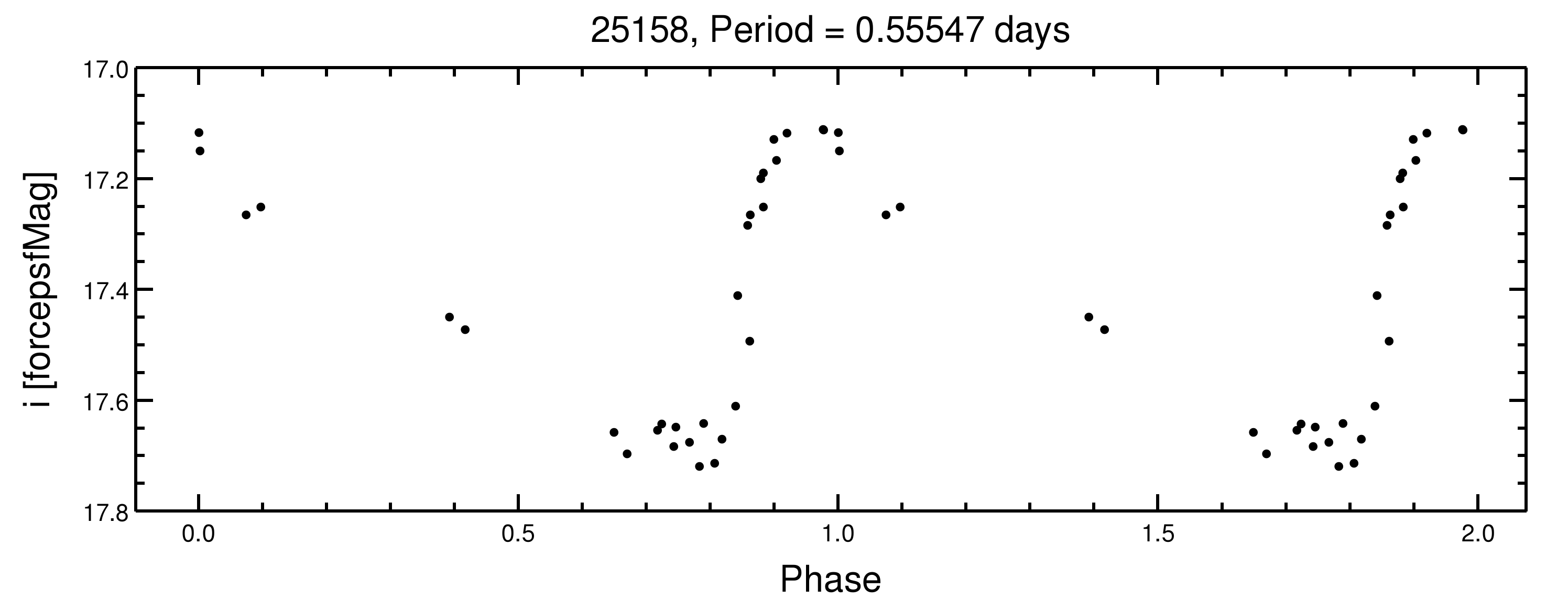}
\caption{Example of a light curve using forced psf photometry.  Star CSS J030521.9+013231 (Catalina Sky Survey), 584630948352256 (GAIA) is an RR Lyrae with period = 0.55547 days. Shown is the i-band forced psf photometry of the star from the 3pi Survey folded with CSS determined period.}
\label{fig:light_curve}
\end{figure}
With the DR2 data it is possible to obtain a light curve for any object in the survey area. Given the position of a given object, the mean magnitudes, number of detections, etc., can be obtained from the MeanObjectView table. By joining with the StackObjectAttributes table it is  possible to obtain the primaryDetection and bestDetection. Note, the bestDetection is corrupted in DR2, but will be fixed in DR2.1. 
By joining on the Detection table the detection parameters can be obtained, and by joining on the ForcedWarpMeasurment table the forced detection parameters can be obtained. Figure \ref{fig:light_curve} shows the i-band forced psf photometry from the 3pi Survey for the Catalina Sky Survey RR Lyrae star 
CSS J030521.9+013231 folded with the CSS determined period = 0.55547 days \cite{2013ApJ...765..154D}.  Additional examples are shown in \citep{flewelling2017}.

\section{The PS1 Science Consortium Science Legacy}
\label{sec:results}

The Pan-STARRS1 Surveys of the PS1 Science Consortium 
have enabled science on topics ranging from Near Earth Objects 
to the most distant quasars. New discoveries will be enabled 
by providing access to the community to the Pan-STARRS1 Archive at 
the Barbara A. Mikulski Archive for Space Telescopes. While Pan-STARRS1 is 
not a space telescope, among other kinds of science these surveys will
advance is a means to reprocess the astrometry of the
Hubble Space Telescope Archive
based on the Pan-STARRS1 extention of the Gaia Reference Frame \cite{magnier2017c}. Below we provide a brief summary of the
legacy science from the PS1 Science Consortium as examples of
the kind of science that can be done with the Pan-STARRS1 Surveys.

A primary goal of the PS1 mission was the Solar System Survey, designed to discover previously unknown Near-Earth Objects (NEOs) and provide additional orbital information
on known bodies. So far, PS1 has been responsible for discovering over 2900 NEOs, including potential targets for both robotic and manned space missions. PS1 survey
data has also led to the discovery of 129 comets and tens of thousands of new main-belt asteroids. A major legacy of both the PS1SC Solar System Survey and the continuing NEO survey (PI Wainscoat) has been the reporting to date of $2.7\times10^7$ astrometric and photometric measurements of moving objects to the IAU Minor Planet Center. A key science result  from this treasure trove has been a determination of the luminosity distribution of NEOs down to diameters of just a few metres by \cite{2017Icar..284..114S}
Additionally, the photometric properties of approximately a quarter of a million main-belt asteroids has been published by \cite{2015Icar..261...34V}
Looking at rarer objects, PS1 has allowed characterisation of the Main-Belt Comet 
(\cite{2015Icar..248..289H}),
and constrained the number of observable asteroid breakups  
(\cite{2015Icar..245....1D}).
Pre-discovery imaging of comet ISON out to Saturns' orbit demonstrated the value of deep solar system surveys in constraining the cometary activity of  inbound long-period comets 
(\cite{2013ApJ...776L..20M}).


The rapid nightly processing of data to search for NEOs yielded objects with motions as slow as 0.05 deg/day, depending on seeing, and yielded numerous Centaurs, making Pan-STARRS1 one of the most prolific discovery telescopes for Centaurs. The cadence of the observations was tuned for discovery of faster moving objects such as NEOs, making it not ideal for discovery of outer solar system objects. Nevertheless, the Pan-STARRS1 dataset is rich in observations of the outer solar system. 
\cite{2016arXiv160704895W}
describe a search for distant solar system objects using the archival PS1 data for the period 2010 Feb 24 to 2015 July 31.  A total of 607 distant solar system objects were identified, 332 being new first observation discoveries, with an additional 24 significantly improving the astrometry of previously designated objects.  While a large number of new objects were found, no new extreme TNOs showing a clustering in their argument of perihelia were found, which, if present, could support the presence of a distant planetary sized perturber in the outer solar system.  
\cite{2016AJ....152..147L}
describe the discovery of five new Neptune Trojans in the PS1 data. Four of these may be primordial, but the fifth is likely a recent capture. 
\cite{2016ApJ...827L..24C}
describe the discovery of a retrograde TNO in the PS1 dataset, and show that this object has similar orbital characteristics to other low semimajor axis high-inclination TNOs and Centaurs, hinting at a common orbital plane.


The scheduling and filter set of Pan-STARRS1 was designed from the
beginning to enable the complete census and study of the
ultracool dwarfs ($T_{\rm eff}<2400$\,K) of the solar neighborhood, as
well as white dwarfs.  The combination of proper motions and
parallaxes allows for relaxed color selections to obtain complete,
volume-limited samples \citep{Deacon11}. Proper motions and colors
also allow for the detection of substellar companions to main sequence
stars, putting constraints on their age and
mass \citep{Deacon14,Deacon2012a,Deacon2012b}; peculiar brown dwarfs
such as low-gravity, young brown dwarfs \citep{Liu13}; members of
nearby stellar structures \citep{Aller16, Best15, Goldman13}. The
depth of the stacked images provides the first colors in the visual
and red part of the spectrum for a large number of ultra-cool stars
and brown dwarfs (Best et al. in prep). The accuracy of both Pan-STARRS\,1 astrometry and photometry also allowed the characterisation of Kepler target stars and the discovery of wide binary companions to planet hosts 
\cite{Deacon2016}.

The panoptic Pan-STARRS1 $3\pi$ survey provides a unique opportunity to map the distribution of stars (e.g. \citealt{2016ApJ...825..140M}) in our own Milky Way and its outskirts and place it in a cosmological context. In particular, it revealed the presence of multiple very faint Milky Way satellites (e.g. \citealt{2015ApJ...813...44L}) and stellar streams likely stemming from the tidal disruption of globular clusters by our Galaxy \citep{2016MNRAS.463.1759B}. Exploiting PS1 as a time-domain survey, the largest and deepest sample of RRLyrae candidate stars was identified \citep{2016ApJ...817...73H}, which provides for unprecedented 3-D mapping of the Milky Way's stellar halo out to $\sim$120 kpc. 
A major goal of Pan-STARRS1 Milky Way science was to map the interstellar dust in 3 dimensions using star colors.  For this purpose, PS1 has three main advantages over SDSS: it is deeper, goes one band redder, and covers more of the low-latitude sky.  The collaboration developed a method to infer the posterior on distance and reddening of each star, and then group stars into angular pixels and estimate the reddening as a function of distance in each \citep{2014ApJ...783..114G}.  They applied this method to PS1 photometry of 800 million stars (some with 2MASS photometry as well) and created a map with 2.4 million angular pixels and 31 distance bins, covering 3/4 of the sky \citep{2015ApJ...810...25G}.  They used a variant of the same technique to produce the largest catalog of molecular cloud distances \citep{2014ApJ...786...29S}.  More recently, PS1 photometry has served as the basis of a new parameterization of the reddening law, and study of its variation in 2 and 3 dimensions \citep{2016ApJ...821...78S,2016arXiv161202818S}.

 
The Pan-Planets survey was a dedicated exo-planet transit survey within the PS1 project. The survey covered an area of 42 sq degrees in the Galactic disk for about 165 hr with the goal
to constrain the occurrence rate of hot Jupiters around M dwarfs. A combination of SED fiting, dust maps, and proper motion information allowed to identify more than 60 000 M dwarfs
in the field. This is the largest sample of low-mass stars observed in a transit survey. With this large sample size, the Pan-Planets survey resulted in an occurrence rate of hot Jupiters of
0.11 (+0.37-0.02) \% in case one of our candidates turns out to be a real detection. If, however, none of our candidates turn out to be true planets, we are able to put an upper limit of 
0.34\% with a 95\% confidence on the hot Jupiter occurrence rate of M dwarfs. This is the best limit for the occurrence rate of hot Jupiters around M stars so far.


The major science goal for the Andromeda monitoring with Pan-STARRS1 (PAndromeda) originally was to identify a large number of gravitational microlensing events towards M31. The final depths and image quality of the survey meant that early 
expectations of event rates were not met. Nevertheless we found 6 events in a subfield of the first year of the survey
\citep{2012AJ....143...89L} and the data are suitable for other PAndromeda stellar science. 
Using only a subset of the PAndromeda data  we identified and classified ~$1700$  Cepheids and analysed their period luminosity/Wesenheit relations
\citep{2013AJ....145..106K}
and we found ~$300$ eclipsing binaries in M31 
\citep{2014ApJ...797...22L}
from which a handful of the brightest ones are suitable to derive an independent M31 distance. 
We furthermore searched for rare variable stars which help to understand stellar structure 
(17 Beat Cepheids; \cite{2013ApJ...777...35L})
and we also identified four new LBVs, i.e. potential supernova progenitors
\citep{2014ApJ...785...11L}. 
The combination of the PAndromeda variability analysis and the HST-PHAT data \citep{2012ApJS..200...18D} turned out to be very powerful. In 
\cite{2015ApJ...799..144K}
we presented the largest M31 (HST) near infrared $J-$band and $H-$band sample at this time (371 Cepheids), studied their near infrared period luminosity relations and showed that the bright part of our sample is well suited for $H_0$ determination using M31 (having no metallicity issues compared to LMC/SMC) for the Cepheid distance ladder.
In summer 2016 we finished completely (re)doing the PAndromeda difference imaging  for the full survey (time and area) by optimzing our pipeline for the data characteristics and by increasing the masking fraction. We hence now have the final PAndromeda data products, i.e. light curves for ``all" variable sources in M31. The analysis of these light curves is ongoing, e.g., the final catalogue and analysis for ~$2700$ PAndromeda Cepheids (750 with HST NIR photometry) 
%
will be made available to the community in 2017 (Kodric et al. 2017, in preparation).

The Medium Deep Survey (MDS, Section\,\ref{subsec:mdeep}) was designed for both deep fields and transient science with multi-colour 
temporal coverage of 70 square degrees in total. In addition the multi-epoch aspect of the 3$\pi$ survey also provided a 
new opportunity for time domain science. The data for the MDS are not in DR1, but they will be released at a future
date. The cadence and filter coverage of the MDS fields was designed to both discover type Ia supernovae (SN~Ia) before
maximum light and to sample their lightcurves sufficiently for distance measurements. 
We discovered $\sim$3000 SNe~Ia within $z \approx 0.8$, and  obtained spectroscopic confirmation for $\sim$500 SNe~Ia. 
\cite{2014ApJ...795...44R,2014ApJ...795...45S} published data from the  first 1.5 years of the MDS, in which 
146 SNe~Ia were used to constrain the dark energy equation-of-state parameter, $w$, to $\sim$7\% 
 An analysis of the full sample is in preparation, and we have undertaken a first step in analyzing the photometric sample \citep{2016arXiv161107042J}. The MDS was also a rich source of exotic transients, with the discovery of 
 high redshift superluminous supernovae. These supernovae are 100 times brighter than normal core-collapse 
 evens and peak at $M < -21$~mag.  They are UV bright and PS1 has discovered some of the most distant SLSNe, including one at $z = 1.566$ 
 \citep{2012ApJ...755L..29B}. 
In a series of papers we studied their  physical parameters 
 \citep{2011ApJ...743..114C,2013ApJ...771...97L,2014MNRAS.437..656M,2016ApJ...831..144L},  
and their host-galaxy environments \citep{2014ApJ...787..138L,2015ApJ...804...90L} and rates 
 \citep{2015MNRAS.448.1206M}. The MDS discovered  two   tidal disruption events, PS1-10jh
 \citep[][, the best studied TDE to date]{2012Natur.485..217G} and PS1-11af \citep{2014ApJ...780...44C},  defined a class of fast-declining transients
 \citep{2014ApJ...794...23D}
and provided an extensive study of type II explosions\citep{2015ApJ...799..208S}. In addition, the 
combination of the MDS and GALEX led to complementary UV data on some transients \citep{2010ApJ...717L..52B,2010ApJ...720L..77G}. 
The 3$\pi$ survey provided discovery and critical lightcurve points for a number of interesting objects. 
Low redshift super-luminous supernovae \citep{2010ApJ...724L..16P,2013ApJ...770..128I,2013Natur.502..346N} 
were either discovered with the  3$\pi$ survey  or had lightcurve data at critical points. The survey also provided the detection of 
a pre-supernova outburst of a type IIn explosion 
\citep{2013ApJ...779L...8F}. Initially we ran a transient search by catalogue matching with SDSS, which provided
the discovery of slowly evolving blue transients at the centres of galaxies \citep{2016MNRAS.463..296L}. 
After the creation of an all sky stack, we progressed to routine difference imaging, leading to 
discovery of some super-luminous supernovae at lower redshift than in MDS
\citep{2016arXiv160401226I,2016ApJ...826...39N}. Pan-STARRS1 is now the world leading disoverer of low-redshift supernovae, 
according to the IAU statistics\footnote{https://wis-tns.weizmann.ac.il/stats-maps}.

The $3\pi$ Steradian Survey is being used for 
a citizen-scientist enabled nearby galaxy survey based on the optical imaging from Pan-STARRS1, but also incorporating multi-wavelength data from the ultraviolet and infrared regimes (GALEX, WISE). This project is called the PS1 Optical Galaxy Survey or POGS\footnote{http://pogs.theskynet.org}
\citep[][Thilker et al. in prep]{2013A&C.....3....1V}. 
Distributed computing resources contributed by tens of thousands of volunteers allow comprehensive pixel-by-pixel spectral energy distribution (SED) fitting for $>100,000$ galaxies, which in turn provides key physical parameters such as the local stellar mass surface density, star formation rate (SFR), and dust attenuation. Sufficiently nearby galaxies are being processed using complete UV-optical-IR SED coverage, whereas distant (but more numerous) galaxies are analyzed with optical only data due to the resolution of ancillary observations. With pixel SED fitting output, the POGS pipeline constrains parametric models of galaxy structure and measures non-parametric morphology indicators in a more meaningful way than ordinarily achieved, by operating on images of estimated physical parameters.  The depth, sky-coverage and time-domain capabilities of PS1 have also been leveraged to conduct various focused studies of galaxy properties, including host galaxy properties of variability 
selected AGN 
\citep{2016ApJ...826...62H}
the structure of outer galactic disks 
\citep{2015ApJ...800..120Z}
and the influence of group environment on the SFR-stellar mass relation 
\cite{Lin14}.

Early data from the 3$\pi$ survey have been used to quantify galaxy angular clustering \citep{2014MNRAS.437..748F}, confirm and determine redshifts for Planck cluster candidates
\citep{Liu15} and detect a large void in front of the CMB Cold Spot \citep{Szapudi15}. Preliminary data from the MDS has been used to find galaxy groups and clusters and investigate the dependence of star formation on environment \citep{Jian14,Lin14}.
 
 
One of the key science goals from the inception of PS1 was the discovery of quasars at the highest redshifts ($z \sim 6$), and to push the redshift barrier of $z =6.4$ (imposed at the time by the choice of the SDSS filters). These high—redshift quasars are thought to be one of the most massive structures that exist in the first Gyr of the Universe. The quasar host galaxies harbour accreting supermassive black holes, which allow detailed studies of key quasar properties, as well as the impact of the quasars on the surrounding intergalactic medium. PS1 has now become the survey in which most of the $z \sim 6$ quasars have been discovered, and a number of papers have resulted from this effort (e.g., first papers by \cite{2012AJ....143..142M} and \cite{2014AJ....148...14B}).
The highest-redshift quasars found by PS1 are discussed in \cite{2015ApJ...801L..11V},
and the most complete catalog of the high-$z$ quasar population is published in 
\cite{2016ApJS..227...11B}.
The latter discusses the properties of 77 newly detected PS1 quasars (out of a total of 124 known) at $z > 5.6$.

\section{Conclusions}
\label{sec:conc}

We have presented an introduction to and overview of the Pan-STARRS1 surveys.
Under the auspices of the Pan-STARRS1 Science Consortium, PS1 observed the entire sky north of Dec$=-30$\,deg (the 3$\pi$ survey) in \grizy\ (to \rps$\simeq23.3$), additional ecliptic fields in \wps\ for a Solar System survey (to \wps$\simeq22.5$ per visit CHECK!), 
several extragalactic deep fields in \grizy\ (to $\rps\simeq 23$ per visit), and some specialized fields (M31, transit survey fields).  These data have been calibrated to 
$\sim12$\,mmag 
internal photometric precision and 
$\sim20$\,mas 
internal astrometric precision.  DR1  consists of the image stacks and associated catalogs for the 3$\pi$ survey, distributed through the MAST system at STScI.  The second data release (DR2, expected January 2019) will distribute the 3$\pi$ time-domain data. Future data releases will release the  Medium Deep data, difference image data, and photometric redshifts from the Photo-Classification Server.

These data have already produced a variety of science, on subjects as varied as asteroids, Milky Way structure, galaxy formation, supernovae and cosmology, but we hope they will find further utility in achieving science goals beyond the scope of  the Pan-STARRS1 Science Consortium.  
The PS1 data are currently being used to provide targeting information for the SDSS-IV Time Domain Spectroscopic Survey  \citep{2015ApJ...806..244M}, and to provide high resolution, deep, multi-colour reference
images for transients in many transient surveys \cite[e.g. PESSTO, ATLAS, ASASSN, GAIA][]{2015A&A...579A..40S,2016ATel.8680....1T,2017MNRAS.464.2672H,2016MNRAS.455..603B}. 
Besides additional science investigations based directly on these data, the data have great legacy value in providing a high-quality network of calibration sources across the sky.  The data are already being used to calibrate the Hyper Suprime-Cam Survey \citep{2015PASJ...67...86T,2016ApJ...832..135C}
and will also be useful in cross-checking the calibration of the northern areas imaged by the Dark Energy Survey and the Large Synoptic Survey Telescope.  But in addition to large-scale surveys, individual 
programmes with relatively small observing fields north of Dec$=-30$\,deg will all have good PS1 calibration sources in the field, observed through exactly the same column as the target sources, allowing simple relative calibration for astrometry and photometry.

The legacy of the PS1 surveys extends beyond just the data.  The experience and lessons learned from designing and executing the PS1 surveys are assisting in the development of future survey projects. The Pan-STARRS2 observatory is now operational. PS1 and PS2 together are now primarily engaged in the Wide Area Survey for NEOs \citep{2018DPS....5031002C}, building up substantial time domain and stack data in the \wps and \ips bands, and significant data in \zps bands during twilight and bright time. Combined with the $u$ and $r$ band Canada France Imaging Survey \citep{2017ApJ...848..128I}, the resulting Ultraviolet Near Infrared Optical Northern Survey (UNIONS) \citep{2019AAS...23331301C}
will be the premier multi-band northern hemisphere survey for years to come. 
Some algorithms and code from PS1 are being used in the development of LSST, and students and postdocs who built careers starting with PS1 are applying their experience to new and larger surveys.  We hope the PS1 surveys will be useful to the astronomical community for many years to come.

\acknowledgments

The Pan-STARRS1 Surveys (PS1) have been made possible through contributions of the Institute for Astronomy, the University of Hawaii, the Pan-STARRS Project Office, the Max-Planck Society and its participating institutes, the Max Planck Institute for Astronomy, Heidelberg, and the Max Planck Institute for Extraterrestrial Physics, Garching, The Johns Hopkins University, Durham University, the University of Edinburgh, Queen's University Belfast, the Harvard-Smithsonian Center for Astrophysics, the Las Cumbres Observatory Global Telescope Network Incorporated, the National Central University of Taiwan, the Space Telescope Science Institute, the National Aeronautics and Space Administration  Grants No.s NNX08AR22G, NNX12AR65G, and NNX14AM74G, the National Science Foundation under Grant No. AST-1238877, the University of Maryland,  Eotvos Lorand University (ELTE), the Los Alamos National Laboratory and the Gordon and Betty Moore foundation. 
Ongoing support has come from the NASA SSO Near Earth Observation Program under grants NNX13AQ47G and NNH17ZDA001N-SSO and by the State of Hawaii. 

%
%

This work has made use of data from the European Space Agency (ESA)
mission {\it Gaia} (\url{http://www.cosmos.esa.int/gaia}), processed by
the {\it Gaia} Data Processing and Analysis Consortium (DPAC,
\url{http://www.cosmos.esa.int/web/gaia/dpac/consortium}). Funding
for the DPAC has been provided by national institutions, in particular
the institutions participating in the {\it Gaia} Multilateral Agreement.

This work has also made use of data from 
the USNO B catalog, a product of the US Naval Observatory, 
the Tycho Catalog, which is a product of the European Space Agency's Hipparcos mission,
and the Two Micron All Sky Survey, 
which is a joint project of the University of Massachusetts and the Infrared Processing and Analysis Center/California Institute of Technology, 
funded by the National Aeronautics and Space Administration and the National Science Foundation.

SJS acknowledges funding from (FP7/2007-2013)/ERC Grant agreement n$^{\rm o}$ [291222]
and STFC grants ST/I001123/1 and ST/L000709/1. 

DT acknowledges funding from the  National Science Foundation under Grant No. AST-1412596 for the POGS program.
%
%

The authors acknowledge the hard work and dedication of the University of Hawaii Institute for Astronomy staff who worked on Pan-STARRS1:
Rick Anderson,
Robert Calder,
Greg Ching,
M. Todd Dukes,
Richard Harris,
Haydn Huntley,
Brooke Gibson,
Jill Kajikara-Kent,
Sifan Kahale,
Chris Kaukali,
Aaron Lee,
Tom Melshiemer,
Louis Robertson,
Donna Roher,
Freddie Ratuta,
Gavin Seo,
Diane Tokumura, 
Robin Uyeshiro, and 
Gail Yamada.

\bibliographystyle{apj}

\bibliography{lib.bib}

\begin{thebibliography}{}
\expandafter\ifx\csname natexlab\endcsname\relax\def\natexlab#1{#1}\fi

\bibitem[{{Afonso} \& {Henning}(2007)}]{2007ASPC..366..326A}
{Afonso}, C., \& {Henning}, T. 2007, in Astronomical Society of the Pacific
  Conference Series, Vol. 366, Transiting Extrapolar Planets Workshop, ed.
  C.~{Afonso}, D.~{Weldrake}, \& T.~{Henning}, 326

\bibitem[{{Alard} \& {Lupton}(1998)}]{1998ApJ...503..325A}
{Alard}, C., \& {Lupton}, R.~H. 1998, \apj, 503, 325

\bibitem[{{Aller} {et~al.}(2016){Aller}, {Liu}, {Magnier}, {Best}, {Kotson},
  {Burgett}, {Chambers}, {Hodapp}, {Flewelling}, {Kaiser}, {Metcalfe}, {Tonry},
  {Wainscoat}, \& {Waters}}]{Aller16}
{Aller}, K.~M., {Liu}, M.~C., {Magnier}, E.~A., {et~al.} 2016, \apj, 821, 120

\bibitem[{{Ba{\~n}ados} {et~al.}(2014){Ba{\~n}ados}, {Venemans}, {Morganson},
  {Decarli}, {Walter}, {Chambers}, {Rix}, {Farina}, {Fan}, {Jiang}, {McGreer},
  {De Rosa}, {Simcoe}, {Wei{\ss}}, {Price}, {Morgan}, {Burgett}, {Greiner},
  {Kaiser}, {Kudritzki}, {Magnier}, {Metcalfe}, {Stubbs}, {Sweeney}, {Tonry},
  {Wainscoat}, \& {Waters}}]{2014AJ....148...14B}
{Ba{\~n}ados}, E., {Venemans}, B.~P., {Morganson}, E., {et~al.} 2014, \aj, 148,
  14

\bibitem[{{Ba{\~n}ados} {et~al.}(2016){Ba{\~n}ados}, {Venemans}, {Decarli},
  {Farina}, {Mazzucchelli}, {Walter}, {Fan}, {Stern}, {Schlafly}, {Chambers},
  {Rix}, {Jiang}, {McGreer}, {Simcoe}, {Wang}, {Yang}, {Morganson}, {De Rosa},
  {Greiner}, {Balokovi{\'c}}, {Burgett}, {Cooper}, {Draper}, {Flewelling},
  {Hodapp}, {Jun}, {Kaiser}, {Kudritzki}, {Magnier}, {Metcalfe}, {Miller},
  {Schindler}, {Tonry}, {Wainscoat}, {Waters}, \& {Yang}}]{2016ApJS..227...11B}
{Ba{\~n}ados}, E., {Venemans}, B.~P., {Decarli}, R., {et~al.} 2016, \apjs, 227,
  11

\bibitem[{{Berger} {et~al.}(2012){Berger}, {Chornock}, {Lunnan}, {Foley},
  {Czekala}, {Rest}, {Leibler}, {Soderberg}, {Roth}, {Narayan}, {Huber},
  {Milisavljevic}, {Sanders}, {Drout}, {Margutti}, {Kirshner}, {Marion},
  {Challis}, {Riess}, {Smartt}, {Burgett}, {Hodapp}, {Heasley}, {Kaiser},
  {Kudritzki}, {Magnier}, {McCrum}, {Price}, {Smith}, {Tonry}, \&
  {Wainscoat}}]{2012ApJ...755L..29B}
{Berger}, E., {Chornock}, R., {Lunnan}, R., {et~al.} 2012, \apjl, 755, L29

\bibitem[{{Bernard} {et~al.}(2016){Bernard}, {Ferguson}, {Schlafly}, {Martin},
  {Rix}, {Bell}, {Finkbeiner}, {Goldman}, {Mart{\'{\i}}nez-Delgado}, {Sesar},
  {Wyse}, {Burgett}, {Chambers}, {Draper}, {Hodapp}, {Kaiser}, {Kudritzki},
  {Magnier}, {Metcalfe}, {Wainscoat}, \& {Waters}}]{2016MNRAS.463.1759B}
{Bernard}, E.~J., {Ferguson}, A.~M.~N., {Schlafly}, E.~F., {et~al.} 2016,
  \mnras, 463, 1759

\bibitem[{{Bertin}(2012)}]{2012ASPC..461..263B}
{Bertin}, E. 2012, in Astronomical Society of the Pacific Conference Series,
  Vol. 461, Astronomical Data Analysis Software and Systems XXI, ed.
  P.~{Ballester}, D.~{Egret}, \& N.~P.~F. {Lorente}, 263

\bibitem[{{Best} {et~al.}(2015){Best}, {Liu}, {Magnier}, {Deacon}, {Aller},
  {Redstone}, {Burgett}, {Chambers}, {Draper}, {Flewelling}, {Hodapp},
  {Kaiser}, {Metcalfe}, {Tonry}, {Wainscoat}, \& {Waters}}]{Best15}
{Best}, W.~M.~J., {Liu}, M.~C., {Magnier}, E.~A., {et~al.} 2015, \apj, 814, 118

\bibitem[{{Blagorodnova} {et~al.}(2016){Blagorodnova}, {Van Velzen},
  {Harrison}, {Koposov}, {Mattila}, {Campbell}, {Walton}, \&
  {Wyrzykowski}}]{2016MNRAS.455..603B}
{Blagorodnova}, N., {Van Velzen}, S., {Harrison}, D.~L., {et~al.} 2016, \mnras,
  455, 603

\bibitem[{{Bohlin} {et~al.}(2001){Bohlin}, {Dickinson}, \&
  {Calzetti}}]{2001AJ....122.2118B}
{Bohlin}, R.~C., {Dickinson}, M.~E., \& {Calzetti}, D. 2001, \aj, 122, 2118

\bibitem[{{Botticella} {et~al.}(2010){Botticella}, {Trundle}, {Pastorello},
  {Rodney}, {Rest}, {Gezari}, {Smartt}, {Narayan}, {Huber}, {Tonry}, {Young},
  {Smith}, {Bresolin}, {Valenti}, {Kotak}, {Mattila}, {Kankare}, {Wood-Vasey},
  {Riess}, {Neill}, {Forster}, {Martin}, {Stubbs}, {Burgett}, {Chambers},
  {Dombeck}, {Flewelling}, {Grav}, {Heasley}, {Hodapp}, {Kaiser}, {Kudritzki},
  {Luppino}, {Lupton}, {Magnier}, {Monet}, {Morgan}, {Onaka}, {Price},
  {Rhoads}, {Siegmund}, {Sweeney}, {Wainscoat}, {Waters}, {Waterson}, \&
  {Wynn-Williams}}]{2010ApJ...717L..52B}
{Botticella}, M.~T., {Trundle}, C., {Pastorello}, A., {et~al.} 2010, \apjl,
  717, L52

\bibitem[{{Carter} \& {Williams}(1973)}]{1973egfs.conf..433C}
{Carter}, W.~E., \& {Williams}, J.~D. 1973, in The Earth's Gravitational Field
  and Secular Variations in Position, ed. P.~V. {Angus-Leppan}, A.~G.
  {Bomford}, J.~C. {Dooley}, \& R.~S. {Mather}, 433

\bibitem[{{Chambers}(2006{\natexlab{a}})}]{2006amos.confE..39C}
{Chambers}, K. 2006{\natexlab{a}}, in The Advanced Maui Optical and Space
  Surveillance Technologies Conference, E39

\bibitem[{{Chambers}(2019)}]{2019AAS...23331301C}
{Chambers}, K. 2019, in American Astronomical Society Meeting Abstracts, Vol.
  233, American Astronomical Society Meeting Abstracts \#233

\bibitem[{{Chambers}(2006{\natexlab{b}})}]{chamberskc2006199843}
{Chambers}, K.~C. 2006{\natexlab{b}}, Mission Concept Statement for PS1,
  doi:10.5281/zenodo.199843

\bibitem[{{Chambers}(2007)}]{chamberskc2007199832}
---. 2007, PS1 Science Goals Statement, doi:10.5281/zenodo.199832

\bibitem[{{Chambers} \& {Denneau}(2008)}]{chamberskc2008199860}
{Chambers}, K.~C., \& {Denneau}, L.~J. 2008, PS1 Design Reference Mission,
  doi:10.5281/zenodo.199860

\bibitem[{{Chambers} {et~al.}(2018){Chambers}, {Onaka}, {Wainscoat}, {Magnier},
  \& {Pan-Starrs Team}}]{2018DPS....5031002C}
{Chambers}, K.~C., {Onaka}, P., {Wainscoat}, R., {Magnier}, E.~A., \&
  {Pan-Starrs Team}. 2018, in AAS/Division for Planetary Sciences Meeting
  Abstracts, Vol.~50, AAS/Division for Planetary Sciences Meeting Abstracts
  \#50, 310.02

\bibitem[{{Chan} {et~al.}(2016){Chan}, {Suyu}, {More}, {Oguri}, {Chiueh},
  {Coupon}, {Hsieh}, {Komiyama}, {Miyazaki}, {Murayama}, {Nishizawa}, {Price},
  {Tait}, {Terai}, {Utsumi}, \& {Wang}}]{2016ApJ...832..135C}
{Chan}, J.~H.~H., {Suyu}, S.~H., {More}, A., {et~al.} 2016, \apj, 832, 135

\bibitem[{{Chen} {et~al.}(2016){Chen}, {Lin}, {Holman}, {Payne}, {Fraser},
  {Lacerda}, {Ip}, {Chen}, {Kudritzki}, {Jedicke}, {Wainscoat}, {Tonry},
  {Magnier}, {Waters}, {Kaiser}, {Wang}, \& {Lehner}}]{2016ApJ...827L..24C}
{Chen}, Y.-T., {Lin}, H.~W., {Holman}, M.~J., {et~al.} 2016, \apjl, 827, L24

\bibitem[{{Chomiuk} {et~al.}(2011){Chomiuk}, {Chornock}, {Soderberg}, {Berger},
  {Chevalier}, {Foley}, {Huber}, {Narayan}, {Rest}, {Gezari}, {Kirshner},
  {Riess}, {Rodney}, {Smartt}, {Stubbs}, {Tonry}, {Wood-Vasey}, {Burgett},
  {Chambers}, {Czekala}, {Flewelling}, {Forster}, {Kaiser}, {Kudritzki},
  {Magnier}, {Martin}, {Morgan}, {Neill}, {Price}, {Roth}, {Sanders}, \&
  {Wainscoat}}]{2011ApJ...743..114C}
{Chomiuk}, L., {Chornock}, R., {Soderberg}, A.~M., {et~al.} 2011, \apj, 743,
  114

\bibitem[{{Chornock} {et~al.}(2014){Chornock}, {Berger}, {Gezari}, {Zauderer},
  {Rest}, {Chomiuk}, {Kamble}, {Soderberg}, {Czekala}, {Dittmann}, {Drout},
  {Foley}, {Fong}, {Huber}, {Kirshner}, {Lawrence}, {Lunnan}, {Marion},
  {Narayan}, {Riess}, {Roth}, {Sanders}, {Scolnic}, {Smartt}, {Smith},
  {Stubbs}, {Tonry}, {Burgett}, {Chambers}, {Flewelling}, {Hodapp}, {Kaiser},
  {Magnier}, {Martin}, {Neill}, {Price}, \& {Wainscoat}}]{2014ApJ...780...44C}
{Chornock}, R., {Berger}, E., {Gezari}, S., {et~al.} 2014, \apj, 780, 44

\bibitem[{{Dalcanton} {et~al.}(2012){Dalcanton}, {Williams}, {Lang}, {Lauer},
  {Kalirai}, {Seth}, {Dolphin}, {Rosenfield}, {Weisz}, {Bell}, {Bianchi},
  {Boyer}, {Caldwell}, {Dong}, {Dorman}, {Gilbert}, {Girardi}, {Gogarten},
  {Gordon}, {Guhathakurta}, {Hodge}, {Holtzman}, {Johnson}, {Larsen}, {Lewis},
  {Melbourne}, {Olsen}, {Rix}, {Rosema}, {Saha}, {Sarajedini}, {Skillman}, \&
  {Stanek}}]{2012ApJS..200...18D}
{Dalcanton}, J.~J., {Williams}, B.~F., {Lang}, D., {et~al.} 2012, \apjs, 200,
  18

\bibitem[{{de Vaucouleurs}(1948)}]{1948AnAp...11..247D}
{de Vaucouleurs}, G. 1948, Annales d'Astrophysique, 11, 247

\bibitem[{{Deacon} {et~al.}(2011{\natexlab{a}}){Deacon}, {Liu}, {Magnier},
  {Bowler}, {Goldman}, {Redstone}, {Burgett}, {Chambers}, {Flewelling},
  {Kaiser}, {Lupton}, {Morgan}, {Price}, {Sweeney}, {Tonry}, {Wainscoat}, \&
  {Waters}}]{2011AJ....142...77D}
{Deacon}, N.~R., {Liu}, M.~C., {Magnier}, E.~A., {et~al.} 2011{\natexlab{a}},
  \aj, 142, 77

\bibitem[{{Deacon} {et~al.}(2011{\natexlab{b}}){Deacon}, {Liu}, {Magnier},
  {Bowler}, {Goldman}, {Redstone}, {Burgett}, {Chambers}, {Flewelling},
  {Kaiser}, {Lupton}, {Morgan}, {Price}, {Sweeney}, {Tonry}, {Wainscoat}, \&
  {Waters}}]{Deacon11}
---. 2011{\natexlab{b}}, \aj, 142, 77

\bibitem[{Deacon {et~al.}(2012{\natexlab{a}})Deacon, Liu, Magnier, Bowler,
  Redstone, Goldman, Burgett, Chambers, Flewelling, Kaiser, Morgan, Price,
  Sweeney, Tonry, Wainscoat, \& Waters}]{Deacon2012a}
Deacon, N.~R., Liu, M.~C., Magnier, E.~A., {et~al.} 2012{\natexlab{a}},
  Astrophys. J., 755, 94

\bibitem[{Deacon {et~al.}(2012{\natexlab{b}})Deacon, Liu, Magnier, Bowler,
  Mann, Redstone, Burgett, Chambers, Hodapp, Kaiser, Kudritzki, Morgan, Price,
  Tonry, \& Wainscoat}]{Deacon2012b}
---. 2012{\natexlab{b}}, Astrophys. J., 757, 100

\bibitem[{{Deacon} {et~al.}(2014){Deacon}, {Liu}, {Magnier}, {Aller}, {Best},
  {Dupuy}, {Bowler}, {Mann}, {Redstone}, {Burgett}, {Chambers}, {Draper},
  {Flewelling}, {Hodapp}, {Kaiser}, {Kudritzki}, {Morgan}, {Metcalfe}, {Price},
  {Tonry}, \& {Wainscoat}}]{Deacon14}
{Deacon}, N.~R., {Liu}, M.~C., {Magnier}, E.~A., {et~al.} 2014, \apj, 792, 119

\bibitem[{Deacon {et~al.}(2016)Deacon, Kraus, Mann, Magnier, Chambers,
  Wainscoat, Tonry, Kaiser, Waters, Flewelling, Hodapp, \&
  Burgett}]{Deacon2016}
Deacon, N.~R., Kraus, A.~L., Mann, A.~W., {et~al.} 2016, Mon. Not. R. Astron.
  Soc., 455, 4212

\bibitem[{{Denneau} {et~al.}(2013){Denneau}, {Jedicke}, {Grav}, {Granvik},
  {Kubica}, {Milani}, {Vere{\v s}}, {Wainscoat}, {Chang}, {Pierfederici},
  {Kaiser}, {Chambers}, {Heasley}, {Magnier}, {Price}, {Myers}, {Kleyna},
  {Hsieh}, {Farnocchia}, {Waters}, {Sweeney}, {Green}, {Bolin}, {Burgett},
  {Morgan}, {Tonry}, {Hodapp}, {Chastel}, {Chesley}, {Fitzsimmons}, {Holman},
  {Spahr}, {Tholen}, {Williams}, {Abe}, {Armstrong}, {Bressi}, {Holmes},
  {Lister}, {McMillan}, {Micheli}, {Ryan}, {Ryan}, \&
  {Scotti}}]{2013PASP..125..357D}
{Denneau}, L., {Jedicke}, R., {Grav}, T., {et~al.} 2013, \pasp, 125, 357

\bibitem[{{Denneau} {et~al.}(2015){Denneau}, {Jedicke}, {Fitzsimmons}, {Hsieh},
  {Kleyna}, {Granvik}, {Micheli}, {Spahr}, {Vere{\v s}}, {Wainscoat},
  {Burgett}, {Chambers}, {Draper}, {Flewelling}, {Huber}, {Kaiser}, {Morgan},
  \& {Tonry}}]{2015Icar..245....1D}
{Denneau}, L., {Jedicke}, R., {Fitzsimmons}, A., {et~al.} 2015, Icarus, 245, 1

\bibitem[{{Drake} {et~al.}(2013){Drake}, {Catelan}, {Djorgovski}, {Torrealba},
  {Graham}, {Mahabal}, {Prieto}, {Donalek}, {Williams}, {Larson},
  {Christensen}, \& {Beshore}}]{2013ApJ...765..154D}
{Drake}, A.~J., {Catelan}, M., {Djorgovski}, S.~G., {et~al.} 2013, \apj, 765,
  154

\bibitem[{{Drout} {et~al.}(2014){Drout}, {Chornock}, {Soderberg}, {Sanders},
  {McKinnon}, {Rest}, {Foley}, {Milisavljevic}, {Margutti}, {Berger},
  {Calkins}, {Fong}, {Gezari}, {Huber}, {Kankare}, {Kirshner}, {Leibler},
  {Lunnan}, {Mattila}, {Marion}, {Narayan}, {Riess}, {Roth}, {Scolnic},
  {Smartt}, {Tonry}, {Burgett}, {Chambers}, {Hodapp}, {Jedicke}, {Kaiser},
  {Magnier}, {Metcalfe}, {Morgan}, {Price}, \& {Waters}}]{2014ApJ...794...23D}
{Drout}, M.~R., {Chornock}, R., {Soderberg}, A.~M., {et~al.} 2014, \apj, 794,
  23

\bibitem[{{Farrow} {et~al.}(2014){Farrow}, {Cole}, {Metcalfe}, {Draper},
  {Norberg}, {Foucaud}, {Burgett}, {Chambers}, {Kaiser}, {Kudritzki},
  {Magnier}, {Price}, {Tonry}, \& {Waters}}]{2014MNRAS.437..748F}
{Farrow}, D.~J., {Cole}, S., {Metcalfe}, N., {et~al.} 2014, \mnras, 437, 748

\bibitem[{{Finkbeiner} {et~al.}(2015){Finkbeiner}, {Schlafly}, {Schlegel},
  {Padmanabhan}, {Juric}, {Burgett}, {Chambers}, {Denneau}, {Draper},
  {Flewelling}, {Hodapp}, {Kaiser}, {Magnier}, {Metcalfe}, {Morgan}, {Price},
  {Stubbs}, \& {Tonry}}]{2015arXiv151201214F}
{Finkbeiner}, D.~P., {Schlafly}, E.~F., {Schlegel}, D.~J., {et~al.} 2015, ArXiv
  e-prints, arXiv:1512.01214

\bibitem[{{Flewelling} {et~al.}(2016){Flewelling}, {Magnier}, {Chambers},
  {Heasley}, {Holmberg}, {Huber}, {Sweeney}, {Waters}, {Chen}, {Farrow},
  {Hasinger}, {Henderson}, {Long}, {Metcalfe}, {Nieto-Santisteban}, {Norberg},
  {Saglia}, {Szalay}, {Rest}, {Thakar}, {Tonry}, {Valenti}, {Werner}, {White},
  {Denneau}, {Draper}, {Jedicke}, {Kudritzki}, {Price}, {Chastel}, {McClean},
  {Postman}, \& {Shiao}}]{flewelling2017}
{Flewelling}, H.~A., {Magnier}, E.~A., {Chambers}, K.~C., {et~al.} 2016, ArXiv
  e-prints, arXiv:1612.05243

\bibitem[{{Fraser} {et~al.}(2013){Fraser}, {Magee}, {Kotak}, {Smartt}, {Smith},
  {Polshaw}, {Drake}, {Boles}, {Lee}, {Burgett}, {Chambers}, {Draper},
  {Flewelling}, {Hodapp}, {Kaiser}, {Kudritzki}, {Magnier}, {Price}, {Tonry},
  {Wainscoat}, \& {Waters}}]{2013ApJ...779L...8F}
{Fraser}, M., {Magee}, M., {Kotak}, R., {et~al.} 2013, \apjl, 779, L8

\bibitem[{{Frei} \& {Gunn}(1994)}]{1994AJ....108.1476F}
{Frei}, Z., \& {Gunn}, J.~E. 1994, \aj, 108, 1476

\bibitem[{{Gaia Collaboration} {et~al.}(2016{\natexlab{a}}){Gaia
  Collaboration}, {Brown}, {Vallenari}, {Prusti}, {de Bruijne}, {Mignard},
  {Drimmel}, {Babusiaux}, {Bailer-Jones}, {Bastian}, \&
  et~al.}]{2016A&A...595A...2G}
{Gaia Collaboration}, {Brown}, A.~G.~A., {Vallenari}, A., {et~al.}
  2016{\natexlab{a}}, \aap, 595, A2

\bibitem[{{Gaia Collaboration} {et~al.}(2016{\natexlab{b}}){Gaia
  Collaboration}, {Prusti}, {de Bruijne}, {Brown}, {Vallenari}, {Babusiaux},
  {Bailer-Jones}, {Bastian}, {Biermann}, {Evans}, \&
  et~al.}]{2016A&A...595A...1G}
{Gaia Collaboration}, {Prusti}, T., {de Bruijne}, J.~H.~J., {et~al.}
  2016{\natexlab{b}}, \aap, 595, A1

\bibitem[{{Gezari} {et~al.}(2010){Gezari}, {Rest}, {Huber}, {Narayan},
  {Forster}, {Neill}, {Martin}, {Valenti}, {Smartt}, {Chornock}, {Berger},
  {Soderberg}, {Mattila}, {Kankare}, {Burgett}, {Chambers}, {Dombeck}, {Grav},
  {Heasley}, {Hodapp}, {Jedicke}, {Kaiser}, {Kudritzki}, {Luppino}, {Lupton},
  {Magnier}, {Monet}, {Morgan}, {Onaka}, {Price}, {Rhoads}, {Siegmund},
  {Stubbs}, {Tonry}, {Wainscoat}, {Waterson}, \&
  {Wynn-Williams}}]{2010ApJ...720L..77G}
{Gezari}, S., {Rest}, A., {Huber}, M.~E., {et~al.} 2010, \apjl, 720, L77

\bibitem[{{Gezari} {et~al.}(2012){Gezari}, {Chornock}, {Rest}, {Huber},
  {Forster}, {Berger}, {Challis}, {Neill}, {Martin}, {Heckman}, {Lawrence},
  {Norman}, {Narayan}, {Foley}, {Marion}, {Scolnic}, {Chomiuk}, {Soderberg},
  {Smith}, {Kirshner}, {Riess}, {Smartt}, {Stubbs}, {Tonry}, {Wood-Vasey},
  {Burgett}, {Chambers}, {Grav}, {Heasley}, {Kaiser}, {Kudritzki}, {Magnier},
  {Morgan}, \& {Price}}]{2012Natur.485..217G}
{Gezari}, S., {Chornock}, R., {Rest}, A., {et~al.} 2012, \nat, 485, 217

\bibitem[{{Goldman} {et~al.}(2013){Goldman}, {R{\"o}ser}, {Schilbach},
  {Magnier}, {Olczak}, {Henning}, {Juri{\'c}}, {Schlafly}, {Chen}, {Platais},
  {Burgett}, {Hodapp}, {Heasley}, {Kudritzki}, {Morgan}, {Price}, {Tonry}, \&
  {Wainscoat}}]{Goldman13}
{Goldman}, B., {R{\"o}ser}, S., {Schilbach}, E., {et~al.} 2013, \aap, 559, A43

\bibitem[{{G{\'o}rski} {et~al.}(2005){G{\'o}rski}, {Hivon}, {Banday},
  {Wandelt}, {Hansen}, {Reinecke}, \& {Bartelmann}}]{2005ApJ...622..759G}
{G{\'o}rski}, K.~M., {Hivon}, E., {Banday}, A.~J., {et~al.} 2005, \apj, 622,
  759

\bibitem[{{Grav} {et~al.}(2011){Grav}, {Jedicke}, {Denneau}, {Chesley},
  {Holman}, \& {Spahr}}]{Grav2011}
{Grav}, T., {Jedicke}, R., {Denneau}, L., {et~al.} 2011, \pasp, 123, 423

\bibitem[{{Green} {et~al.}(2014){Green}, {Schlafly}, {Finkbeiner}, {Juri{\'c}},
  {Rix}, {Burgett}, {Chambers}, {Draper}, {Flewelling}, {Kudritzki}, {Magnier},
  {Martin}, {Metcalfe}, {Tonry}, {Wainscoat}, \&
  {Waters}}]{2014ApJ...783..114G}
{Green}, G.~M., {Schlafly}, E.~F., {Finkbeiner}, D.~P., {et~al.} 2014, \apj,
  783, 114

\bibitem[{{Green} {et~al.}(2015){Green}, {Schlafly}, {Finkbeiner}, {Rix},
  {Martin}, {Burgett}, {Draper}, {Flewelling}, {Hodapp}, {Kaiser}, {Kudritzki},
  {Magnier}, {Metcalfe}, {Price}, {Tonry}, \&
  {Wainscoat}}]{2015ApJ...810...25G}
---. 2015, \apj, 810, 25

\bibitem[{{Heasley}(2008)}]{2008AIPC.1082..352H}
{Heasley}, J.~N. 2008, in American Institute of Physics Conference Series, Vol.
  1082, American Institute of Physics Conference Series, ed. C.~A.~L.
  {Bailer-Jones}, 352--358

\bibitem[{{Heinis} {et~al.}(2016){Heinis}, {Gezari}, {Kumar}, {Burgett},
  {Flewelling}, {Huber}, {Kaiser}, {Wainscoat}, \&
  {Waters}}]{2016ApJ...826...62H}
{Heinis}, S., {Gezari}, S., {Kumar}, S., {et~al.} 2016, \apj, 826, 62

\bibitem[{{Hernitschek} {et~al.}(2016){Hernitschek}, {Schlafly}, {Sesar},
  {Rix}, {Hogg}, {Ivezi{\'c}}, {Grebel}, {Bell}, {Martin}, {Burgett},
  {Flewelling}, {Hodapp}, {Kaiser}, {Magnier}, {Metcalfe}, {Wainscoat}, \&
  {Waters}}]{2016ApJ...817...73H}
{Hernitschek}, N., {Schlafly}, E.~F., {Sesar}, B., {et~al.} 2016, \apj, 817, 73

\bibitem[{{Hodapp} {et~al.}(2004{\natexlab{a}}){Hodapp}, {Siegmund}, {Kaiser},
  {Chambers}, {Laux}, {Morgan}, \& {Mannery}}]{2004SPIE.5489..667H}
{Hodapp}, K.~W., {Siegmund}, W.~A., {Kaiser}, N., {et~al.} 2004{\natexlab{a}},
  in \procspie, Vol. 5489, Ground-based Telescopes, ed. J.~M. {Oschmann}, Jr.,
  667--678

\bibitem[{{Hodapp} {et~al.}(2004{\natexlab{b}}){Hodapp}, {Kaiser}, {Aussel},
  {Burgett}, {Chambers}, {Chun}, {Dombeck}, {Douglas}, {Hafner}, {Heasley},
  {Hoblitt}, {Hude}, {Isani}, {Jedicke}, {Jewitt}, {Laux}, {Luppino}, {Lupton},
  {Maberry}, {Magnier}, {Mannery}, {Monet}, {Morgan}, {Onaka}, {Price}, {Ryan},
  {Siegmund}, {Szapudi}, {Tonry}, {Wainscoat}, \&
  {Waterson}}]{2004AN....325..636H}
{Hodapp}, K.~W., {Kaiser}, N., {Aussel}, H., {et~al.} 2004{\natexlab{b}},
  Astronomische Nachrichten, 325, 636

\bibitem[{{H{\o}g} {et~al.}(2000){H{\o}g}, {Fabricius}, {Makarov}, {Urban},
  {Corbin}, {Wycoff}, {Bastian}, {Schwekendiek}, \&
  {Wicenec}}]{2000A&A...355L..27H}
{H{\o}g}, E., {Fabricius}, C., {Makarov}, V.~V., {et~al.} 2000, \aap, 355, L27

\bibitem[{{Holoien} {et~al.}(2017){Holoien}, {Stanek}, {Kochanek}, {Shappee},
  {Prieto}, {Brimacombe}, {Bersier}, {Bishop}, {Dong}, {Brown}, {Danilet},
  {Simonian}, {Basu}, {Beacom}, {Falco}, {Pojmanski}, {Skowron}, {Wo{\'z}niak},
  {{\'A}vila}, {Conseil}, {Contreras}, {Cruz}, {Fern{\'a}ndez}, {Koff}, {Guo},
  {Herczeg}, {Hissong}, {Hsiao}, {Jose}, {Kiyota}, {Long}, {Monard},
  {Nicholls}, {Nicolas}, \& {Wiethoff}}]{2017MNRAS.464.2672H}
{Holoien}, T.~W.-S., {Stanek}, K.~Z., {Kochanek}, C.~S., {et~al.} 2017, \mnras,
  464, arXiv:1604.00396

\bibitem[{{Hsieh} {et~al.}(2015){Hsieh}, {Denneau}, {Wainscoat},
  {Sch{\"o}rghofer}, {Bolin}, {Fitzsimmons}, {Jedicke}, {Kleyna}, {Micheli},
  {Vere{\v s}}, {Kaiser}, {Chambers}, {Burgett}, {Flewelling}, {Hodapp},
  {Magnier}, {Morgan}, {Price}, {Tonry}, \& {Waters}}]{2015Icar..248..289H}
{Hsieh}, H.~H., {Denneau}, L., {Wainscoat}, R.~J., {et~al.} 2015, Icarus, 248,
  289

\bibitem[{{Huber} {et~al.}(2015){Huber}, {Carter Chambers}, {Flewelling},
  {Smartt}, {Smith}, \& {Wright}}]{2015IAUGA..2258303H}
{Huber}, M., {Carter Chambers}, K., {Flewelling}, H., {et~al.} 2015, IAU
  General Assembly, 22, 58303

\bibitem[{{Ibata} {et~al.}(2017){Ibata}, {McConnachie}, {Cuillandre}, {Fantin},
  {Haywood}, {Martin}, {Bergeron}, {Beckmann}, {Bernard}, {Bonifacio},
  {Caffau}, {Carlberg}, {C{\^o}t{\'e}}, {Cabanac}, {Chapman}, {Duc}, {Durret},
  {Famaey}, {Fabbro}, {Gwyn}, {Hammer}, {Hill}, {Hudson}, {Lan{\c c}on},
  {Lewis}, {Malhan}, {di Matteo}, {McCracken}, {Mei}, {Mellier}, {Navarro},
  {Pires}, {Pritchet}, {Reyl{\'e}}, {Richer}, {Robin}, {S{\'a}nchez-Janssen},
  {Sawicki}, {Scott}, {Scottez}, {Spekkens}, {Starkenburg}, {Thomas}, \&
  {Venn}}]{2017ApJ...848..128I}
{Ibata}, R.~A., {McConnachie}, A., {Cuillandre}, J.-C., {et~al.} 2017, \apj,
  848, 128

\bibitem[{{Inserra} {et~al.}(2013){Inserra}, {Smartt}, {Jerkstrand}, {Valenti},
  {Fraser}, {Wright}, {Smith}, {Chen}, {Kotak}, {Pastorello}, {Nicholl},
  {Bresolin}, {Kudritzki}, {Benetti}, {Botticella}, {Burgett}, {Chambers},
  {Ergon}, {Flewelling}, {Fynbo}, {Geier}, {Hodapp}, {Howell}, {Huber},
  {Kaiser}, {Leloudas}, {Magill}, {Magnier}, {McCrum}, {Metcalfe}, {Price},
  {Rest}, {Sollerman}, {Sweeney}, {Taddia}, {Taubenberger}, {Tonry},
  {Wainscoat}, {Waters}, \& {Young}}]{2013ApJ...770..128I}
{Inserra}, C., {Smartt}, S.~J., {Jerkstrand}, A., {et~al.} 2013, \apj, 770, 128

\bibitem[{{Inserra} {et~al.}(2016){Inserra}, {Smartt}, {Gall}, {Leloudas},
  {Chen}, {Schulze}, {Jerkstarnd}, {Nicholl}, {Anderson}, {Arcavi}, {Benetti},
  {Cartier}, {Childress}, {Della Valle}, {Flewelling}, {Fraser}, {Gal-Yam},
  {Gutierrez}, {Hosseinzadeh}, {Howell}, {Huber}, {Kankare}, {Magnier},
  {Maguire}, {McCully}, {Prajs}, {Primak}, {Scalzo}, {Schmidt}, {Smith},
  {Tucker}, {Valenti}, {Wilman}, {Young}, \& {Yuan}}]{2016arXiv160401226I}
{Inserra}, C., {Smartt}, S.~J., {Gall}, E.~E.~E., {et~al.} 2016, ArXiv
  e-prints, arXiv:1604.01226

\bibitem[{{Jian} {et~al.}(2014){Jian}, {Lin}, {Chiueh}, {Lin}, {Liu}, {Merson},
  {Baugh}, {Huang}, {Chen}, {Foucaud}, {Murphy}, {Cole}, {Burgett}, \&
  {Kaiser}}]{Jian14}
{Jian}, H.-Y., {Lin}, L., {Chiueh}, T., {et~al.} 2014, \apj, 788, 109

\bibitem[{{Jones} {et~al.}(2016){Jones}, {Scolnic}, {Riess}, {Kessler}, {Rest},
  {Kirshner}, {Berger}, {Ortega}, {Foley}, {Chornock}, {Challis}, {Burgett},
  {Chambers}, {Draper}, {Flewelling}, {Huber}, {Kaiser}, {Kudritzki},
  {Metcalfe}, {Wainscoat}, \& {Waters}}]{2016arXiv161107042J}
{Jones}, D.~O., {Scolnic}, D.~M., {Riess}, A.~G., {et~al.} 2016, ArXiv
  e-prints, arXiv:1611.07042

\bibitem[{{Juric}(2011)}]{2011AAS...21743319J}
{Juric}, M. 2011, in Bulletin of the American Astronomical Society, Vol.~43,
  American Astronomical Society Meeting Abstracts \#217, 433.19

\bibitem[{{Kaiser} {et~al.}(1995){Kaiser}, {Squires}, \&
  {Broadhurst}}]{1995ApJ...449..460K}
{Kaiser}, N., {Squires}, G., \& {Broadhurst}, T. 1995, \apj, 449, 460

\bibitem[{{Kaiser} {et~al.}(2000){Kaiser}, {Tonry}, \&
  {Luppino}}]{2000PASP..112..768K}
{Kaiser}, N., {Tonry}, J.~L., \& {Luppino}, G.~A. 2000, \pasp, 112, 768

\bibitem[{{Kaiser} {et~al.}(2002){Kaiser}, {Aussel}, {Burke}, {Boesgaard},
  {Chambers}, {Chun}, {Heasley}, {Hodapp}, {Hunt}, {Jedicke}, {Jewitt},
  {Kudritzki}, {Luppino}, {Maberry}, {Magnier}, {Monet}, {Onaka}, {Pickles},
  {Rhoads}, {Simon}, {Szalay}, {Szapudi}, {Tholen}, {Tonry}, {Waterson}, \&
  {Wick}}]{2002SPIE.4836..154K}
{Kaiser}, N., {Aussel}, H., {Burke}, B.~E., {et~al.} 2002, in \procspie, Vol.
  4836, Survey and Other Telescope Technologies and Discoveries, ed. J.~A.
  {Tyson} \& S.~{Wolff}, 154--164

\bibitem[{{Kaiser} {et~al.}(2010){Kaiser}, {Burgett}, {Chambers}, {Denneau},
  {Heasley}, {Jedicke}, {Magnier}, {Morgan}, {Onaka}, \& {Tonry}}]{kaiser2010}
{Kaiser}, N., {Burgett}, W., {Chambers}, K., {et~al.} 2010, in \procspie, Vol.
  7733, Ground-based and Airborne Telescopes III, 77330E

\bibitem[{{Kodric} {et~al.}(2013){Kodric}, {Riffeser}, {Hopp}, {Seitz},
  {Koppenhoefer}, {Bender}, {Goessl}, {Snigula}, {Lee}, {Ngeow}, {Chambers},
  {Magnier}, {Price}, {Burgett}, {Hodapp}, {Kaiser}, \&
  {Kudritzki}}]{2013AJ....145..106K}
{Kodric}, M., {Riffeser}, A., {Hopp}, U., {et~al.} 2013, \aj, 145, 106

\bibitem[{{Kodric} {et~al.}(2015){Kodric}, {Riffeser}, {Seitz}, {Snigula},
  {Hopp}, {Lee}, {Goessl}, {Koppenhoefer}, {Bender}, \&
  {Gieren}}]{2015ApJ...799..144K}
{Kodric}, M., {Riffeser}, A., {Seitz}, S., {et~al.} 2015, \apj, 799, 144

\bibitem[{{Kron}(1980)}]{1980ApJS...43..305K}
{Kron}, R.~G. 1980, \apjs, 43, 305

\bibitem[{{Laevens} {et~al.}(2015){Laevens}, {Martin}, {Bernard}, {Schlafly},
  {Sesar}, {Rix}, {Bell}, {Ferguson}, {Slater}, {Sweeney}, {Wyse}, {Huxor},
  {Burgett}, {Chambers}, {Draper}, {Hodapp}, {Kaiser}, {Magnier}, {Metcalfe},
  {Tonry}, {Wainscoat}, \& {Waters}}]{2015ApJ...813...44L}
{Laevens}, B.~P.~M., {Martin}, N.~F., {Bernard}, E.~J., {et~al.} 2015, \apj,
  813, 44

\bibitem[{{Lawrence} {et~al.}(2016){Lawrence}, {Bruce}, {MacLeod}, {Gezari},
  {Elvis}, {Ward}, {Smartt}, {Smith}, {Wright}, {Fraser}, {Marshall}, {Kaiser},
  {Burgett}, {Magnier}, {Tonry}, {Chambers}, {Wainscoat}, {Waters}, {Price},
  {Metcalfe}, {Valenti}, {Kotak}, {Mead}, {Inserra}, {Chen}, \&
  {Soderberg}}]{2016MNRAS.463..296L}
{Lawrence}, A., {Bruce}, A.~G., {MacLeod}, C., {et~al.} 2016, \mnras, 463, 296

\bibitem[{{Lee} {et~al.}(2012){Lee}, {Riffeser}, {Koppenhoefer}, {Seitz},
  {Bender}, {Hopp}, {G{\"o}ssl}, {Saglia}, {Snigula}, {Sweeney}, {Burgett},
  {Chambers}, {Grav}, {Heasley}, {Hodapp}, {Kaiser}, {Magnier}, {Morgan},
  {Price}, {Stubbs}, {Tonry}, \& {Wainscoat}}]{2012AJ....143...89L}
{Lee}, C.-H., {Riffeser}, A., {Koppenhoefer}, J., {et~al.} 2012, \aj, 143, 89

\bibitem[{{Lee} {et~al.}(2013){Lee}, {Kodric}, {Seitz}, {Riffeser},
  {Koppenhoefer}, {Bender}, {Hopp}, {G{\"o}ssl}, {Snigula}, {Burgett},
  {Chambers}, {Flewelling}, {Hodapp}, {Kaiser}, {Kudritzki}, {Price}, {Tonry},
  \& {Wainscoat}}]{2013ApJ...777...35L}
{Lee}, C.-H., {Kodric}, M., {Seitz}, S., {et~al.} 2013, \apj, 777, 35

\bibitem[{{Lee} {et~al.}(2014{\natexlab{a}}){Lee}, {Seitz}, {Kodric},
  {Riffeser}, {Koppenhoefer}, {Bender}, {Snigula}, {Hopp}, {G{\"o}ssl},
  {Bianchi}, {Price}, {Fraser}, {Burgett}, {Chambers}, {Draper}, {Flewelling},
  {Kaiser}, {Kudritzki}, \& {Magnier}}]{2014ApJ...785...11L}
{Lee}, C.-H., {Seitz}, S., {Kodric}, M., {et~al.} 2014{\natexlab{a}}, \apj,
  785, 11

\bibitem[{{Lee} {et~al.}(2014{\natexlab{b}}){Lee}, {Koppenhoefer}, {Seitz},
  {Bender}, {Riffeser}, {Kodric}, {Hopp}, {Snigula}, {G{\"o}ssl}, {Kudritzki},
  {Burgett}, {Chambers}, {Hodapp}, {Kaiser}, \& {Waters}}]{2014ApJ...797...22L}
{Lee}, C.-H., {Koppenhoefer}, J., {Seitz}, S., {et~al.} 2014{\natexlab{b}},
  \apj, 797, 22

\bibitem[{{Lin} {et~al.}(2016){Lin}, {Chen}, {Holman}, {Ip}, {Payne},
  {Lacerda}, {Fraser}, {Gerdes}, {Bieryla}, {Sie}, {Chen}, {Burgett},
  {Denneau}, {Jedicke}, {Kaiser}, {Magnier}, {Tonry}, {Wainscoat}, \&
  {Waters}}]{2016AJ....152..147L}
{Lin}, H.~W., {Chen}, Y.-T., {Holman}, M.~J., {et~al.} 2016, \aj, 152, 147

\bibitem[{{Lin} {et~al.}(2014){Lin}, {Jian}, {Foucaud}, {Norberg}, {Bower},
  {Cole}, {Arnalte-Mur}, {Chen}, {Coupon}, {Hsieh}, {Heinis}, {Phleps}, {Chen},
  {Lee}, {Burgett}, {Chambers}, {Denneau}, {Draper}, {Flewelling}, {Hodapp},
  {Huber}, {Kaiser}, {Kudritzki}, {Magnier}, {Metcalfe}, {Price}, {Tonry},
  {Wainscoat}, \& {Waters}}]{Lin14}
{Lin}, L., {Jian}, H.-Y., {Foucaud}, S., {et~al.} 2014, \apj, 782, 33

\bibitem[{{Lindegren} {et~al.}(2016){Lindegren}, {Lammers}, {Bastian},
  {Hern{\'a}ndez}, {Klioner}, {Hobbs}, {Bombrun}, {Michalik}, {Ramos-Lerate},
  {Butkevich}, {Comoretto}, {Joliet}, {Holl}, {Hutton}, {Parsons},
  {Steidelm{\"u}ller}, {Abbas}, {Altmann}, {Andrei}, {Anton}, {Bach},
  {Barache}, {Becciani}, {Berthier}, {Bianchi}, {Biermann}, {Bouquillon},
  {Bourda}, {Br{\"u}semeister}, {Bucciarelli}, {Busonero}, {Carlucci},
  {Casta{\~n}eda}, {Charlot}, {Clotet}, {Crosta}, {Davidson}, {de Felice},
  {Drimmel}, {Fabricius}, {Fienga}, {Figueras}, {Fraile}, {Gai}, {Garralda},
  {Geyer}, {Gonz{\'a}lez-Vidal}, {Guerra}, {Hambly}, {Hauser}, {Jordan},
  {Lattanzi}, {Lenhardt}, {Liao}, {L{\"o}ffler}, {McMillan}, {Mignard}, {Mora},
  {Morbidelli}, {Portell}, {Riva}, {Sarasso}, {Serraller}, {Siddiqui}, {Smart},
  {Spagna}, {Stampa}, {Steele}, {Taris}, {Torra}, {van Reeven}, {Vecchiato},
  {Zschocke}, {de Bruijne}, {Gracia}, {Raison}, {Lister}, {Marchant},
  {Messineo}, {Soffel}, {Osorio}, {de Torres}, \&
  {O'Mullane}}]{2016A&A...595A...4L}
{Lindegren}, L., {Lammers}, U., {Bastian}, U., {et~al.} 2016, \aap, 595, A4

\bibitem[{{Liu} {et~al.}(2015){Liu}, {Hennig}, {Desai}, {Hoyle},
  {Koppenhoefer}, {Mohr}, {Paech}, {Burgett}, {Chambers}, {Cole}, {Draper},
  {Kaiser}, {Metcalfe}, {Morgan}, {Price}, {Stubbs}, {Tonry}, {Wainscoat}, \&
  {Waters}}]{Liu15}
{Liu}, J., {Hennig}, C., {Desai}, S., {et~al.} 2015, \mnras, 449, 3370

\bibitem[{{Liu} {et~al.}(2013{\natexlab{a}}){Liu}, {Magnier}, {Deacon},
  {Allers}, {Dupuy}, {Kotson}, {Aller}, {Burgett}, {Chambers}, {Draper},
  {Hodapp}, {Jedicke}, {Kaiser}, {Kudritzki}, {Metcalfe}, {Morgan}, {Price},
  {Tonry}, \& {Wainscoat}}]{2013ApJ...777L..20L}
{Liu}, M.~C., {Magnier}, E.~A., {Deacon}, N.~R., {et~al.} 2013{\natexlab{a}},
  \apjl, 777, L20

\bibitem[{{Liu} {et~al.}(2013{\natexlab{b}}){Liu}, {Magnier}, {Deacon},
  {Allers}, {Dupuy}, {Kotson}, {Aller}, {Burgett}, {Chambers}, {Draper},
  {Hodapp}, {Jedicke}, {Kaiser}, {Kudritzki}, {Metcalfe}, {Morgan}, {Price},
  {Tonry}, \& {Wainscoat}}]{Liu13}
---. 2013{\natexlab{b}}, \apjl, 777, L20

\bibitem[{{Lunnan} {et~al.}(2013){Lunnan}, {Chornock}, {Berger},
  {Milisavljevic}, {Drout}, {Sanders}, {Challis}, {Czekala}, {Foley}, {Fong},
  {Huber}, {Kirshner}, {Leibler}, {Marion}, {McCrum}, {Narayan}, {Rest},
  {Roth}, {Scolnic}, {Smartt}, {Smith}, {Soderberg}, {Stubbs}, {Tonry},
  {Burgett}, {Chambers}, {Kudritzki}, {Magnier}, \&
  {Price}}]{2013ApJ...771...97L}
{Lunnan}, R., {Chornock}, R., {Berger}, E., {et~al.} 2013, \apj, 771, 97

\bibitem[{{Lunnan} {et~al.}(2014){Lunnan}, {Chornock}, {Berger}, {Laskar},
  {Fong}, {Rest}, {Sanders}, {Challis}, {Drout}, {Foley}, {Huber}, {Kirshner},
  {Leibler}, {Marion}, {McCrum}, {Milisavljevic}, {Narayan}, {Scolnic},
  {Smartt}, {Smith}, {Soderberg}, {Tonry}, {Burgett}, {Chambers}, {Flewelling},
  {Hodapp}, {Kaiser}, {Magnier}, {Price}, \& {Wainscoat}}]{2014ApJ...787..138L}
---. 2014, \apj, 787, 138

\bibitem[{{Lunnan} {et~al.}(2015){Lunnan}, {Chornock}, {Berger}, {Rest},
  {Fong}, {Scolnic}, {Jones}, {Soderberg}, {Challis}, {Drout}, {Foley},
  {Huber}, {Kirshner}, {Leibler}, {Marion}, {McCrum}, {Milisavljevic},
  {Narayan}, {Sanders}, {Smartt}, {Smith}, {Tonry}, {Burgett}, {Chambers},
  {Flewelling}, {Kudritzki}, {Wainscoat}, \& {Waters}}]{2015ApJ...804...90L}
---. 2015, \apj, 804, 90

\bibitem[{{Lunnan} {et~al.}(2016){Lunnan}, {Chornock}, {Berger},
  {Milisavljevic}, {Jones}, {Rest}, {Fong}, {Fransson}, {Margutti}, {Drout},
  {Blanchard}, {Challis}, {Cowperthwaite}, {Foley}, {Kirshner}, {Morrell},
  {Riess}, {Roth}, {Scolnic}, {Smartt}, {Smith}, {Villar}, {Chambers},
  {Draper}, {Huber}, {Kaiser}, {Kudritzki}, {Magnier}, {Metcalfe}, \&
  {Waters}}]{2016ApJ...831..144L}
---. 2016, \apj, 831, 144

\bibitem[{{Lupton} {et~al.}(1999){Lupton}, {Gunn}, \&
  {Szalay}}]{1999AJ....118.1406L}
{Lupton}, R.~H., {Gunn}, J.~E., \& {Szalay}, A.~S. 1999, \aj, 118, 1406

\bibitem[{{Magnier} {et~al.}(2013){Magnier}, {Schlafly}, {Finkbeiner}, {Juric},
  {Tonry}, {Burgett}, {Chambers}, {Flewelling}, {Kaiser}, {Kudritzki},
  {Morgan}, {Price}, {Sweeney}, \& {Stubbs}}]{2013ApJS..205...20M}
{Magnier}, E.~A., {Schlafly}, E., {Finkbeiner}, D., {et~al.} 2013, \apjs, 205,
  20

\bibitem[{{Magnier} {et~al.}(2016{\natexlab{a}}){Magnier}, {Chambers},
  {Flewelling}, {Hoblitt}, {Huber}, {Price}, {Sweeney}, {Waters}, {Denneau},
  {Draper}, {Hodapp}, {Jedicke}, {Kudritzki}, {Metcalfe}, {Stubbs}, \&
  {Wainscoast}}]{magnier2017a}
{Magnier}, E.~A., {Chambers}, K.~C., {Flewelling}, H.~A., {et~al.}
  2016{\natexlab{a}}, ArXiv e-prints, arXiv:1612.05240

\bibitem[{{Magnier} {et~al.}(2016{\natexlab{b}}){Magnier}, {Schlafly},
  {Finkbeiner}, {Tonry}, {Goldman}, {R{\"o}ser}, {Schilbach}, {Chambers},
  {Flewelling}, {Huber}, {Price}, {Sweeney}, {Waters}, {Denneau}, {Draper},
  {Hodapp}, {Jedicke}, {Kudritzki}, {Metcalfe}, {Stubbs}, \&
  {Wainscoast}}]{magnier2017c}
{Magnier}, E.~A., {Schlafly}, E.~F., {Finkbeiner}, D.~P., {et~al.}
  2016{\natexlab{b}}, ArXiv e-prints, arXiv:1612.05242

\bibitem[{{Magnier} {et~al.}(2016{\natexlab{c}}){Magnier}, {Sweeney},
  {Chambers}, {Flewelling}, {Huber}, {Price}, {Waters}, {Denneau}, {Draper},
  {Jedicke}, {Hodapp}, {Kudritzki}, {Metcalfe}, {Stubbs}, \&
  {Wainscoast}}]{magnier2017b}
{Magnier}, E.~A., {Sweeney}, W.~E., {Chambers}, K.~C., {et~al.}
  2016{\natexlab{c}}, ArXiv e-prints, arXiv:1612.05244

\bibitem[{{Magnier} {et~al.}(2018){Magnier}, {Tonry}, {Finkbeiner}, {Schlafly},
  {Burgett}, {Chambers}, {Flewelling}, {Hodapp}, {Kaiser}, {Kudritzki},
  {Metcalfe}, {Wainscoat}, \& {Waters}}]{2018PASP..130f5002M}
{Magnier}, E.~A., {Tonry}, J.~L., {Finkbeiner}, D., {et~al.} 2018, \pasp, 130,
  065002

\bibitem[{{McCrum} {et~al.}(2014){McCrum}, {Smartt}, {Kotak}, {Rest},
  {Jerkstrand}, {Inserra}, {Rodney}, {Chen}, {Howell}, {Huber}, {Pastorello},
  {Tonry}, {Bresolin}, {Kudritzki}, {Chornock}, {Berger}, {Smith},
  {Botticella}, {Foley}, {Fraser}, {Milisavljevic}, {Nicholl}, {Riess},
  {Stubbs}, {Valenti}, {Wood-Vasey}, {Wright}, {Young}, {Drout}, {Czekala},
  {Burgett}, {Chambers}, {Draper}, {Flewelling}, {Hodapp}, {Kaiser}, {Magnier},
  {Metcalfe}, {Price}, {Sweeney}, \& {Wainscoat}}]{2014MNRAS.437..656M}
{McCrum}, M., {Smartt}, S.~J., {Kotak}, R., {et~al.} 2014, \mnras, 437, 656

\bibitem[{{McCrum} {et~al.}(2015){McCrum}, {Smartt}, {Rest}, {Smith}, {Kotak},
  {Rodney}, {Young}, {Chornock}, {Berger}, {Foley}, {Fraser}, {Wright},
  {Scolnic}, {Tonry}, {Urata}, {Huang}, {Pastorello}, {Botticella}, {Valenti},
  {Mattila}, {Kankare}, {Farrow}, {Huber}, {Stubbs}, {Kirshner}, {Bresolin},
  {Burgett}, {Chambers}, {Draper}, {Flewelling}, {Jedicke}, {Kaiser},
  {Magnier}, {Metcalfe}, {Morgan}, {Price}, {Sweeney}, {Wainscoat}, \&
  {Waters}}]{2015MNRAS.448.1206M}
{McCrum}, M., {Smartt}, S.~J., {Rest}, A., {et~al.} 2015, \mnras, 448, 1206

\bibitem[{{Meech} {et~al.}(2013){Meech}, {Yang}, {Kleyna}, {Ansdell}, {Chiang},
  {Hainaut}, {Vincent}, {Boehnhardt}, {Fitzsimmons}, {Rector}, {Riesen},
  {Keane}, {Reipurth}, {Hsieh}, {Michaud}, {Milani}, {Bryssinck}, {Ligustri},
  {Trabatti}, {Tozzi}, {Mottola}, {Kuehrt}, {Bhatt}, {Sahu}, {Lisse},
  {Denneau}, {Jedicke}, {Magnier}, \& {Wainscoat}}]{2013ApJ...776L..20M}
{Meech}, K.~J., {Yang}, B., {Kleyna}, J., {et~al.} 2013, \apjl, 776, L20

\bibitem[{{Metcalfe} {et~al.}(2013){Metcalfe}, {Farrow}, {Cole}, {Draper},
  {Norberg}, {Burgett}, {Chambers}, {Denneau}, {Flewelling}, {Kaiser},
  {Kudritzki}, {Magnier}, {Morgan}, {Price}, {Sweeney}, {Tonry}, {Wainscoat},
  \& {Waters}}]{2013MNRAS.435.1825M}
{Metcalfe}, N., {Farrow}, D.~J., {Cole}, S., {et~al.} 2013, \mnras, 435, 1825

\bibitem[{{Monet} {et~al.}(2003){Monet}, {Levine}, {Canzian}, {Ables}, {Bird},
  {Dahn}, {Guetter}, {Harris}, {Henden}, {Leggett}, {Levison}, {Luginbuhl},
  {Martini}, {Monet}, {Munn}, {Pier}, {Rhodes}, {Riepe}, {Sell}, {Stone},
  {Vrba}, {Walker}, {Westerhout}, {Brucato}, {Reid}, {Schoening}, {Hartley},
  {Read}, \& {Tritton}}]{2003AJ....125..984M}
{Monet}, D.~G., {Levine}, S.~E., {Canzian}, B., {et~al.} 2003, \aj, 125, 984

\bibitem[{{Morgan} \& {Kaiser}(2008)}]{2008SPIE.7012E..1KM}
{Morgan}, J.~S., \& {Kaiser}, N. 2008, in \procspie, Vol. 7012, Ground-based
  and Airborne Telescopes II, 70121K

\bibitem[{{Morgan} {et~al.}(2012){Morgan}, {Kaiser}, {Moreau}, {Anderson}, \&
  {Burgett}}]{2012SPIE.8444E..0HM}
{Morgan}, J.~S., {Kaiser}, N., {Moreau}, V., {Anderson}, D., \& {Burgett}, W.
  2012, in \procspie, Vol. 8444, Ground-based and Airborne Telescopes IV,
  84440H

\bibitem[{{Morganson} {et~al.}(2012){Morganson}, {De Rosa}, {Decarli},
  {Walter}, {Chambers}, {McGreer}, {Fan}, {Burgett}, {Flewelling}, {Greiner},
  {Hodapp}, {Kaiser}, {Magnier}, {Price}, {Rix}, {Sweeney}, \&
  {Waters}}]{2012AJ....143..142M}
{Morganson}, E., {De Rosa}, G., {Decarli}, R., {et~al.} 2012, \aj, 143, 142

\bibitem[{{Morganson} {et~al.}(2015){Morganson}, {Green}, {Anderson}, {Ruan},
  {Myers}, {Eracleous}, {Kelly}, {Badenes}, {Ba{\~n}ados}, {Blanton},
  {Bershady}, {Borissova}, {Nielsen Brandt}, {Burgett}, {Chambers}, {Draper},
  {Davenport}, {Flewelling}, {Garnavich}, {Hawley}, {Hodapp}, {Isler},
  {Kaiser}, {Kinemuchi}, {Kudritzki}, {Metcalfe}, {Morgan}, {P{\^a}ris},
  {Parvizi}, {Poleski}, {Price}, {Salvato}, {Shanks}, {Schlafly}, {Schneider},
  {Shen}, {Stassun}, {Tonry}, {Walter}, \& {Waters}}]{2015ApJ...806..244M}
{Morganson}, E., {Green}, P.~J., {Anderson}, S.~F., {et~al.} 2015, \apj, 806,
  244

\bibitem[{{Morganson} {et~al.}(2016){Morganson}, {Conn}, {Rix}, {Bell},
  {Burgett}, {Chambers}, {Dolphin}, {Draper}, {Flewelling}, {Hodapp}, {Kaiser},
  {Magnier}, {Martin}, {Martinez-Delgado}, {Metcalfe}, {Schlafly}, {Slater},
  {Wainscoat}, \& {Waters}}]{2016ApJ...825..140M}
{Morganson}, E., {Conn}, B., {Rix}, H.-W., {et~al.} 2016, \apj, 825, 140

\bibitem[{{Nicholl} {et~al.}(2013){Nicholl}, {Smartt}, {Jerkstrand}, {Inserra},
  {McCrum}, {Kotak}, {Fraser}, {Wright}, {Chen}, {Smith}, {Young}, {Sim},
  {Valenti}, {Howell}, {Bresolin}, {Kudritzki}, {Tonry}, {Huber}, {Rest},
  {Pastorello}, {Tomasella}, {Cappellaro}, {Benetti}, {Mattila}, {Kankare},
  {Kangas}, {Leloudas}, {Sollerman}, {Taddia}, {Berger}, {Chornock}, {Narayan},
  {Stubbs}, {Foley}, {Lunnan}, {Soderberg}, {Sanders}, {Milisavljevic},
  {Margutti}, {Kirshner}, {Elias-Rosa}, {Morales-Garoffolo}, {Taubenberger},
  {Botticella}, {Gezari}, {Urata}, {Rodney}, {Riess}, {Scolnic}, {Wood-Vasey},
  {Burgett}, {Chambers}, {Flewelling}, {Magnier}, {Kaiser}, {Metcalfe},
  {Morgan}, {Price}, {Sweeney}, \& {Waters}}]{2013Natur.502..346N}
{Nicholl}, M., {Smartt}, S.~J., {Jerkstrand}, A., {et~al.} 2013, \nat, 502, 346

\bibitem[{{Nicholl} {et~al.}(2016){Nicholl}, {Berger}, {Smartt}, {Margutti},
  {Kamble}, {Alexander}, {Chen}, {Inserra}, {Arcavi}, {Blanchard}, {Cartier},
  {Chambers}, {Childress}, {Chornock}, {Cowperthwaite}, {Drout}, {Flewelling},
  {Fraser}, {Gal-Yam}, {Galbany}, {Harmanen}, {Holoien}, {Hosseinzadeh},
  {Howell}, {Huber}, {Jerkstrand}, {Kankare}, {Kochanek}, {Lin}, {Lunnan},
  {Magnier}, {Maguire}, {McCully}, {McDonald}, {Metzger}, {Milisavljevic},
  {Mitra}, {Reynolds}, {Saario}, {Shappee}, {Smith}, {Valenti}, {Villar},
  {Waters}, \& {Young}}]{2016ApJ...826...39N}
{Nicholl}, M., {Berger}, E., {Smartt}, S.~J., {et~al.} 2016, \apj, 826, 39

\bibitem[{{Obermeier} {et~al.}(2016){Obermeier}, {Koppenhoefer}, {Saglia},
  {Henning}, {Bender}, {Kodric}, {Deacon}, {Riffeser}, {Burgett}, {Chambers},
  {Draper}, {Flewelling}, {Hodapp}, {Kaiser}, {Kudritzki}, {Magnier},
  {Metcalfe}, {Price}, {Sweeney}, {Wainscoat}, \&
  {Waters}}]{2016A&A...587A..49O}
{Obermeier}, C., {Koppenhoefer}, J., {Saglia}, R.~P., {et~al.} 2016, \aap, 587,
  A49

\bibitem[{{Oke} \& {Gunn}(1983)}]{1983ApJ...266..713O}
{Oke}, J.~B., \& {Gunn}, J.~E. 1983, \apj, 266, 713

\bibitem[{{O'Mullane} {et~al.}(2005){O'Mullane}, {Li}, {Nieto-Santisteban},
  {Szalay}, {Thakar}, \& {Gray}}]{casjobs}
{O'Mullane}, W., {Li}, N., {Nieto-Santisteban}, M., {et~al.} 2005, Batch is
  back: CasJobs, serving multi-TB data on the Web, doi:10.1109/ICWS.2005.29

\bibitem[{{Onaka} {et~al.}(2012){Onaka}, {Rae}, {Isani}, {Tonry}, {Lee},
  {Uyeshiro}, {Robertson}, \& {Ching}}]{2012SPIE.8453E..0KO}
{Onaka}, P., {Rae}, C., {Isani}, S., {et~al.} 2012, in \procspie, Vol. 8453,
  High Energy, Optical, and Infrared Detectors for Astronomy V, 84530K

\bibitem[{{Onaka} {et~al.}(2008){Onaka}, {Tonry}, {Isani}, {Lee}, {Uyeshiro},
  {Rae}, {Robertson}, \& {Ching}}]{2008SPIE.7014E..0DO}
{Onaka}, P., {Tonry}, J.~L., {Isani}, S., {et~al.} 2008, in \procspie, Vol.
  7014, Ground-based and Airborne Instrumentation for Astronomy II, 70140D

\bibitem[{{Pastorello} {et~al.}(2010){Pastorello}, {Smartt}, {Botticella},
  {Maguire}, {Fraser}, {Smith}, {Kotak}, {Magill}, {Valenti}, {Young},
  {Gezari}, {Bresolin}, {Kudritzki}, {Howell}, {Rest}, {Metcalfe}, {Mattila},
  {Kankare}, {Huang}, {Urata}, {Burgett}, {Chambers}, {Dombeck}, {Flewelling},
  {Grav}, {Heasley}, {Hodapp}, {Kaiser}, {Luppino}, {Lupton}, {Magnier},
  {Monet}, {Morgan}, {Onaka}, {Price}, {Rhoads}, {Siegmund}, {Stubbs},
  {Sweeney}, {Tonry}, {Wainscoat}, {Waterson}, {Waters}, \&
  {Wynn-Williams}}]{2010ApJ...724L..16P}
{Pastorello}, A., {Smartt}, S.~J., {Botticella}, M.~T., {et~al.} 2010, \apjl,
  724, L16

\bibitem[{{Petrosian}(1976)}]{1976ApJ...209L...1P}
{Petrosian}, V. 1976, \apjl, 209, L1

\bibitem[{{Plazas} {et~al.}(2014){Plazas}, {Bernstein}, \&
  {Sheldon}}]{2014PASP..126..750P}
{Plazas}, A.~A., {Bernstein}, G.~M., \& {Sheldon}, E.~S. 2014, \pasp, 126, 750

\bibitem[{{Rest} {et~al.}(2014){Rest}, {Scolnic}, {Foley}, {Huber}, {Chornock},
  {Narayan}, {Tonry}, {Berger}, {Soderberg}, {Stubbs}, {Riess}, {Kirshner},
  {Smartt}, {Schlafly}, {Rodney}, {Botticella}, {Brout}, {Challis}, {Czekala},
  {Drout}, {Hudson}, {Kotak}, {Leibler}, {Lunnan}, {Marion}, {McCrum},
  {Milisavljevic}, {Pastorello}, {Sanders}, {Smith}, {Stafford}, {Thilker},
  {Valenti}, {Wood-Vasey}, {Zheng}, {Burgett}, {Chambers}, {Denneau}, {Draper},
  {Flewelling}, {Hodapp}, {Kaiser}, {Kudritzki}, {Magnier}, {Metcalfe},
  {Price}, {Sweeney}, {Wainscoat}, \& {Waters}}]{2014ApJ...795...44R}
{Rest}, A., {Scolnic}, D., {Foley}, R.~J., {et~al.} 2014, \apj, 795, 44

\bibitem[{{Saglia} {et~al.}(2012){Saglia}, {Tonry}, {Bender}, {Greisel},
  {Seitz}, {Senger}, {Snigula}, {Phleps}, {Wilman}, {Bailer-Jones}, {Klement},
  {Rix}, {Smith}, {Green}, {Burgett}, {Chambers}, {Heasley}, {Kaiser},
  {Magnier}, {Morgan}, {Price}, {Stubbs}, \& {Wainscoat}}]{2012ApJ...746..128S}
{Saglia}, R.~P., {Tonry}, J.~L., {Bender}, R., {et~al.} 2012, \apj, 746, 128

\bibitem[{{Sanders} {et~al.}(2015){Sanders}, {Soderberg}, {Gezari},
  {Betancourt}, {Chornock}, {Berger}, {Foley}, {Challis}, {Drout}, {Kirshner},
  {Lunnan}, {Marion}, {Margutti}, {McKinnon}, {Milisavljevic}, {Narayan},
  {Rest}, {Kankare}, {Mattila}, {Smartt}, {Huber}, {Burgett}, {Draper},
  {Hodapp}, {Kaiser}, {Kudritzki}, {Magnier}, {Metcalfe}, {Morgan}, {Price},
  {Tonry}, {Wainscoat}, \& {Waters}}]{2015ApJ...799..208S}
{Sanders}, N.~E., {Soderberg}, A.~M., {Gezari}, S., {et~al.} 2015, \apj, 799,
  208

\bibitem[{{Schlafly} {et~al.}(2016{\natexlab{a}}){Schlafly}, {Peek},
  {Finkbeiner}, \& {Green}}]{2016arXiv161202818S}
{Schlafly}, E.~F., {Peek}, J.~E.~G., {Finkbeiner}, D.~P., \& {Green}, G.~M.
  2016{\natexlab{a}}, ArXiv e-prints, arXiv:1612.02818

\bibitem[{{Schlafly} {et~al.}(2012){Schlafly}, {Finkbeiner}, {Juri{\'c}},
  {Magnier}, {Burgett}, {Chambers}, {Grav}, {Hodapp}, {Kaiser}, {Kudritzki},
  {Martin}, {Morgan}, {Price}, {Rix}, {Stubbs}, {Tonry}, \&
  {Wainscoat}}]{2012ApJ...756..158S}
{Schlafly}, E.~F., {Finkbeiner}, D.~P., {Juri{\'c}}, M., {et~al.} 2012, \apj,
  756, 158

\bibitem[{{Schlafly} {et~al.}(2014){Schlafly}, {Green}, {Finkbeiner}, {Rix},
  {Bell}, {Burgett}, {Chambers}, {Draper}, {Hodapp}, {Kaiser}, {Magnier},
  {Martin}, {Metcalfe}, {Price}, \& {Tonry}}]{2014ApJ...786...29S}
{Schlafly}, E.~F., {Green}, G., {Finkbeiner}, D.~P., {et~al.} 2014, \apj, 786,
  29

\bibitem[{{Schlafly} {et~al.}(2016{\natexlab{b}}){Schlafly}, {Meisner},
  {Stutz}, {Kainulainen}, {Peek}, {Tchernyshyov}, {Rix}, {Finkbeiner}, {Covey},
  {Green}, {Bell}, {Burgett}, {Chambers}, {Draper}, {Flewelling}, {Hodapp},
  {Kaiser}, {Magnier}, {Martin}, {Metcalfe}, {Wainscoat}, \&
  {Waters}}]{2016ApJ...821...78S}
{Schlafly}, E.~F., {Meisner}, A.~M., {Stutz}, A.~M., {et~al.}
  2016{\natexlab{b}}, \apj, 821, 78

\bibitem[{{Schunov{\'a}-Lilly} {et~al.}(2017){Schunov{\'a}-Lilly}, {Jedicke},
  {Vere{\v s}}, {Denneau}, \& {Wainscoat}}]{2017Icar..284..114S}
{Schunov{\'a}-Lilly}, E., {Jedicke}, R., {Vere{\v s}}, P., {Denneau}, L., \&
  {Wainscoat}, R.~J. 2017, Icarus, 284, 114

\bibitem[{{Scolnic} {et~al.}(2014){Scolnic}, {Rest}, {Riess}, {Huber}, {Foley},
  {Brout}, {Chornock}, {Narayan}, {Tonry}, {Berger}, {Soderberg}, {Stubbs},
  {Kirshner}, {Rodney}, {Smartt}, {Schlafly}, {Botticella}, {Challis},
  {Czekala}, {Drout}, {Hudson}, {Kotak}, {Leibler}, {Lunnan}, {Marion},
  {McCrum}, {Milisavljevic}, {Pastorello}, {Sanders}, {Smith}, {Stafford},
  {Thilker}, {Valenti}, {Wood-Vasey}, {Zheng}, {Burgett}, {Chambers},
  {Denneau}, {Draper}, {Flewelling}, {Hodapp}, {Kaiser}, {Kudritzki},
  {Magnier}, {Metcalfe}, {Price}, {Sweeney}, {Wainscoat}, \&
  {Waters}}]{2014ApJ...795...45S}
{Scolnic}, D., {Rest}, A., {Riess}, A., {et~al.} 2014, \apj, 795, 45

\bibitem[{{Scolnic} {et~al.}(2015){Scolnic}, {Casertano}, {Riess}, {Rest},
  {Schlafly}, {Foley}, {Finkbeiner}, {Tang}, {Burgett}, {Chambers}, {Draper},
  {Flewelling}, {Hodapp}, {Huber}, {Kaiser}, {Kudritzki}, {Magnier},
  {Metcalfe}, \& {Stubbs}}]{2015ApJ...815..117S}
{Scolnic}, D., {Casertano}, S., {Riess}, A., {et~al.} 2015, \apj, 815, 117

\bibitem[{{S{\'e}rsic}(1963)}]{1963BAAA....6...41S}
{S{\'e}rsic}, J.~L. 1963, Boletin de la Asociacion Argentina de Astronomia La
  Plata Argentina, 6, 41

\bibitem[{{Skrutskie} {et~al.}(2006){Skrutskie}, {Cutri}, {Stiening},
  {Weinberg}, {Schneider}, {Carpenter}, {Beichman}, {Capps}, {Chester},
  {Elias}, {Huchra}, {Liebert}, {Lonsdale}, {Monet}, {Price}, {Seitzer},
  {Jarrett}, {Kirkpatrick}, {Gizis}, {Howard}, {Evans}, {Fowler}, {Fullmer},
  {Hurt}, {Light}, {Kopan}, {Marsh}, {McCallon}, {Tam}, {Van Dyk}, \&
  {Wheelock}}]{2006AJ....131.1163S}
{Skrutskie}, M.~F., {Cutri}, R.~M., {Stiening}, R., {et~al.} 2006, \aj, 131,
  1163

\bibitem[{{Smartt} {et~al.}(2015){Smartt}, {Valenti}, {Fraser}, {Inserra},
  {Young}, {Sullivan}, {Pastorello}, {Benetti}, {Gal-Yam}, {Knapic},
  {Molinaro}, {Smareglia}, {Smith}, {Taubenberger}, {Yaron}, {Anderson},
  {Ashall}, {Balland}, {Baltay}, {Barbarino}, {Bauer}, {Baumont}, {Bersier},
  {Blagorodnova}, {Bongard}, {Botticella}, {Bufano}, {Bulla}, {Cappellaro},
  {Campbell}, {Cellier-Holzem}, {Chen}, {Childress}, {Clocchiatti},
  {Contreras}, {Dall'Ora}, {Danziger}, {de Jaeger}, {De Cia}, {Della Valle},
  {Dennefeld}, {Elias-Rosa}, {Elman}, {Feindt}, {Fleury}, {Gall},
  {Gonzalez-Gaitan}, {Galbany}, {Morales Garoffolo}, {Greggio}, {Guillou},
  {Hachinger}, {Hadjiyska}, {Hage}, {Hillebrandt}, {Hodgkin}, {Hsiao}, {James},
  {Jerkstrand}, {Kangas}, {Kankare}, {Kotak}, {Kromer}, {Kuncarayakti},
  {Leloudas}, {Lundqvist}, {Lyman}, {Hook}, {Maguire}, {Manulis}, {Margheim},
  {Mattila}, {Maund}, {Mazzali}, {McCrum}, {McKinnon}, {Moreno-Raya},
  {Nicholl}, {Nugent}, {Pain}, {Pignata}, {Phillips}, {Polshaw}, {Pumo},
  {Rabinowitz}, {Reilly}, {Romero-Ca{\~n}izales}, {Scalzo}, {Schmidt},
  {Schulze}, {Sim}, {Sollerman}, {Taddia}, {Tartaglia}, {Terreran},
  {Tomasella}, {Turatto}, {Walker}, {Walton}, {Wyrzykowski}, {Yuan}, \&
  {Zampieri}}]{2015A&A...579A..40S}
{Smartt}, S.~J., {Valenti}, S., {Fraser}, M., {et~al.} 2015, \aap, 579, A40

\bibitem[{{Smartt} {et~al.}(2016){Smartt}, {Chambers}, {Smith}, {Huber},
  {Young}, {Cappellaro}, {Wright}, {Coughlin}, {Schultz}, {Denneau},
  {Flewelling}, {Heinze}, {Magnier}, {Primak}, {Rest}, {Sherstyuk}, {Stalder},
  {Stubbs}, {Tonry}, {Waters}, {Willman}, {Anderson}, {Baltay}, {Botticella},
  {Campbell}, {Dennefeld}, {Chen}, {Della Valle}, {Elias-Rosa}, {Fraser},
  {Inserra}, {Kankare}, {Kotak}, {Kupfer}, {Harmanen}, {Galbany}, {Gal-Yam},
  {Le Guillou}, {Lyman}, {Maguire}, {Mitra}, {Nicholl}, {Olivares E},
  {Rabinowitz}, {Razza}, {Sollerman}, {Smith}, {Terreran}, {Valenti}, {Gibson},
  \& {Goggia}}]{2016MNRAS.462.4094S}
{Smartt}, S.~J., {Chambers}, K.~C., {Smith}, K.~W., {et~al.} 2016, \mnras, 462,
  4094

\bibitem[{{Stalder} {et~al.}(2009){Stalder}, {Chambers}, \&
  {Vacca}}]{2009ApJS..185..124S}
{Stalder}, B., {Chambers}, K.~C., \& {Vacca}, W.~D. 2009, \apjs, 185, 124

\bibitem[{{Stoughton} {et~al.}(2002){Stoughton}, {Lupton}, {Bernardi},
  {Blanton}, {Burles}, {Castander}, {Connolly}, {Eisenstein}, {Frieman},
  {Hennessy}, {Hindsley}, {Ivezi{\'c}}, {Kent}, {Kunszt}, {Lee}, {Meiksin},
  {Munn}, {Newberg}, {Nichol}, {Nicinski}, {Pier}, {Richards}, {Richmond},
  {Schlegel}, {Smith}, {Strauss}, {SubbaRao}, {Szalay}, {Thakar}, {Tucker},
  {Vanden Berk}, {Yanny}, {Adelman}, {Anderson}, {Anderson}, {Annis},
  {Bahcall}, {Bakken}, {Bartelmann}, {Bastian}, {Bauer}, {Berman},
  {B{\"o}hringer}, {Boroski}, {Bracker}, {Briegel}, {Briggs}, {Brinkmann},
  {Brunner}, {Carey}, {Carr}, {Chen}, {Christian}, {Colestock}, {Crocker},
  {Csabai}, {Czarapata}, {Dalcanton}, {Davidsen}, {Davis}, {Dehnen},
  {Dodelson}, {Doi}, {Dombeck}, {Donahue}, {Ellman}, {Elms}, {Evans}, {Eyer},
  {Fan}, {Federwitz}, {Friedman}, {Fukugita}, {Gal}, {Gillespie}, {Glazebrook},
  {Gray}, {Grebel}, {Greenawalt}, {Greene}, {Gunn}, {de Haas}, {Haiman},
  {Haldeman}, {Hall}, {Hamabe}, {Hansen}, {Harris}, {Harris}, {Harvanek},
  {Hawley}, {Hayes}, {Heckman}, {Helmi}, {Henden}, {Hogan}, {Hogg}, {Holmgren},
  {Holtzman}, {Huang}, {Hull}, {Ichikawa}, {Ichikawa}, {Johnston}, {Kauffmann},
  {Kim}, {Kimball}, {Kinney}, {Klaene}, {Kleinman}, {Klypin}, {Knapp},
  {Korienek}, {Krolik}, {Kron}, {Krzesi{\'n}ski}, {Lamb}, {Leger},
  {Limmongkol}, {Lindenmeyer}, {Long}, {Loomis}, {Loveday}, {MacKinnon},
  {Mannery}, {Mantsch}, {Margon}, {McGehee}, {McKay}, {McLean}, {Menou},
  {Merelli}, {Mo}, {Monet}, {Nakamura}, {Narayanan}, {Nash}, {Neilsen},
  {Newman}, {Nitta}, {Odenkirchen}, {Okada}, {Okamura}, {Ostriker}, {Owen},
  {Pauls}, {Peoples}, {Peterson}, {Petravick}, {Pope}, {Pordes}, {Postman},
  {Prosapio}, {Quinn}, {Rechenmacher}, {Rivetta}, {Rix}, {Rockosi}, {Rosner},
  {Ruthmansdorfer}, {Sandford}, {Schneider}, {Scranton}, {Sekiguchi}, {Sergey},
  {Sheth}, {Shimasaku}, {Smee}, {Snedden}, {Stebbins}, {Stubbs}, {Szapudi},
  {Szkody}, {Szokoly}, {Tabachnik}, {Tsvetanov}, {Uomoto}, {Vogeley}, {Voges},
  {Waddell}, {Walterbos}, {Wang}, {Watanabe}, {Weinberg}, {White}, {White},
  {Wilhite}, {Wolfe}, {Yasuda}, {York}, {Zehavi}, \&
  {Zheng}}]{2002AJ....123..485S}
{Stoughton}, C., {Lupton}, R.~H., {Bernardi}, M., {et~al.} 2002, \aj, 123, 485

\bibitem[{{Stubbs} {et~al.}(2010){Stubbs}, {Doherty}, {Cramer}, {Narayan},
  {Brown}, {Lykke}, {Woodward}, \& {Tonry}}]{2010ApJS..191..376S}
{Stubbs}, C.~W., {Doherty}, P., {Cramer}, C., {et~al.} 2010, \apjs, 191, 376

\bibitem[{{Szapudi} {et~al.}(2015){Szapudi}, {Kov{\'a}cs}, {Granett}, {Frei},
  {Silk}, {Burgett}, {Cole}, {Draper}, {Farrow}, {Kaiser}, {Magnier},
  {Metcalfe}, {Morgan}, {Price}, {Tonry}, \& {Wainscoat}}]{Szapudi15}
{Szapudi}, I., {Kov{\'a}cs}, A., {Granett}, B.~R., {et~al.} 2015, \mnras, 450,
  288

\bibitem[{{Thakar} \& {Li}(2008)}]{10.1109/MCSE.2008.6}
{Thakar}, A.~R., \& {Li}, N. 2008, Computing in Science \& Engineering, 10, 18

\bibitem[{{Thakar} {et~al.}(2003){Thakar}, {Szalay}, {Vandenberg}, {Gray}, \&
  {Stoughton}}]{2003ASPC..295..217T}
{Thakar}, A.~R., {Szalay}, A.~S., {Vandenberg}, J.~V., {Gray}, J., \&
  {Stoughton}, A.~S. 2003, in Astronomical Society of the Pacific Conference
  Series, Vol. 295, Astronomical Data Analysis Software and Systems XII, ed.
  H.~E. {Payne}, R.~I. {Jedrzejewski}, \& R.~N. {Hook}, 217

\bibitem[{{Toba} {et~al.}(2015){Toba}, {Nagao}, {Strauss}, {Aoki}, {Goto},
  {Imanishi}, {Kawaguchi}, {Terashima}, {Ueda}, {Bosch}, {Bundy}, {Doi},
  {Inami}, {Komiyama}, {Lupton}, {Matsuhara}, {Matsuoka}, {Miyazaki},
  {Morokuma}, {Nakata}, {Oi}, {Onoue}, {Oyabu}, {Price}, {Tait}, {Takata},
  {Tanaka}, {Terai}, {Turner}, {Uchida}, {Usuda}, {Utsumi}, {Yamada}, \&
  {Wang}}]{2015PASJ...67...86T}
{Toba}, Y., {Nagao}, T., {Strauss}, M.~A., {et~al.} 2015, \pasj, 67, 86

\bibitem[{{Tonry} {et~al.}(1997){Tonry}, {Burke}, \&
  {Schechter}}]{1997PASP..109.1154T}
{Tonry}, J., {Burke}, B.~E., \& {Schechter}, P.~L. 1997, \pasp, 109, 1154

\bibitem[{{Tonry} {et~al.}(2016){Tonry}, {Denneau}, {Stalder}, {Heinze},
  {Sherstyuk}, {Rest}, {Smith}, \& {Smartt}}]{2016ATel.8680....1T}
{Tonry}, J., {Denneau}, L., {Stalder}, B., {et~al.} 2016, The Astronomer's
  Telegram, 8680

\bibitem[{{Tonry} {et~al.}(2006){Tonry}, {Onaka}, {Luppino}, \&
  {Isani}}]{2006amos.confE..47T}
{Tonry}, J., {Onaka}, P., {Luppino}, G., \& {Isani}, S. 2006, in The Advanced
  Maui Optical and Space Surveillance Technologies Conference, E47

\bibitem[{{Tonry} {et~al.}(2008){Tonry}, {Burke}, {Isani}, {Onaka}, \&
  {Cooper}}]{2008SPIE.7021E..05T}
{Tonry}, J.~L., {Burke}, B.~E., {Isani}, S., {Onaka}, P.~M., \& {Cooper}, M.~J.
  2008, in \procspie, Vol. 7021, High Energy, Optical, and Infrared Detectors
  for Astronomy III, 702105

\bibitem[{{Tonry} {et~al.}(2012{\natexlab{a}}){Tonry}, {Stubbs}, {Kilic},
  {Flewelling}, {Deacon}, {Chornock}, {Berger}, {Burgett}, {Chambers},
  {Kaiser}, {Kudritzki}, {Hodapp}, {Magnier}, {Morgan}, {Price}, \&
  {Wainscoat}}]{2012ApJ...745...42T}
{Tonry}, J.~L., {Stubbs}, C.~W., {Kilic}, M., {et~al.} 2012{\natexlab{a}},
  \apj, 745, 42

\bibitem[{{Tonry} {et~al.}(2012{\natexlab{b}}){Tonry}, {Stubbs}, {Lykke},
  {Doherty}, {Shivvers}, {Burgett}, {Chambers}, {Hodapp}, {Kaiser},
  {Kudritzki}, {Magnier}, {Morgan}, {Price}, \&
  {Wainscoat}}]{2012ApJ...750...99T}
{Tonry}, J.~L., {Stubbs}, C.~W., {Lykke}, K.~R., {et~al.} 2012{\natexlab{b}},
  \apj, 750, 99

\bibitem[{{Venemans} {et~al.}(2015){Venemans}, {Ba{\~n}ados}, {Decarli},
  {Farina}, {Walter}, {Chambers}, {Fan}, {Rix}, {Schlafly}, {McMahon},
  {Simcoe}, {Stern}, {Burgett}, {Draper}, {Flewelling}, {Hodapp}, {Kaiser},
  {Magnier}, {Metcalfe}, {Morgan}, {Price}, {Tonry}, {Waters}, {AlSayyad},
  {Banerji}, {Chen}, {Gonz{\'a}lez-Solares}, {Greiner}, {Mazzucchelli},
  {McGreer}, {Miller}, {Reed}, \& {Sullivan}}]{2015ApJ...801L..11V}
{Venemans}, B.~P., {Ba{\~n}ados}, E., {Decarli}, R., {et~al.} 2015, \apjl, 801,
  L11

\bibitem[{{Vere{\v s}} {et~al.}(2015){Vere{\v s}}, {Jedicke}, {Fitzsimmons},
  {Denneau}, {Granvik}, {Bolin}, {Chastel}, {Wainscoat}, {Burgett}, {Chambers},
  {Flewelling}, {Kaiser}, {Magnier}, {Morgan}, {Price}, {Tonry}, \&
  {Waters}}]{2015Icar..261...34V}
{Vere{\v s}}, P., {Jedicke}, R., {Fitzsimmons}, A., {et~al.} 2015, Icarus, 261,
  34

\bibitem[{{Vinsen} \& {Thilker}(2013)}]{2013A&C.....3....1V}
{Vinsen}, K., \& {Thilker}, D. 2013, Astronomy and Computing, 3, 1

\bibitem[{{Waters} {et~al.}(2016){Waters}, {Magnier}, {Price}, {Chambers},
  {Draper}, {Flewelling}, {Hodapp}, {Huber}, {Jedicke}, {Kaiser}, {Kudritzki},
  {Lupton}, {Metcalfe}, {Rest}, {Sweeney}, {Tonry}, {Wainscoat}, {Wood-Vasey},
  \& {Builders}}]{waters2017}
{Waters}, C.~Z., {Magnier}, E.~A., {Price}, P.~A., {et~al.} 2016, ArXiv
  e-prints, arXiv:1612.05245

\bibitem[{{Weryk} {et~al.}(2016){Weryk}, {Lilly}, {Chastel}, {Denneau},
  {Jedicke}, {Magnier}, {Wainscoat}, {Chambers}, {Flewelling}, {Huber},
  {Waters}, \& {PS1 Builders}}]{2016arXiv160704895W}
{Weryk}, R.~J., {Lilly}, E., {Chastel}, S., {et~al.} 2016, ArXiv e-prints,
  arXiv:1607.04895

\bibitem[{{Wright} {et~al.}(2015){Wright}, {Smartt}, {Smith}, {Miller},
  {Kotak}, {Rest}, {Burgett}, {Chambers}, {Flewelling}, {Hodapp}, {Huber},
  {Jedicke}, {Kaiser}, {Metcalfe}, {Price}, {Tonry}, {Wainscoat}, \&
  {Waters}}]{2015MNRAS.449..451W}
{Wright}, D.~E., {Smartt}, S.~J., {Smith}, K.~W., {et~al.} 2015, \mnras, 449,
  451

\bibitem[{{York} {et~al.}(2000){York}, {Adelman}, {Anderson}, {Anderson},
  {Annis}, {Bahcall}, {Bakken}, {Barkhouser}, {Bastian}, {Berman}, {Boroski},
  {Bracker}, {Briegel}, {Briggs}, {Brinkmann}, {Brunner}, {Burles}, {Carey},
  {Carr}, {Castander}, {Chen}, {Colestock}, {Connolly}, {Crocker}, {Csabai},
  {Czarapata}, {Davis}, {Doi}, {Dombeck}, {Eisenstein}, {Ellman}, {Elms},
  {Evans}, {Fan}, {Federwitz}, {Fiscelli}, {Friedman}, {Frieman}, {Fukugita},
  {Gillespie}, {Gunn}, {Gurbani}, {de Haas}, {Haldeman}, {Harris}, {Hayes},
  {Heckman}, {Hennessy}, {Hindsley}, {Holm}, {Holmgren}, {Huang}, {Hull},
  {Husby}, {Ichikawa}, {Ichikawa}, {Ivezi{\'c}}, {Kent}, {Kim}, {Kinney},
  {Klaene}, {Kleinman}, {Kleinman}, {Knapp}, {Korienek}, {Kron}, {Kunszt},
  {Lamb}, {Lee}, {Leger}, {Limmongkol}, {Lindenmeyer}, {Long}, {Loomis},
  {Loveday}, {Lucinio}, {Lupton}, {MacKinnon}, {Mannery}, {Mantsch}, {Margon},
  {McGehee}, {McKay}, {Meiksin}, {Merelli}, {Monet}, {Munn}, {Narayanan},
  {Nash}, {Neilsen}, {Neswold}, {Newberg}, {Nichol}, {Nicinski}, {Nonino},
  {Okada}, {Okamura}, {Ostriker}, {Owen}, {Pauls}, {Peoples}, {Peterson},
  {Petravick}, {Pier}, {Pope}, {Pordes}, {Prosapio}, {Rechenmacher}, {Quinn},
  {Richards}, {Richmond}, {Rivetta}, {Rockosi}, {Ruthmansdorfer}, {Sandford},
  {Schlegel}, {Schneider}, {Sekiguchi}, {Sergey}, {Shimasaku}, {Siegmund},
  {Smee}, {Smith}, {Snedden}, {Stone}, {Stoughton}, {Strauss}, {Stubbs},
  {SubbaRao}, {Szalay}, {Szapudi}, {Szokoly}, {Thakar}, {Tremonti}, {Tucker},
  {Uomoto}, {Vanden Berk}, {Vogeley}, {Waddell}, {Wang}, {Watanabe},
  {Weinberg}, {Yanny}, {Yasuda}, \& {SDSS Collaboration}}]{2000AJ....120.1579Y}
{York}, D.~G., {Adelman}, J., {Anderson}, Jr., J.~E., {et~al.} 2000, \aj, 120,
  1579

\bibitem[{{Zheng} {et~al.}(2015){Zheng}, {Thilker}, {Heckman}, {Meurer},
  {Burgett}, {Chambers}, {Huber}, {Kaiser}, {Magnier}, {Metcalfe}, {Price},
  {Tonry}, {Wainscoat}, \& {Waters}}]{2015ApJ...800..120Z}
{Zheng}, Z., {Thilker}, D.~A., {Heckman}, T.~M., {et~al.} 2015, \apj, 800, 120

\end{thebibliography}


\end{document}